%% file: DiffLumiEPJC.tex
\RequirePackage{fix-cm}
\documentclass[twocolumn,epjc3]{svjour3}  
\pdfoutput=1
\smartqed  
\RequirePackage{graphicx}
\RequirePackage{mathptmx}      
\include{packages}
\include{definitions}
\journalname{Eur. Phys. J. C}

\begin{document}

\title{Luminosity Spectrum Reconstruction at Linear Colliders
}


\author{St{\'e}phane Poss\thanksref{e1,addr1}
        \and
        Andr{\'e} Sailer\thanksref{e2,addr1} 
}

\thankstext{e1}{e-mail: stephane.poss@cern.ch}
\thankstext{e2}{e-mail: andre.sailer@cern.ch}


\institute{CERN, 1211 Geneva 23, Switzerland\label{addr1}
}

\date{Received: date / Accepted: date}

\maketitle

\begin{abstract}
\noindent
A good knowledge of the luminosity spectrum is mandatory for many measurements
at future \epem colliders. As the beam-parameters determining the luminosity
spectrum cannot be measured precisely, the luminosity spectrum has to be
measured through a gauge process with the detector. The measured distributions,
used to reconstruct the spectrum, depend on Initial State Radiation,
cross-section, and Final State Radiation. To extract the basic luminosity
spectrum, a parametric model of the luminosity spectrum is created, in this case
the spectrum at the 3~TeV Compact Linear Collider (CLIC). The model is used within a reweighting technique to
extract the luminosity spectrum from measured Bhabha event observables, taking
all relevant effects into account. The centre-of-mass energy spectrum is
reconstructed within 5\% over the full validity range of the model. The
reconstructed spectrum does not result in a significant bias or systematic
uncertainty in the exemplary physics benchmark process of smuon pair production.
\end{abstract}

\input{introduction}

\input{lumispectrum}
\input{beamenergyspread}
\input{beamstrahlung}
\input{model}
\input{fit}
\input{initialElectronFit}
\input{bhabhaFit}
\input{physperfs}
\input{summary}


\bibliographystyle{spphys}
\bibliography{DiffLumiBib}

\end{document}

%% file: packages.tex
\usepackage{pdf14}
\usepackage{xspace}
\usepackage{upgreek}
\usepackage[captionskip=-7pt
           ]{subfig}
\usepackage{amsmath}
\usepackage{mathtools}
\usepackage{booktabs}
\usepackage[point]{rccol}
\usepackage{xfrac}
\usepackage{rotating}


%% file: definitions.tex
\newcommand{\sideCwidth}{0.33\linewidth}
\newcommand{\halfwidth}{0.33\textwidth}

\newcommand{\dd}[1]{\ensuremath{\mathrm{d}#1}} 
\newcommand{\conv}{\ensuremath{\otimes}}

\newcommand{\snom}{\ensuremath{s_{\mathrm{nom}}}\xspace}
\newcommand{\roots}{\ensuremath{\sqrt{s}}\xspace}
\newcommand{\rootsnom}{\ensuremath{\sqrt{\smash[b]{\snom}}}\xspace}
\newcommand{\rootsprime}{\ensuremath{\sqrt{s^{\scriptscriptstyle\smash\prime}}}\xspace}

\newcommand{\rootsaco}{\ensuremath{\sqrt{\smash[b]{s^{\raisebox{-0.8pt}{$\scriptscriptstyle\smash\prime$}}_{\smash[t]{\mathrm{acol}}}}}}\xspace}
\newcommand{\rootsvec}{\ensuremath{\sqrt{\smash[b]{s^{\raisebox{-0.8pt}{$\scriptscriptstyle\smash\prime$}}_{\smash[t]{\mathrm{4\text{-}vec}}}}}}\xspace}

\newcommand{\lumispec}[1]{\ensuremath{\mathcal{L}\bigl(#1\bigr)}\xspace}
\newcommand{\lumispecNoArg}{\ensuremath{\mathcal{L}}\xspace}

\newcommand{\lumispecscale}[1]{\ensuremath{\mathcal{L}_{\mathrm{scaled}}\bigl(#1\bigr)}\xspace}

\newcommand{\Lint}{\ensuremath{L_\mathrm{int}}\xspace}
\newcommand{\Neff}[1]{\ensuremath{N_{\mathrm{eff}}(#1)}\xspace}

\newcommand{\sigmaeffNoArg}{\ensuremath{\sigma_{\mathrm{Eff}}}\xspace}

\newcommand{\degrees}{\ensuremath{^\circ}\xspace}
\newcommand{\picob}{\ensuremath{\mathrm{pb}}\xspace}
\newcommand{\fbinv}{\ensuremath{\mathrm{fb}^{-1}}\xspace}

\newcommand{\microrad}{\ensuremath{\upmu\mathrm{rad}}\xspace}

\newcommand{\gauss}{\ensuremath{g}}
\newcommand{\bes}{\ensuremath{\mathrm{BES}}\xspace}
\newcommand{\betaD}{\ensuremath{b}\xspace}

\newcommand{\Gauss}[1]{\ensuremath{\gauss\bigl(#1\bigr)}}

\newcommand{\Bes}[1]{\ensuremath{\bes\bigl(#1\bigr)}\xspace}

\newcommand{\BetaD}[1]{\ensuremath{\betaD\bigl(#1\bigr)}\xspace}

\newcommand{\blimit}[1]{\ensuremath{\beta_{\mathrm{Limit}}^{#1}}\xspace}

\newcommand{\BetaGauss}[1]{\ensuremath{\left(\betaD \conv \gauss \right) (#1)}\xspace}
\newcommand{\betagauss}[1]{\ensuremath{\mathrm{BG}\bigl( #1 \bigr)}\xspace}

\newcommand{\betabes}[1]{\ensuremath{\mathrm{BB}\bigl( #1 \bigr)}\xspace}
\newcommand{\BetaBes}[1]{\ensuremath{\left( \betaD \conv \bes \right) (#1)}\xspace}

\newcommand{\nBeta}{N_{\mathrm{Beta}}\xspace}

\newcommand{\varEv}{\ensuremath{i}\xspace}
\newcommand{\varBin}{\ensuremath{j}\xspace}

\newcommand{\nGuinea}{\ensuremath{N_{\mathrm{GP}}}\xspace}
\newcommand{\dGuinea}{\ensuremath{\sigma_{\mathrm{GP}}}\xspace}

\newcommand{\nModel}{\ensuremath{N_{\mathrm{Model}}}\xspace}
\newcommand{\dModel}{\ensuremath{\sigma_{\mathrm{Model}}}\xspace}

\newcommand{\peak}{\emph{Peak}\xspace}
\newcommand{\arms}{\emph{Arms}\xspace}
\newcommand{\body}{\emph{Body}\xspace}

\newcommand{\scaleFac}{\ensuremath{f_{S}}\xspace}

\newcommand{\parset}[1]{\ensuremath{[p]^{#1}}}
\newcommand{\pregion}[1]{\ensuremath{p_{\mathrm{region}}^{#1}}\xspace}

\newcommand{\ele}{1}
\newcommand{\pos}{2}

\newcommand{\xe}{\ensuremath{x_{\ele}}\xspace}
\newcommand{\xp}{\ensuremath{x_{\pos}}\xspace}
\newcommand{\xep}{\ensuremath{x_{\ele,\pos}}\xspace}
\newcommand{\xs}{\ensuremath{x}\xspace}
\newcommand{\xStrahlung}[1]{\ensuremath{x^{#1}_{\mathrm{Strahl}}}\xspace}
\newcommand{\xSpread}[1]{\ensuremath{{x^{#1}_{\mathrm{Spread}}}}\xspace}
\newcommand{\xGauss}[1]{\ensuremath{{x^{#1}_{\mathrm{G}}}}\xspace}

\newcommand{\pepeak}{\ensuremath{\parset{\ele}_{\mathrm{Peak}}}\xspace}
\newcommand{\pppeak}{\ensuremath{\parset{\pos}_{\mathrm{Peak}}}\xspace}

\newcommand{\pearmA}{\ensuremath{\parset{\ele}_{\mathrm{Arm1}}}\xspace}
\newcommand{\pparmA}{\ensuremath{\parset{\pos}_{\mathrm{Arm1}}}\xspace}

\newcommand{\pearmB}{\ensuremath{\parset{\ele}_{\mathrm{Arm2}}}\xspace}
\newcommand{\pparmB}{\ensuremath{\parset{\pos}_{\mathrm{Arm2}}}\xspace}

\newcommand{\pebody}{\ensuremath{\parset{\ele}_{\smash[b]{\mathrm{Body}}}}\xspace}
\newcommand{\ppbody}{\ensuremath{\parset{\pos}_{\smash[b]{\mathrm{Body}}}}\xspace}

\newcommand{\EBeam}{\ensuremath{E_{\mathrm{Beam}}}\xspace}
\newcommand{\EParticle}{\ensuremath{E_{\mathrm{Particle}}}\xspace}

\newcommand{\tabt}[1]{\multicolumn{1}{c}{#1}}
\newcommand{\tabtt}[1]{\multicolumn{2}{c}{#1}}
\newcommand{\tabttt}[1]{\multicolumn{3}{c}{#1}}

\newcommand{\tabempty}{\multicolumn{1}{c}{}}

\newcommand{\Subref}[1]{\protect\subref{#1}}

\newcommand{\guineapig}{\textsc{Guin\-ea\-Pig}\xspace}
\newcommand{\model}{\emph{Model}\xspace}
\newcommand{\bhwide}{\textsc{Bhwide}\xspace}

\newcommand{\geant}{\textsc{Geant4}\xspace}
\newcommand{\minuit}{\textsc{Minuit}\xspace}
\newcommand{\Root}{\textsc{Root}\xspace}

\newcommand{\beamstrahlungpara}[2]{\ensuremath{\eta^{#1}_{#2}}\xspace}
\newcommand{\beamspreadpara}[2]{\ensuremath{\omega^{#1}_{#2}}\xspace}

\newcommand{\xmin}{\ensuremath{x_{\min}}\xspace}
\newcommand{\xmax}{\ensuremath{x_{\max}}\xspace}

\newcommand{\pPeak}{\ensuremath{p_{\mathrm{Peak}}}\xspace}
\newcommand{\pArmA}{\ensuremath{p_{\mathrm{Arm1}}}\xspace}
\newcommand{\pArmB}{\ensuremath{p_{\mathrm{Arm2}}}\xspace}
\newcommand{\pBody}{\ensuremath{p_{{\mathrm{Body}}}}\xspace}
\newcommand{\betaA}{\ensuremath{\mathrm{a}}}
\newcommand{\betaB}{\ensuremath{\mathrm{b}}}
\newcommand{\aarmAA}{\ensuremath{\beamstrahlungpara{\betaA}{\mathrm{Arm1}}}\xspace}
\newcommand{\aarmAB}{\ensuremath{\beamstrahlungpara{\betaB}{\mathrm{Arm1}}}\xspace}
\newcommand{\bPeakAA}{\ensuremath{\beamspreadpara{\betaA}{\mathrm{Peak1}}}\xspace}
\newcommand{\bPeakAB}{\ensuremath{\beamspreadpara{\betaB}{\mathrm{Peak1}}}\xspace}
\newcommand{\aarmBA}{\ensuremath{\beamstrahlungpara{\betaA}{\mathrm{Arm2}} }\xspace}
\newcommand{\aarmBB}{\ensuremath{\beamstrahlungpara{\betaB}{\mathrm{Arm2}} }\xspace}
\newcommand{\bPeakBA}{\ensuremath{\beamspreadpara  {\betaA}{\mathrm{Peak2}}}\xspace}
\newcommand{\bPeakBB}{\ensuremath{\beamspreadpara  {\betaB}{\mathrm{Peak2}}}\xspace}
\newcommand{\abodyAA}{\ensuremath{\beamstrahlungpara{\betaA}{\mathrm{Body1}}}\xspace}
\newcommand{\abodyAB}{\ensuremath{\beamstrahlungpara{\betaB}{\mathrm{Body1}}}\xspace}
\newcommand{\abodyBA}{\ensuremath{\beamstrahlungpara{\betaA}{\mathrm{Body2}}}\xspace}
\newcommand{\abodyBB}{\ensuremath{\beamstrahlungpara{\betaB}{\mathrm{Body2}}}\xspace}
\newcommand{\barmAA}{\ensuremath{\beamspreadpara{\betaA}{\mathrm{Arm1}}}\xspace}
\newcommand{\barmAB}{\ensuremath{\beamspreadpara{\betaB}{\mathrm{Arm1}}}\xspace}
\newcommand{\barmBA}{\ensuremath{\beamspreadpara{\betaA}{\mathrm{Arm2}}}\xspace}
\newcommand{\barmBB}{\ensuremath{\beamspreadpara{\betaB}{\mathrm{Arm2}}}\xspace}

\newcommand{\ndf}{\ensuremath{\mathrm{ndf}}\xspace}
\newcommand{\chindf}{\ensuremath{\chi^{2}/\ndf}\xspace}
\newcommand{\chisquare}{\ensuremath{\chi^{2}}\xspace}

\newcommand{\gghadron}{\mbox{\ensuremath{\upgamma\upgamma\to\mathrm{hadron}}}\xspace}

\makeatletter
\newcommand\figcaption{\def\@captype{figure}\caption}
\newcommand\tabcaption{\def\@captype{table}\caption}
\makeatother

\newcommand{\spectraPhrase}{Basic luminosity spectrum from \guineapig compared
  to the \model after the fit. The integral over the full range is normalized.\xspace}

\newcommand{\ratioPhrase}{Relative difference between \guineapig
  and the \model after the fit. The horizontal error bars mark the statistical uncertainty from
  \guineapig, the additional error bars mark the uncertainty from the
  parameters.\xspace}

\newcommand{\Emu}{\ensuremath{E_{\upmu}}\xspace}
\newcommand{\EL}[1]{\ensuremath{E_\mathrm{L}\bigl(#1\bigr)}\xspace}
\newcommand{\EH}[1]{\ensuremath{E_\mathrm{H}\bigl(#1\bigr)}\xspace}

\newcommand{\Sigmax}[1]{\ensuremath{\sigma_{x}\bigl(#1\bigr)}\xspace}

\newcommand{\Sigmasmu}[1]{\ensuremath{\sigma_{\tilde{\upmu}\tilde{\upmu}}\bigl(#1\bigr)}\xspace}

\newcommand{\msmu}{\ensuremath{m_{\tilde{\upmu}^{\pm}}}\xspace}
\newcommand{\mneutr}{\ensuremath{m_{\tilde{\upchi}^{0}}}\xspace}
\newcommand{\Uni}[1]{\ensuremath{\mathrm{U}\bigl(#1\bigr)}\xspace}
\newcommand{\Det}[1]{\ensuremath{\mathrm{D}\bigl(#1\bigr)}\xspace}
\newcommand{\ISR}[1]{\ensuremath{\mathrm{ISR}\bigl(#1\bigr)}\xspace}
\newcommand{\FSR}[1]{\ensuremath{\mathrm{FSR}\bigl(#1\bigr)}\xspace}

\newcommand{\epem}{\ensuremath{\mathrm{e}^{+}\mathrm{e}^{-}}\xspace}

\newcommand{\mplus}[1]{\ensuremath{m^{+}_{#1}}\xspace}
\newcommand{\mminus}[1]{\ensuremath{m^{-}_{#1}}\xspace}



%% file: introduction.tex
\section{Introduction}
\label{sec:introduction}

Small, nanometre-sized beams are necessary to reach the required luminosity at
future linear colliders. Together with the high energy, the small beams cause
large electromagnetic fields during the bunch crossing. These intense fields at
the interaction point squeeze the beams. This so-called \emph{pinch effect}
increases the instantaneous luminosity. However, the deflection of the particles
also leads to the emission of Beamstrahlung photons -- which reduce the nominal
energy of colliding particles -- and produces collisions below the nominal
centre-of-mass
energy~\cite{chen_beam,Chen:242895,Schroeder:216344,schulte1996}. The resulting
spectrum of centre-of-mass energies is traditionally called the \emph{luminosity
  spectrum}\footnote{The luminosity spectrum is a dimensionless probability
  density function that is mathematically equivalent to the use of electron structure
  functions and parton density functions.} \cite{schulte1996,frary91:proc,toomi1998,shibata200712}.

The knowledge of the shape of this luminosity spectrum is mandatory for
the precision measurements in which a cross-section has to be known. While
the cross-section depends on the centre-of-mass energy, the observables measured
in the lab frame also depend on the difference in energy of the colliding
electrons\footnote{Unless explicitly stated, \emph{electron} always refers to both
  electrons and positrons.}, which determines the Lo\-rentz boost of the system.

Unlike the electron structure functions (i.e, Initial State Radiation (ISR)) -- which
can be calculated precisely -- the beam-beam forces, and therefore the
Beamstrahlung, highly depend on the geometry of the colliding bunches. The
actual beam-beam interaction taking place at the interaction point cannot be
precisely simulated, because the geometry of the bunches cannot be measured.
Therefore, the luminosity spectrum at the interaction point has to be measured
using a physics channel with well known properties, e.g., Bhabha scattering.

The observables measured in the events are affected by detector resolutions. The
distributions used for the reconstruction of the luminosity spectrum are also dependent
on the cross-section of the process, and Initial and Final State Radiation
(FSR). All effects have to be taken into account for the reconstruction of the
luminosity spectrum.

It was pointed out by Frary and Miller~\cite{frary91:proc} that a precise
reconstruction of the peak of the luminosity spectrum, necessary for a top-quark
threshold scan, can only be achieved with a measurement of the angles of the
outgoing electrons from Bhabha scattering. The angles of the two particles are
the most precisely measurable observable~\cite{frary91:proc}. The angles of the
outgoing electrons -- or rather the acollinearity between the two particles --
are sufficient to extract a relative centre-of-mass energy, which gives access
to the luminosity spectrum.

Toomi et al.~\cite{toomi1998} showed that the reconstruction of a parameterised
luminosity spectrum is possible using a template fit. Their parameterisation
only used three parameters to describe the effective centre-of-mass energy
spectrum. However, as the boost of the initial system and correlation between
the energies of the two particles cannot be neglec\-ted~\cite{moenig:DiffLumi}, a
description of the energies of the pairs of colliding particles is necessary.
Correlations exist between the two particle energies because the probability to
emit beamstrahlung depends on the distance travelled in the field of the
opposite bunch, and the field strength depends on the position inside the
bunch. As two particles can only collide, when they are in similar position in
their respective bunches the energy between two particles is correlated.

The relative centre-of-mass energy, that is reconstructed from the acollinearity,
is equal to unity for back-to-back particles, and always smaller than unity
for larger acollinearity, regardless whether one of the particles has a higher
or lower energy than nominal. Therefore, Shibata et al.~\cite{shibata200712} proposed to
calculate the distribution of the four-vectors of the Bhabha electrons and
extract the luminosity spectrum with the iterative Expectation--Maximisation algorithm.
They have, however, considered neither detector resolutions, nor Initial and
Final State Radiation.  For a full description of the outgoing Bhabha electrons,
the luminosity spectrum would have to be weighted with the Bhabha cross-section
and convoluted with the detector resolutions, which would require a huge
computational effort, when using their method.

A reconstruction of the energy of the particle pairs was done for the 500~GeV
ILC~\cite{sailer}. The acollinearity and the energies of the electrons
measured in the calorimeter were used in a reweighting fit to reconstruct the
luminosity spectrum. The parameterisation -- necessary for the reweighting fit --
accounted for the correlation between the two beams and the beam-energy spread.

This paper follows the approach of the 500~GeV ILC study~\cite{sailer}, extends
it, and applies it to the luminosity spectrum of the 3~TeV
CLIC~\cite{CLICCDR_vol1}, which is the most challenging luminosity
spectrum. This paper is structured as follows: in
Section~\ref{sec:lumin-spectr-bhabha} the basic and cross-section scaled
luminosity spectrum are defined. The Bhabha scattering and observables used for
the reconstruction are also introduced. In Section~\ref{sec:modell-lumin-spectr}
the model of the luminosity spectrum, required to perform a reweighting
fit, is constructed. The reweighting technique is explained in
Section~\ref{sec:reweighting-fit}, and in Section~\ref{sec:lumin-spectr-reconst}
it is applied to first validate the model against the luminosity spectrum at the
3~TeV CLIC; then all the relevant effects leading to the measured observables
are included, and the luminosity spectrum is reconstructed from these
distributions. In Section~\ref{sec:physperfs} the impact of the reconstructed
luminosity spectrum on the measurement of the masses of supersymmetric particles
in a CLIC benchmark process is estimated. The paper closes with a summary,
conclusions, and outlook in Section~\ref{sec:sco}.


%% file: lumispectrum.tex
\section{Luminosity Spectrum, Bhabha Scattering, and the Measurement}
\label{sec:lumin-spectr-bhabha}

The nominal centre-of-mass energy \rootsnom of a collider with two beams with the
nominal beam energy \EBeam is given by \mbox{$\rootsnom = 2\EBeam$}. If the two interacting
particles carry only a fraction of the nominal beam energy
$\xep=E_{1,2}/\EBeam$, the effective centre-of-mass energy is
\begin{equation}
\label{eq:sprime}
  \rootsprime = 2 \EBeam \sqrt{\smash[b]{\xe\xp}}.
\end{equation}
The \emph{basic} luminosity spectrum \lumispec{\xs} describes either the distribution of the
fraction of centre-of-mass energies $\xs=\rootsprime/\rootsnom$.
or the distribution of the fraction of energies of
colliding particles \lumispec{\xe,\xp} prior to hard collisions and
prior to Initial State Radiation. The two functions are connected via the
integral along the lines of constant centre-of-mass energies, given by
Equation~\eqref{eq:sprime}. Therefore,
\begin{equation}
  \lumispec{\xs} =  \iint\limits_{0}^{~~\xs_{\max}}\!\dd{\xe}\dd{\xp}\,\delta\bigl(\xs - \sqrt{\xe\xp}\bigr)\lumispec{\xe,\xp}.
\end{equation}

The luminosity spectrum affects all centre-of-mass energy dependent observables.
For example, the luminosity spectrum has to be used to predict the inclusive
(i.e., observed) cross-section $\sigma_{\mathrm{Eff}}^{\mathrm{Machine}}$ at
the machine. The principle is the same as for the parton density functions at
hadron machines, except that the luminosity spectrum depends on the machine and
not only on the colliding particles.  To calculate the effective cross-section
the differential cross-section is weighted with the luminosity spectrum, 
either with the one-dimensional luminosity spectrum
\begin{align}
\label{eq:sigmaeff}
  \sigmaeffNoArg^{\mathrm{Machine}}  = & \int\limits_{0}^{\xs_{\max}}\dd{\xs} \; \lumispec{\xs} \sigma\bigl(\xs\rootsnom\bigr) ,\\
  \intertext{or for the two-dimensional luminosity spectrum}
  \sigmaeffNoArg^{\mathrm{Machine}}  = & \iint\limits_{0}^{~~x_{\max}}\!\dd{\xe}\dd{\xp}\;\lumispec{\xe,\xp} \sigma \bigl(\sqrt{\xe\xp\snom}\bigr).
\end{align}

Bhabha scattering is the process of choice for luminosity measurements. It can
be calculated with high precision and has a large cross-section. To first order,
the differential Bhabha cross-section is~\cite{schmueser95}
\begin{equation}
  \label{eq:bhabha}
  \frac{\dd{\sigma_{\mathrm{Bhabha}}}}{{\dd{\theta}}} = \frac{2\pi\alpha^{2}}{s}\frac{\sin{\theta}}{\sin^{4}\left(\theta/2\right)},
\end{equation}
where $\alpha$ is the fine-structure constant and $\theta$ the polar scattering
angle. 

Because cross-sections $\Sigmax{\rootsprime}$ depend on the cen\-tre-of-mass
energy, any process used to reconstruct the basic luminosity spectrum will
inherently contain a \emph{scaled} luminosity spectrum
\begin{equation}
  \label{eq:Lscaled}
  \lumispecscale{\xs} =
  \frac{\Sigmax{\xs\rootsnom}}{\int_{\xs^{\prime}_{\min}}^{\xs^{\prime}_{\max}} \dd{\xs^{\prime}}\Sigmax{\xs^{\prime}\rootsnom}} \lumispec{\xs}.
\end{equation}
This means that it is not enough to reconstruct the observed centre-of-mass
energy spectrum, the luminosity spectrum has to be extracted from the observed
spectrum.

The observed centre-of-mass energy is further affected by Initial State
Radiation. It is impossible to distinguish between energy loss from Initial State Radiation and
Beamstrahlung on an event-by-event basis. Initial State Radiation and
Beamstrahlung have to be disentangled statistically.

Finally, the scattered particles are recorded in the detector, where their
properties are reconstructed with\-in the limits of the resolution of the
respective sub-detectors.

\subsection{The Basic Luminosity Spectrum}

\begin{figure*}[tbp]\sidecaption \centering
  \subfloat[]{\label{fig:zVSe}\includegraphics[width=\sideCwidth]{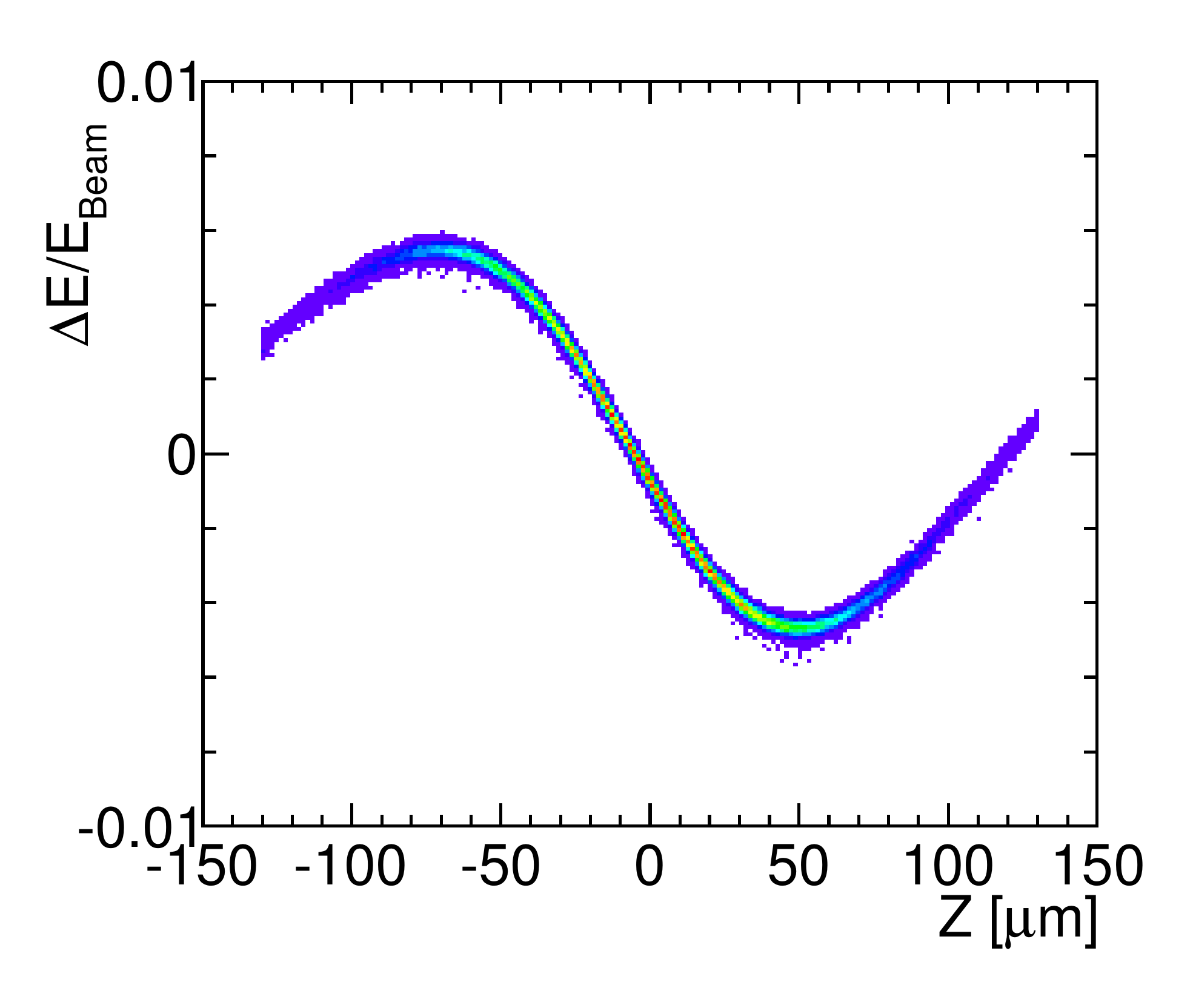}}
  \subfloat[]{\label{fig:bdsE}\includegraphics[width=\sideCwidth]{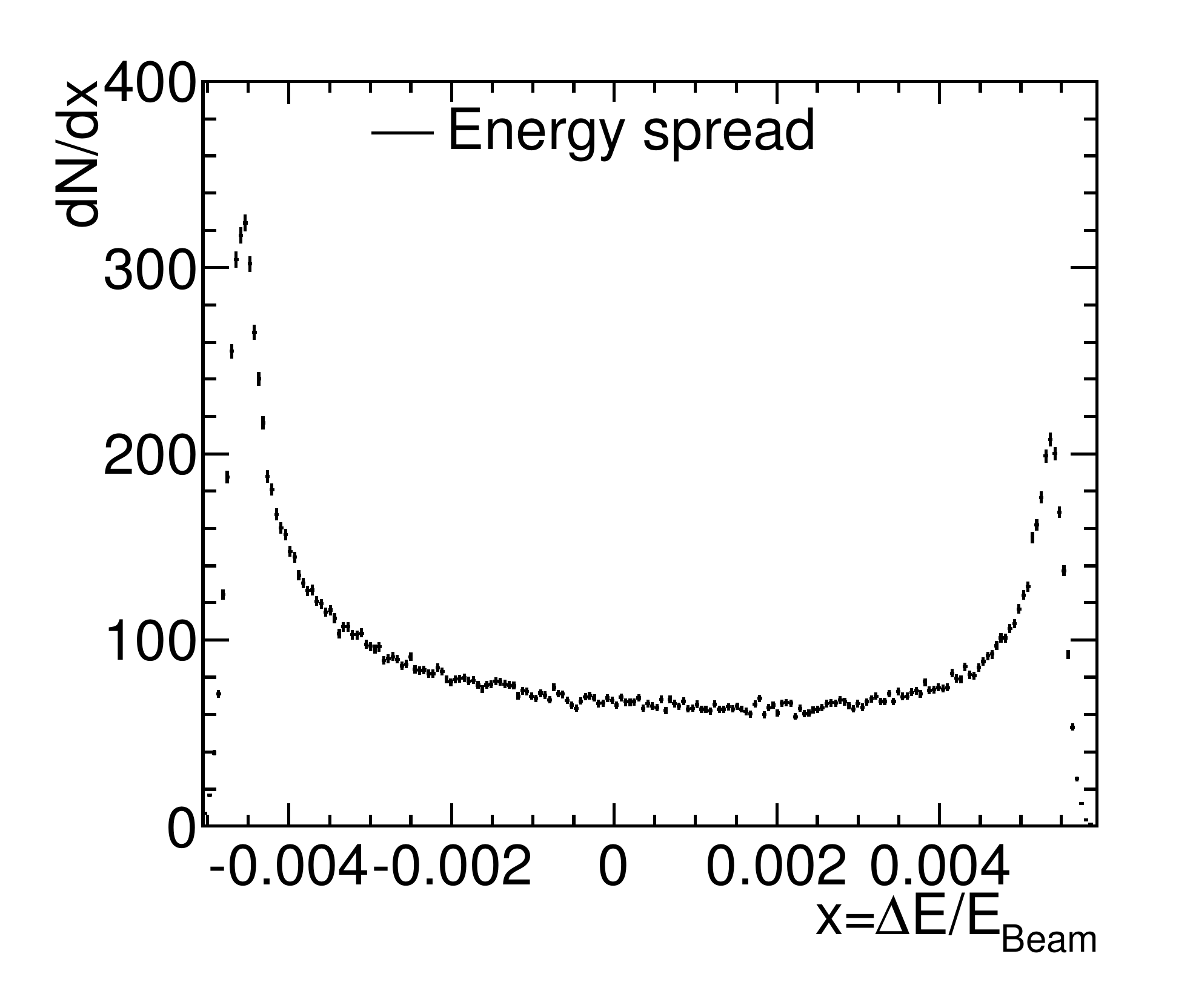}}
  \caption{Energy distribution of the CLIC beams~\cite{Schulte:572820}. \Subref{fig:zVSe}~Dependence
    of the particle energy on the longitudinal position of the particles along
    the length of the bunch, where the beam travels towards the left.
    \Subref{fig:bdsE}~The energy distribution of all particles.} \label{fig:clicbeams}
\end{figure*}

The luminosity spectrum distribution can be seen as a convolution of the beam-energy spread, which is
inherent to the accelerator, and the Beamstrahlung due to the Beam-Beam effects.
Figure~\ref{fig:clicbeams} shows the beam-energy spread of the 3~TeV CLIC machine. It
is obtained from a simulation of the main linear accelerator and the beam
delivery system~\cite{Schulte:572820}.

The energy of a particle depends on its longitudinal position in the
bunch (Figure~\ref{fig:zVSe}). Due to intra-bunch wakefields, particles in the front of
the bunch gain more energy from the RF cavities than particles in the back of
the bunch~\cite{Fischer:217238}. This leads to the two distinct peaks near the
minimal and the maximal value of the beam-energy spread (Figure~\ref{fig:bdsE}). The
energy spread is not following a Gaussian distribution. 
The dependence of the particle energy on the longitudinal position also leads to
larger correlations between the particle energies.  The precise shape of the
beam-energy spread depends on the RF-phase and the bunch
length~\cite{Fischer:217238}. To avoid a loss in the luminosity, these
parameters are not allowed to vary freely and have to be precisely
controlled~\cite{CLICCDR_vol1}. Therefore, it can be assumed that a limited knowledge of the shape of
beam-energy spread is available.

The distribution of particles is used as the input to the beam-beam simulation.
The simulation of the beam-beam effects is done with
\guineapig~\cite{schulte1996}. During the bunch crossing the intense
electromagnetic fields -- due to the opposing bunches -- deflect the beam
particles and cause Beamstrahlung.

\begin{figure*}[tbp]\sidecaption
  \centering
  \subfloat[]{\label{fig:lumi2d}\includegraphics[width=\sideCwidth]{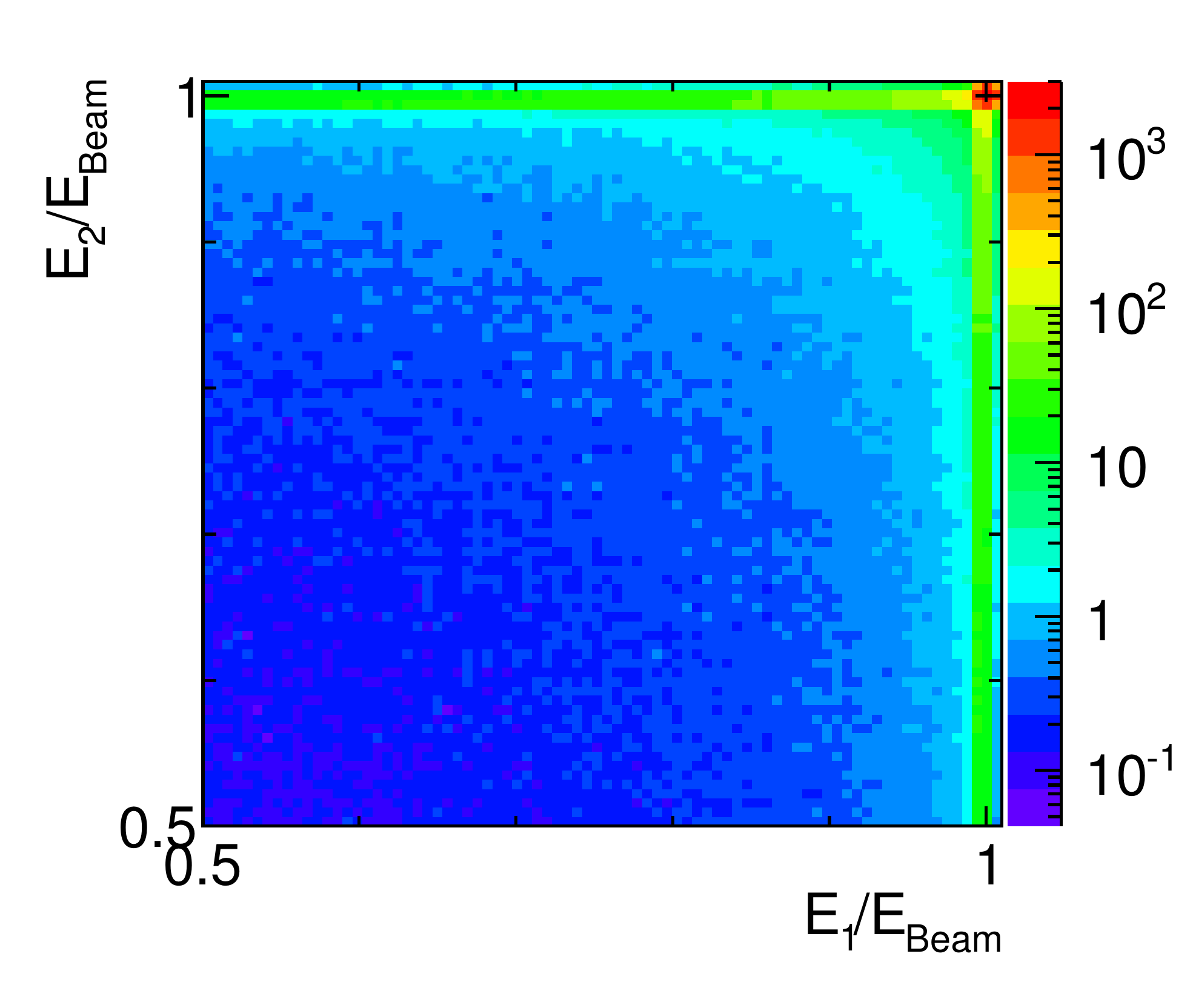}}
  \subfloat[]{\label{fig:2dZoom}\includegraphics[width=\sideCwidth]{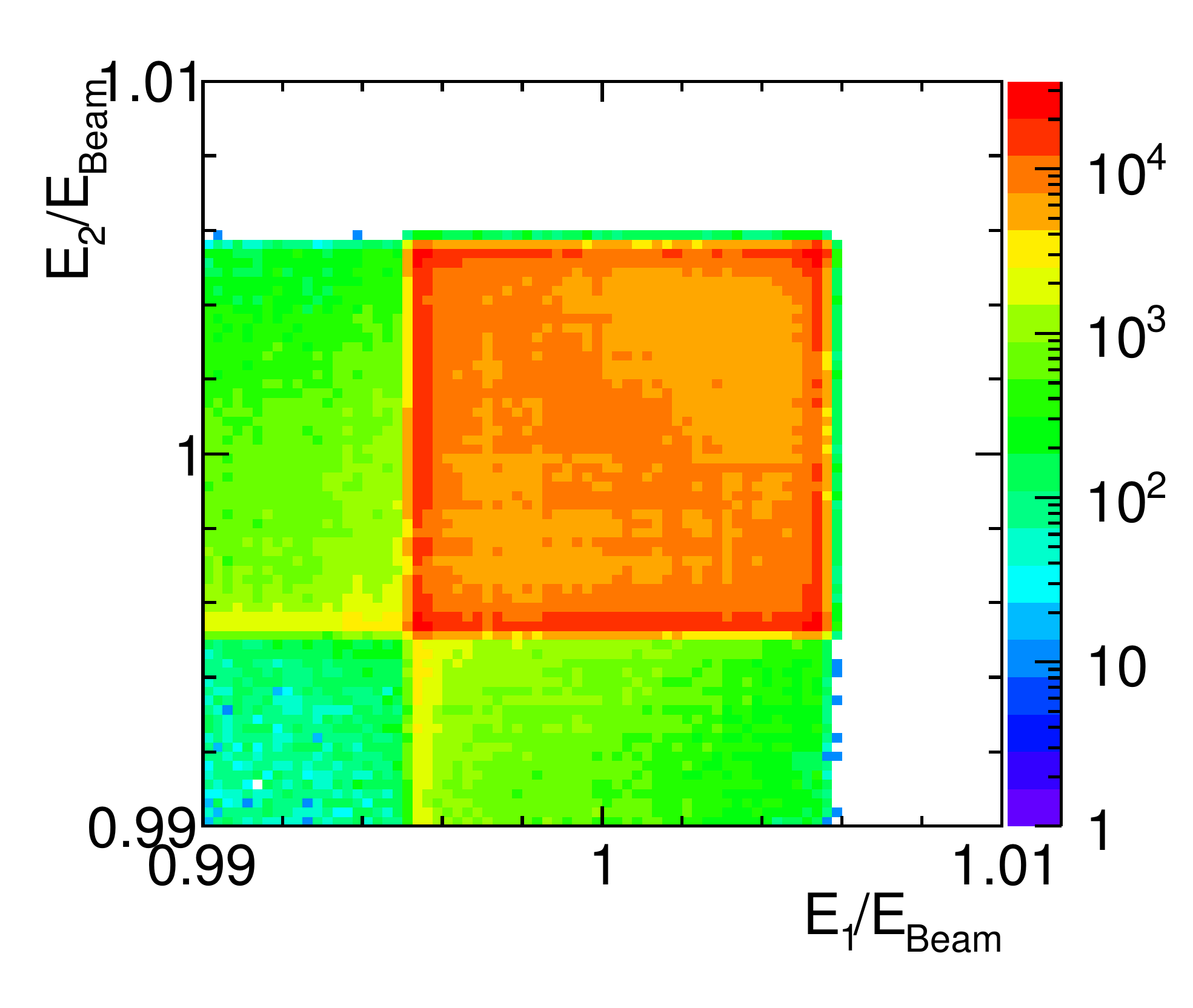}}
  \caption{\Subref{fig:lumi2d}~Energy spectrum of colliding particles as simulated with \guineapig
    for 3~TeV CLIC. \Subref{fig:2dZoom}~Zoom of the luminosity spectrum around the
    nominal beam energies. }
  \label{fig:2dSpectra}
\end{figure*}

Figure~\ref{fig:2dSpectra} shows the full range and the region around the
maximal energy of the two-dimensional luminosity spectrum. The square region in
the distribution of the two energies is due to the beam-energy spread (see
Figure~\ref{fig:bdsE}). Events with $\xe < 0.995$ or $\xp < 0.995$ were
significantly affected by the Beamstrahlung.

\begin{figure*}[tbp]\sidecaption
  \centering
  \subfloat[]{\label{fig:lumiFull}\includegraphics[width=\sideCwidth]{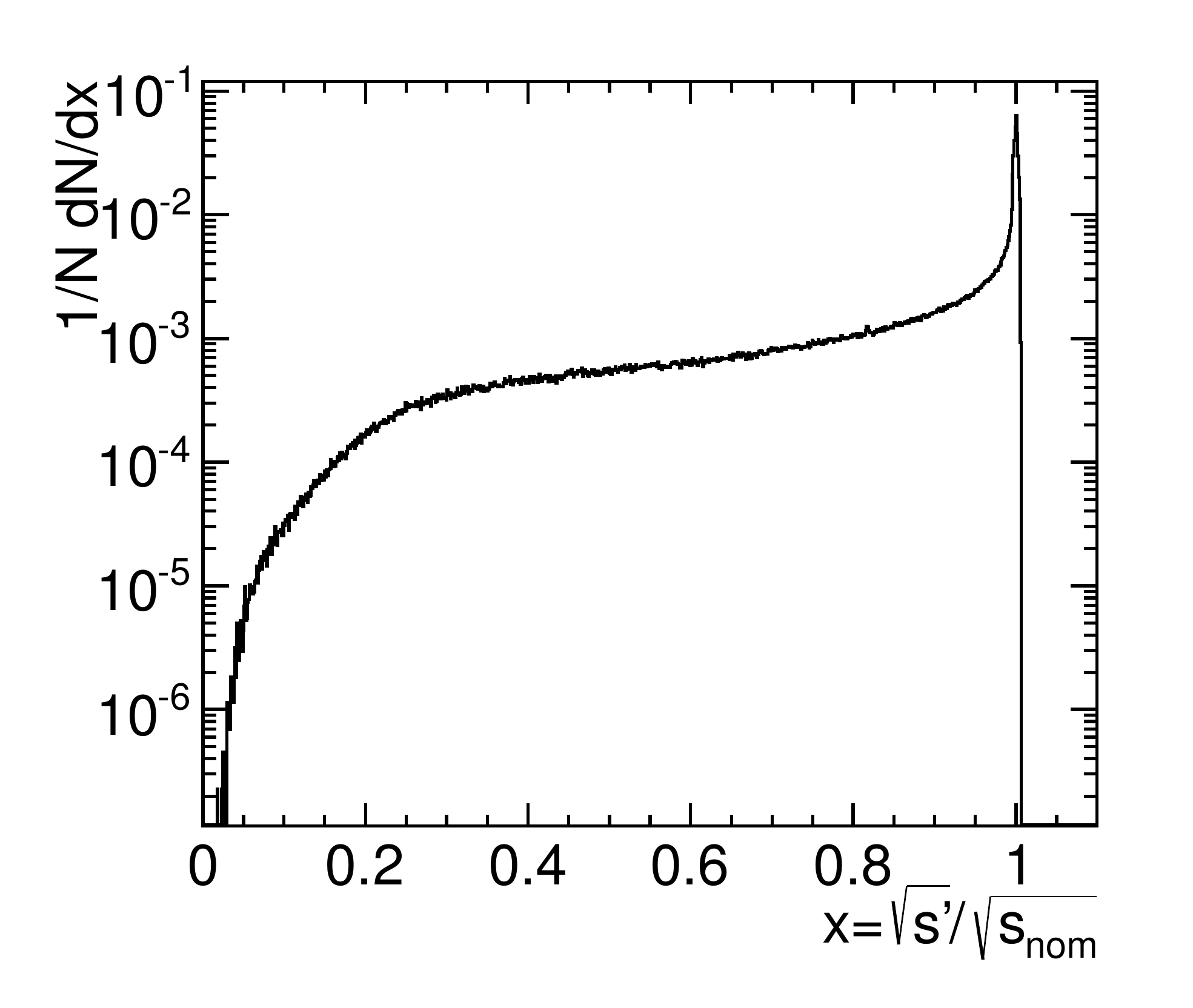}}
  \subfloat[]{\label{fig:lumiPeak}\includegraphics[width=\sideCwidth]{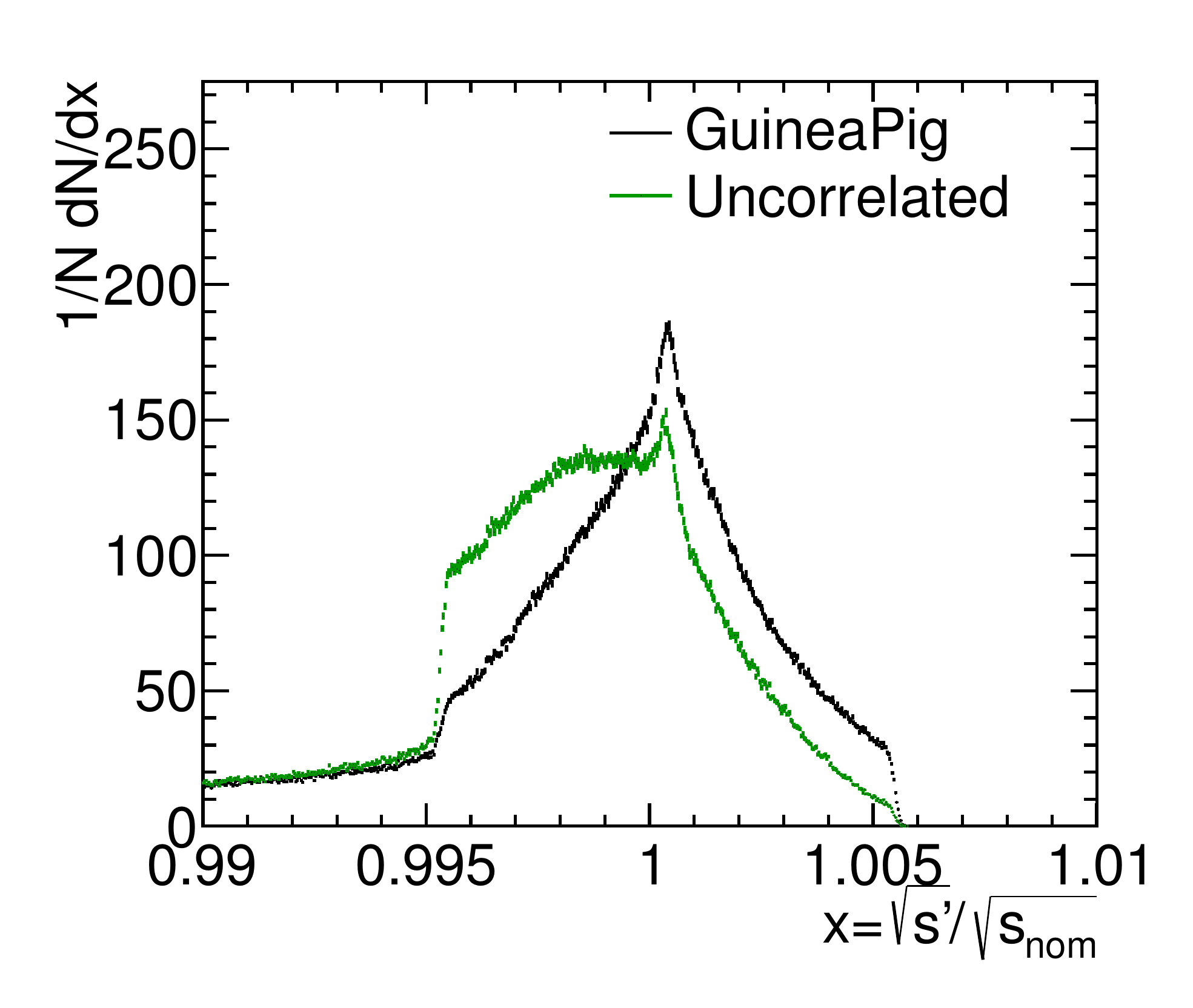}}
  \caption{\Subref{fig:lumiFull}~Luminosity spectrum for 3~TeV CLIC as simulated
    with \guineapig. \Subref{fig:lumiPeak}~The peak of the luminosity spectrum
    with and without correlated particle energies.}
  \label{fig:lumi1d}
\end{figure*}

Figure~\ref{fig:lumiFull} shows the basic luminosity spectrum with respect to the
effective centre-of-mass energy \rootsprime. The spectrum possesses a peak
around the nominal centre-of-mass energy and a long tail down to less than
5\% of the nominal centre-of-mass energy. Figure~\ref{fig:lumiPeak} shows the peak of the luminosity spectrum
as it is produced by \guineapig. Because the beam-energy spread is not Normally distributed, the
centre-of-mass energy peak is not Gaussian either. Figure~\ref{fig:lumiPeak} also shows a
spectrum obtained by randomly pairing the energies of two particles, i.e.,
removing the correlation between the energies of the two beams. There is a clear
difference between the two cases. If the correlation between the particle
energies is not taken into account, the luminosity spectrum cannot be
reconstructed properly.

To describe the beam-energy spread -- and anchor the luminosity spectrum -- the
absolute energy of the beam has to be known. The average beam energy can be measured on a level of
0.04\%~\cite{CLICCDR_vol1} with a dipole and beam position monitor in the beam
delivery system of the accelerator. If the distribution itself can be measured
as well is still under study.

\subsection{Cross-Section-Scaled Luminosity Spectrum}
\label{sec:fitGenXsec}

The observed events are distributed according to the scaled
luminosity spectrum (Equation~\eqref{eq:Lscaled}), thus the events
obtained from \guineapig have to be either weighted with very large weights, or
sampled with an accept--reject method~\cite{neumann51:random} to obtain events
with constant weights. Because large event-weights are undesirable, the
accept--reject method is chosen.

The 3~TeV CLIC luminosity spectrum extends over more than three orders of magnitude of the
Bhabha cross-section, meaning the accept--reject method is very
inefficient. If a very large number of events for the basic luminosity spectrum
were available, the scaled luminosity spectrum could be directly sampled from
them. To avoid storing the large number of basic events the accept--reject
method is directly added in \guineapig.

\begin{figure*}[tbp]\sidecaption
  \centering
  \includegraphics[width=\sideCwidth]{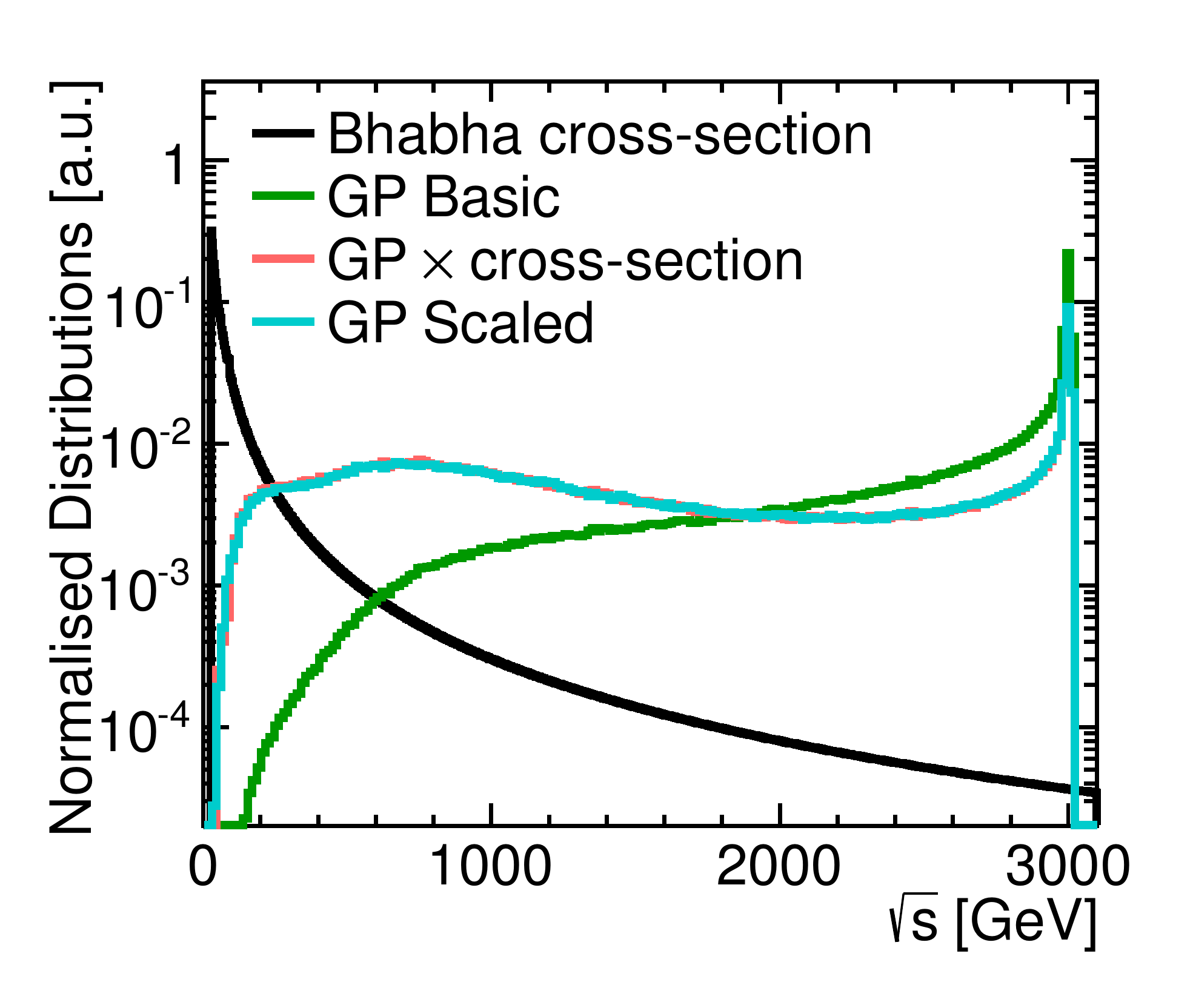}
  \caption{Bhabha cross-section from \bhwide
    for electrons with a polar angle $7\degrees < \theta < 173\degrees$,
    and the \emph{basic} luminosity spectrum from \guineapig
    (GP Basic), the luminosity spectrum scaled by bin-wise multiplication with a normalised
    Bhabha cross-section (GP $\times$ cross-section), and the luminosity spectrum scaled with an
    accept--reject method in \guineapig (GP Scaled). The lines of `GP $\times$
    cross-section' and `GP Scaled' are overlapping.}
  \label{fig:bhabha}
\end{figure*}

The differential cross-section of the Bhabha scattering has to be known for the
accept--reject method. Instead of using Equation~\eqref{eq:bhabha} to calculate
the Bhabha cross-section, it is estimated with \bhwide~\cite{Jadach:309289}.
\bhwide includes higher-order effects and Initial State Radiation. Only events
with the electron and positron polar angle inside the tracking acceptance
($7\degrees<\theta<173\degrees$) are accepted. The cross-section is estimated at
precise centre-of-mass energies from 10~GeV to 3000~GeV without any luminosity
spectrum. Figure~\ref{fig:bhabha} shows the cross-section as given by \bhwide.

Figure~\ref{fig:bhabha} shows the basic luminosity spectrum obtained with \guineapig,
the bin-wise multiplication of the luminosity spectrum with the cross-section,
and the scaled luminosity spectrum from \guineapig with the cross-section used
in the accept--reject method. The last two curves are nearly identical showing
that the modified \guineapig produces a properly scaled luminosity spectrum with
equally weighted events.

\subsection{Generation of Bhabha Events}
\label{sec:gener-bhabha-events}

The different luminosity spectra are used in the Bhabha generator to create the
events which are observed in the detector. The Bhabha events are generated with
\bhwide, where the energies of the initial electron and positron can be defined
on an event-by-event basis, as implemented by Rimbault et
al.~\cite{beambeamimpact}. The polar angle $\theta$ of the final state electrons
must be \mbox{$7\degrees < \theta < 173\degrees$} to ensure they will be
observable in the tracker. \bhwide produces also Initial and Final State
Radiation photons and accounts for their effects during the Bhabha scattering.

The cross-section -- including the luminosity spectrum -- for events with a
centre-of-mass energy above 1.5~TeV is 11~\picob, which results in more than 1
million events for an integrated luminosity of 100~\fbinv. The expectation for the 3~TeV CLIC is an
integrated luminosity of 500~\fbinv per year~\cite{cdrvol2}.

\subsection{Observables and Detector Resolutions}\label{sec:observables-resolutions}

Three observables, which can be extracted from the final state electrons, are
used for the reconstruction of the luminosity spectrum: the relative
centre-of-mass energy calculated from the polar angles of the outgoing electron
and positron $\rootsaco/\rootsnom$, the energy of the electron $E_{\ele}$, and the
energy of the positron $E_{\pos}$.  The relative centre-of-mass energy
reconstructed from the acollinearity of the final state electrons
is~\cite{moenig:DiffLumi,sailer}
\begin{equation}
  \label{eq:acollinearity}
  \frac{\rootsaco}{\rootsnom} = \sqrt{\frac{\sin(\theta_{\ele})+\sin(\theta_{\pos})+\sin(\theta_{\ele}+\theta_{\pos})}{\sin(\theta_{\ele})+\sin(\theta_{\pos})-\sin(\theta_{\ele}+\theta_{\pos})}},
\end{equation}
where $\theta_\ele$ is the polar angle of the electron and $\theta_\pos$ that of
the positron\footnote{Strictly speaking, $\theta_{\ele}$ and $\theta_{\pos}$ are
  the angles with respect the positive or negative z-axis. The angles
  have to fulfil $\theta_{\ele}+\theta_{\pos} > \pi$ by construction.}. \rootsaco is equal to the
effective centre-of-mass energy \rootsprime, if only one of the particles
radiated photons -- Beamstrahlung or Initial State Radiation. If both the
electron and the positron radiated photons, the reconstructed centre-of-mass
energy \rootsaco will be larger than \rootsprime.

The \geant simulation of tens of millions of electrons is too time-consuming. To
include the detector effects, resolution functions of the energy and angles have
been obtained from fully simulated and reconstructed Bhabha events using the
CLIC\_ILD\_CDR detector model~\cite{muennichsailer2011}. The dominating
beam-induced background, the \gghadron events~\cite{barklow2011}, was accounted for.

\begin{figure*}[tbp]\sidecaption
  \centering
  \subfloat[]{\label{fig:eresoloverlay}\includegraphics[width=\sideCwidth]{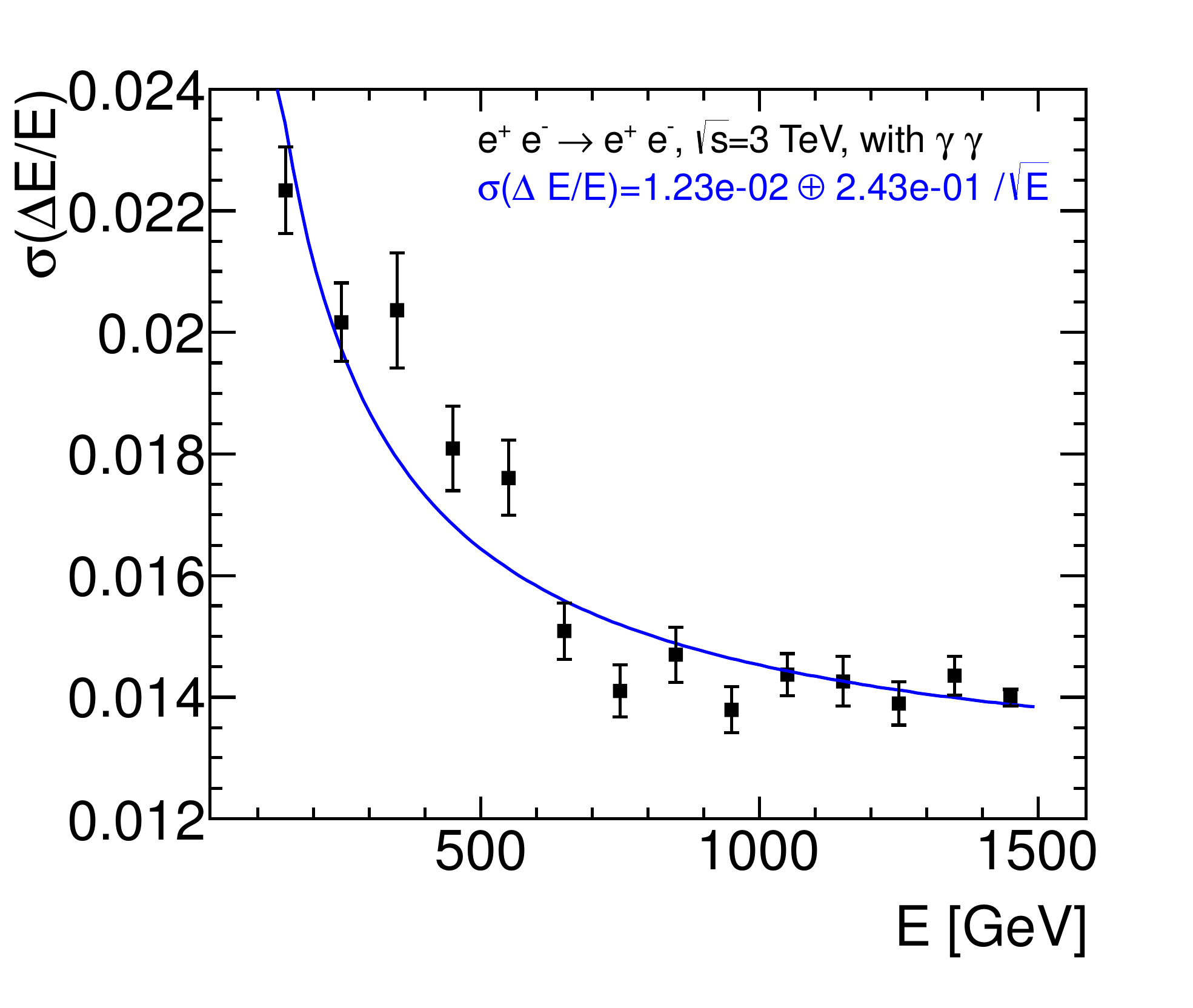}}
  \subfloat[]{\label{fig:thetaresol}\includegraphics[width=\sideCwidth]{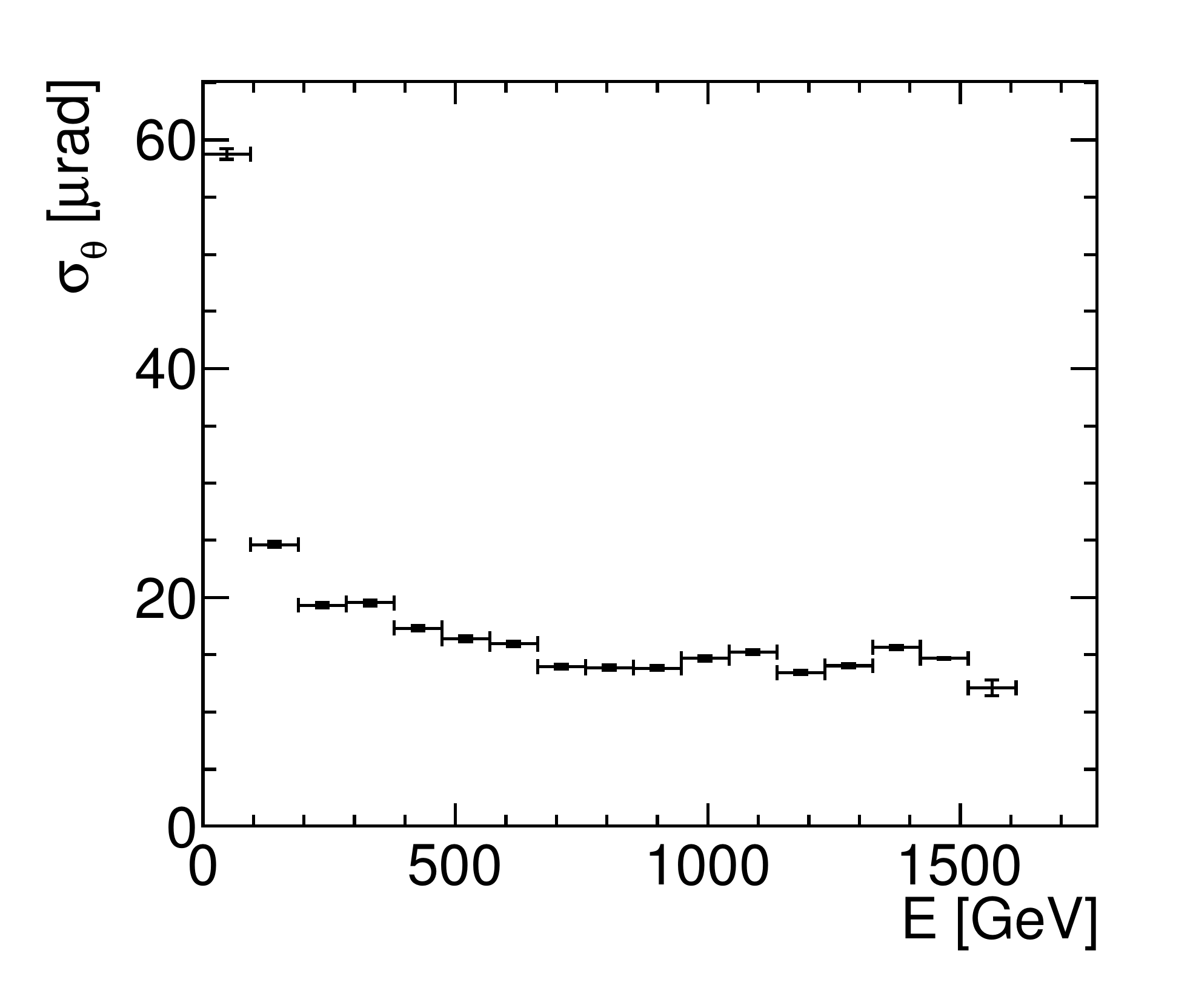}}
  \caption{Detector resolutions obtained through full reconstruction of Bhabha
    events with overlay of $\gamma\gamma\to\mathrm{hadrons}$ events:
    \Subref{fig:eresoloverlay}~Energy resolution of outgoing electrons and
    resolution function according to Equation~\eqref{eq:caloRes}. \Subref{fig:thetaresol}
    Angular resolution for electrons.}
  \label{fig:energyresol}
\end{figure*}

The rate of electrons produced in Bhabha scattering falls with an increasing
polar angle $\theta$ (cf.\ Equation~\eqref{eq:bhabha}) and the events will be predominantly
at small polar angles. Because the magnetic field is nearly collinear to those
tracks, their curvature does not allow for an accurate measurement of the
momentum. Therefore, the energy is reconstructed using only the calorimeter
information. The tracking information is used to measure the angles.  The energy
resolution is shown in Figure~\ref{fig:eresoloverlay}.  It is modelled in the analysis
with
\begin{equation}
\label{eq:caloRes}
  \frac{\sigma_{E}}{E} = \frac{24.3\%}{\sqrt{E/\mathrm{GeV}}} \oplus 1.23\%,
\end{equation}
obtained from the reconstruction with overlaid CLIC 3~TeV \gghadron background. The angular
resolution needed for the computation of the relative centre-of-mass energy
depends on the energy, shown in Figure~\ref{fig:thetaresol}. The angular resolution is
better than 20~\microrad for particle energies above 200~GeV. 

\begin{figure*}[tbp]\sidecaption
  \centering%
  \begin{minipage}{0.66\linewidth}
    \subfloat[]{\label{fig:bhwideenergy}\includegraphics[width=0.5\linewidth]{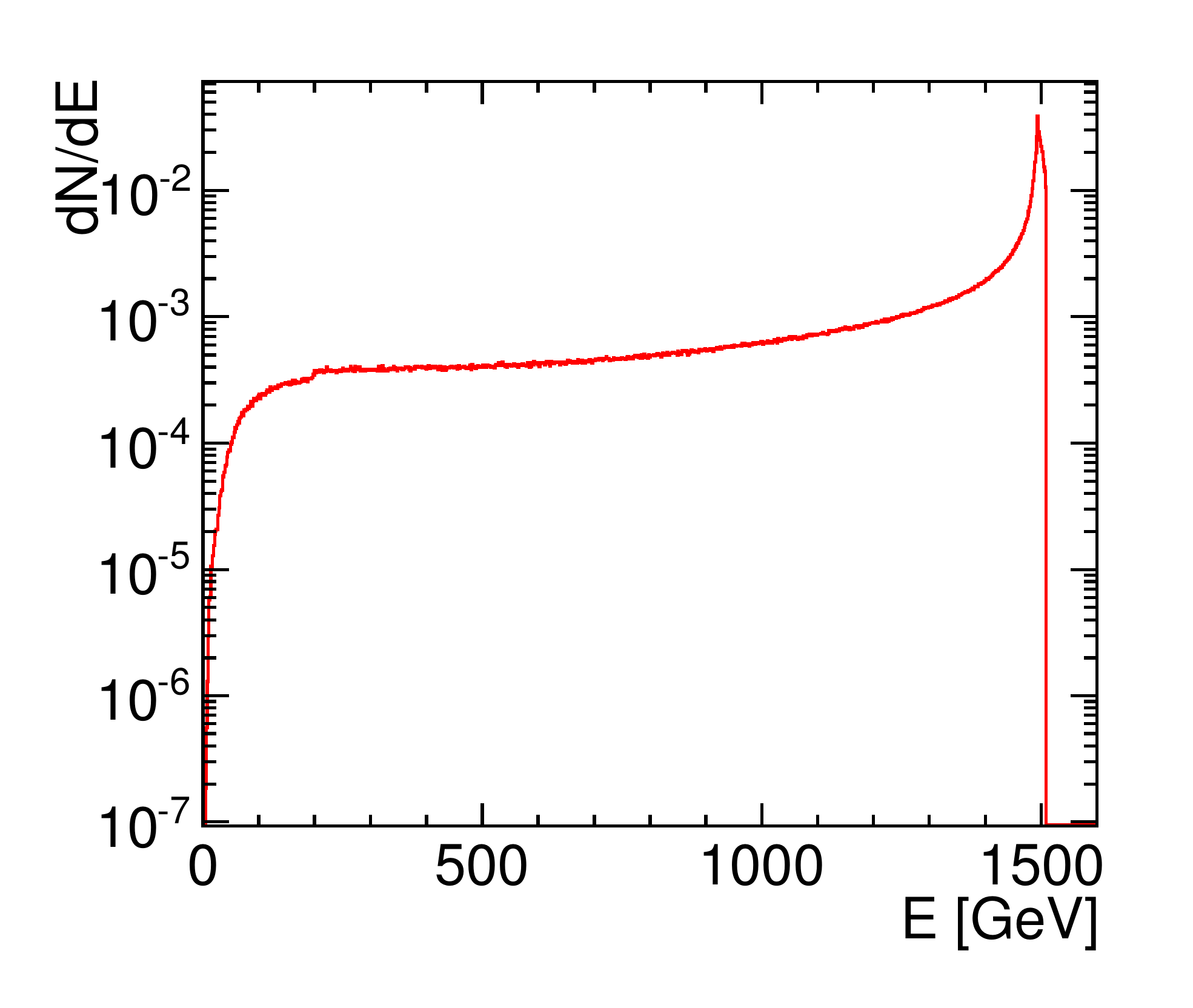}}
    \subfloat[]{\label{fig:bhwideacol}\includegraphics[width=0.5\linewidth]{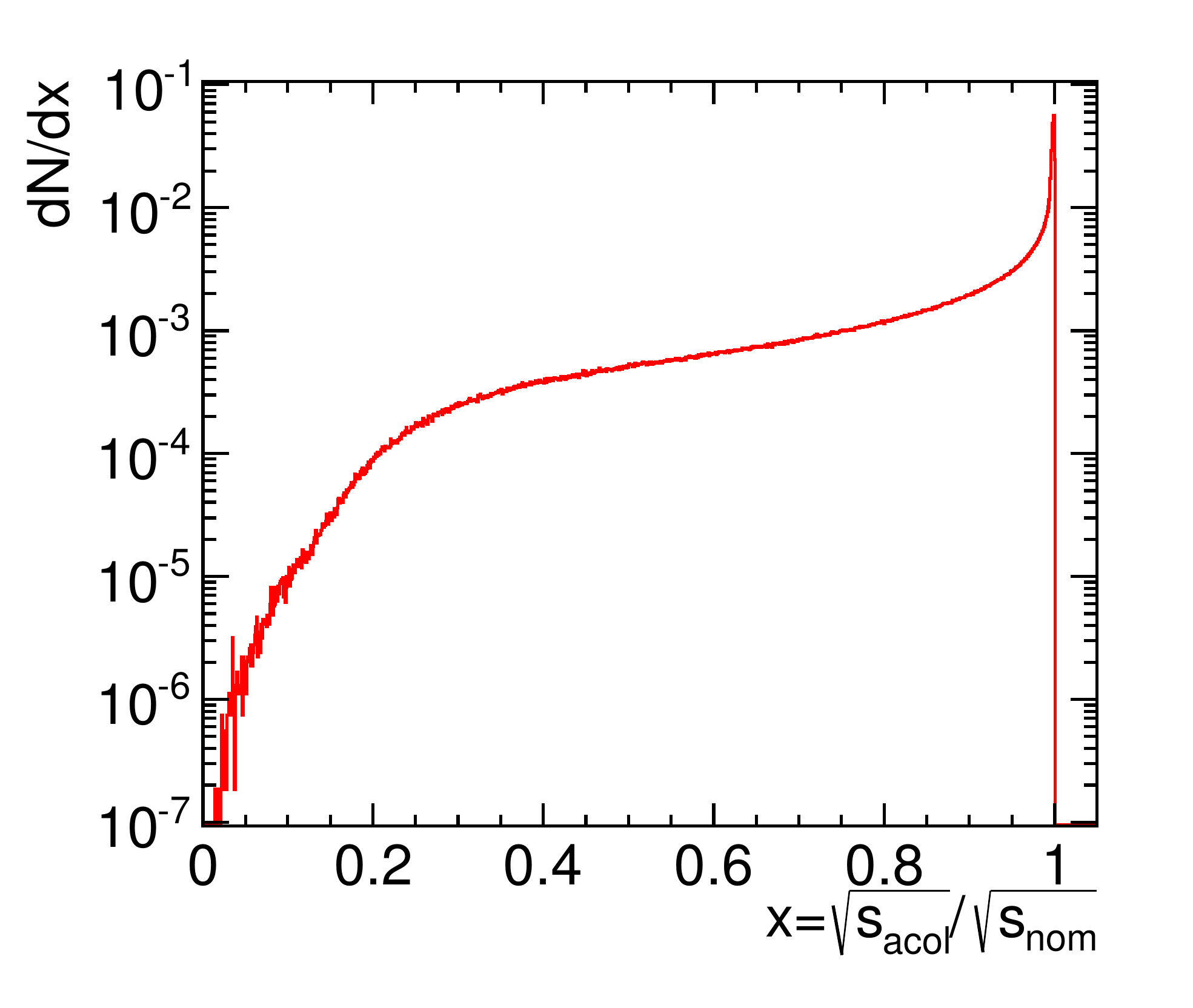}}\\
    \subfloat[]{\label{fig:bhwideEsmeared}\includegraphics[width=0.5\linewidth]{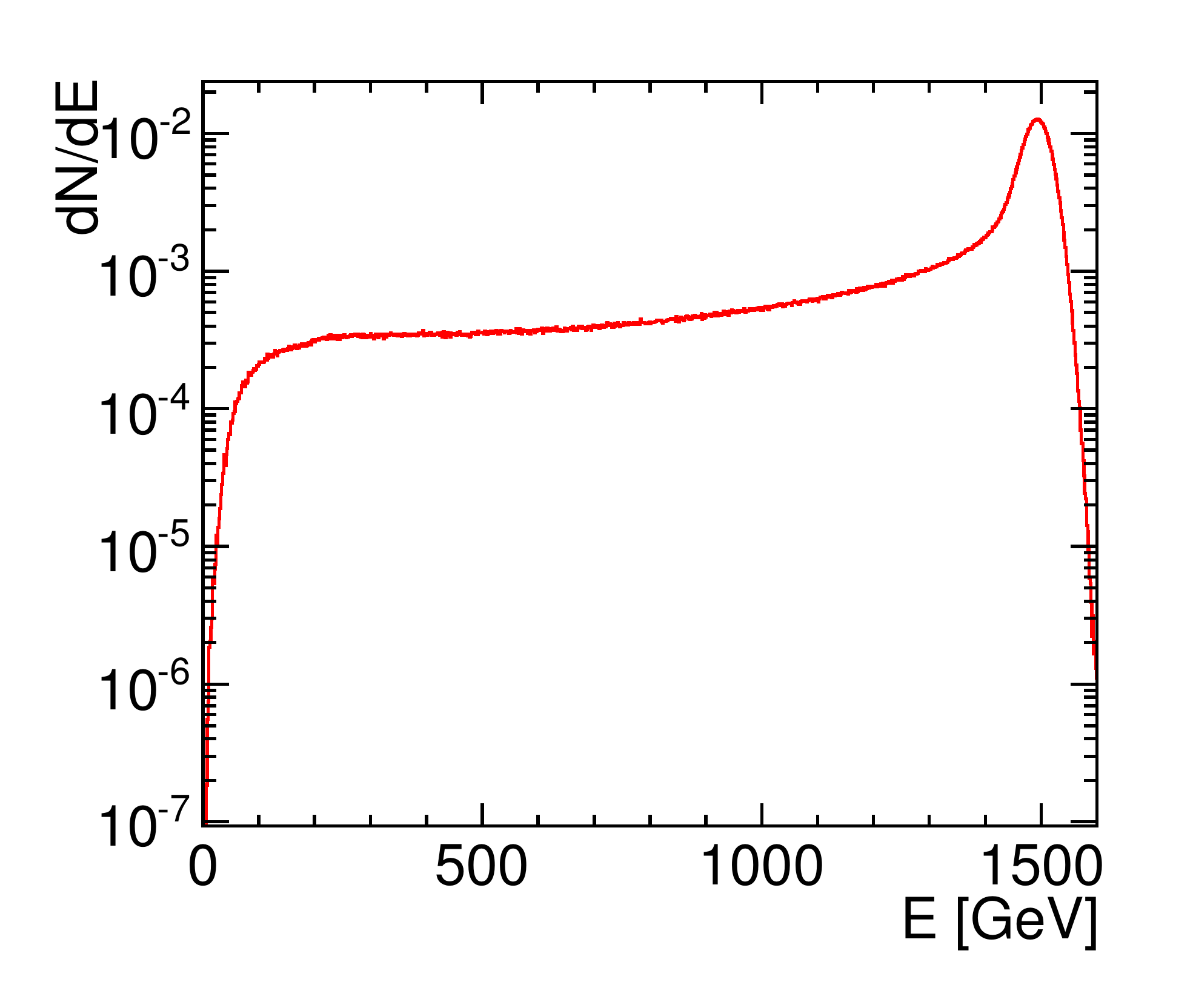}}
    \subfloat[]{\label{fig:bhwideAcolsmeared}\includegraphics[width=0.5\linewidth]{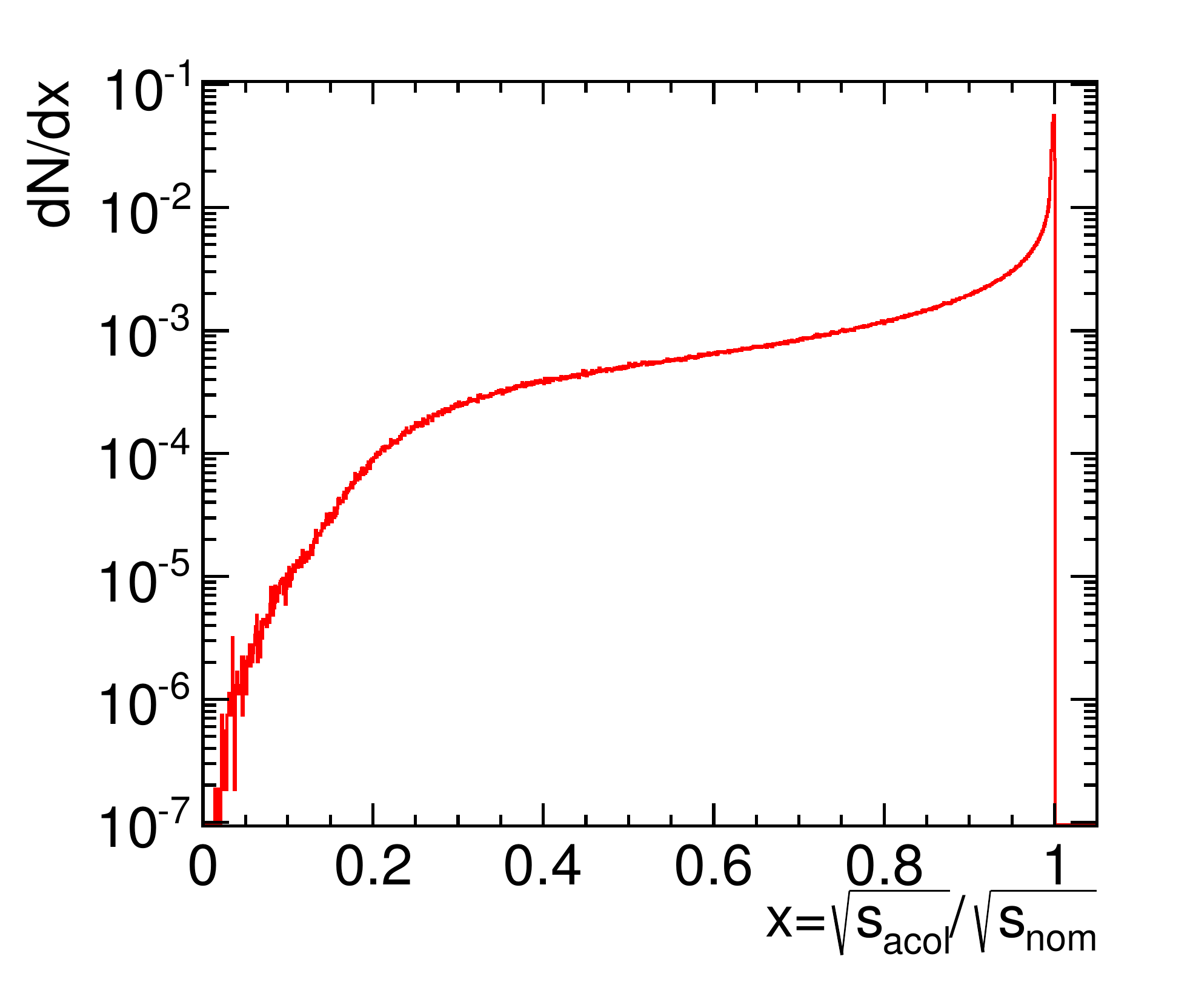}}
  \end{minipage}
  \caption{Energy of the final state electrons and relative
    centre-of-mass energy of the final state
    before~(\Subref{fig:bhwideenergy}, \Subref{fig:bhwideacol}) and
    after~(\Subref{fig:bhwideEsmeared}, \Subref{fig:bhwideAcolsmeared}) application of
    the resolution functions.}
  \label{fig:bhwidesmearedEandacol}
\end{figure*}

The distributions of the particle energies and the relative centre-of-mass
energy are shown in Figure~\ref{fig:bhwidesmearedEandacol}.
Figures~\ref{fig:bhwideenergy} and \ref{fig:bhwideacol} show the distributions before and
Figure~\ref{fig:bhwideEsmeared} and \ref{fig:bhwideAcolsmeared} after the application of the
resolution effects via four-vector smearing. The relative centre-of-mass energy
is hardly affected by the resolution, due to the high angular resolution of the
tracking detectors. The energy of the particles is much more affected by the
detector resolution.

%% file: beamenergyspread.tex
\section{Modelling the Luminosity Spectrum}
\label{sec:modell-lumin-spectr}

For the reconstruction of the basic luminosity spectrum with the reweighting fit,
a model or parameterisation of the luminosity spectrum is needed. If the beam
energy were not affected by beam-energy spread or Beamstrahlung, it could simply
be described by a Dirac delta-distribution
\begin{equation}
  \label{eq:deltaFunc}
  \EBeam(x) = \delta(x-1),
\end{equation}
with the random variates for this function \mbox{$x_{\EBeam}=1$}. The nominal
energy is modified by the contributions from beam-energy spread $\Delta\xSpread{}$ and
Beamstrahlung $\Delta\xStrahlung{}$ which change the beam energy away from its initial
value. The functions describing the two contributions will therefore describe
the difference to the nominal value, and the final random variate will be
\begin{equation}
\label{eq:randomVariatesX}
  x_{\mathrm{Final}} = x_{\EBeam} + \Delta\xSpread{} + \Delta\xStrahlung{}.
\end{equation}
Thus, the functions for the beam-energy spread and Beamstrahlung should
describe the \emph{energy change} of particles due to the respective effect.

\begin{figure}[tbp]\centering
  \includegraphics[width=\halfwidth]{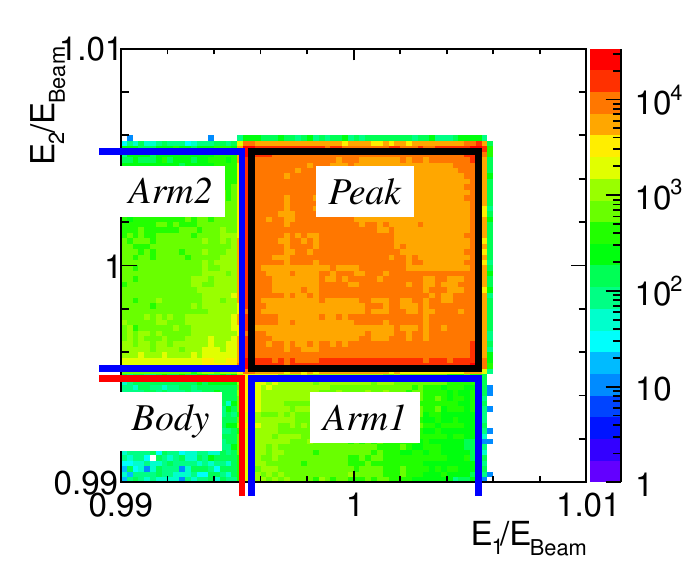}
  \caption{Two-dimensional luminosity spectrum where the different regions have
    been labelled. The beam-energy spread is included and in principle
    all the regions are overlapping.}
  \label{fig:2dregions}
\end{figure}

The luminosity spectrum model will be built by describing the two particle
energies. Using \xe and \xp as defined in Section~\ref{sec:lumin-spectr-bhabha}, the
simplest description of a two-dimensional model is $\lumispec{\xe,\xp} =
f(\xe)f(\xp)$. However, a purely factorising Ansatz is insufficient to describe
the correlation between the particle energies. Therefore, the two-dimensional
energy distribution is divided into four regions (as shown in Figure~\ref{fig:2dregions}): one
region where neither particle radiated Beamstrahlung (called the `\peak');
two regions where one or the other particle radiated Beamstrahlung (called the
`\arms'); and one region where both particles radiated Beamstrahlung
(called the `\body'). This separation is only determined by whether a
particle produced Beamstrahlung or not and ignores the beam-energy spread for
the moment. The result of this division is a piecewise function
\begin{equation}
  \label{eq:piecewise}
  \lumispec{\xe,\xp}=
  \begin{cases}
    f_{\mathrm{Peak}}, & \text{for } \xe=1 \mathrm{~and~} \xp=1 \\
    f_{\mathrm{Arm1}}, & \text{for } \xe=1 \mathrm{~and~} \xp<1 \\
    f_{\mathrm{Arm2}}, & \text{for } \xe<1 \mathrm{~and~} \xp=1 \\
    f_{\mathrm{Body}}, & \text{for } \xe<1 \mathrm{~and~} \xp<1.
  \end{cases}
\end{equation}
For each region, the resulting particle energies are described by a product of
the functions for the two particles $f_{\mathrm{Region}} = f_{\mathrm{Region}}^{\ele}(\xe) \cdot f_{\mathrm{Region}}^{\pos}(\xp)$,
and the individual functions are constructed from convolutions of the beam-energy spread and
Beamstrahlung functions, depending on the region.

\subsection{Parameterisation of the Beam-Energy Spread}
\label{sec:defining-the-bds-model}

\begin{figure*}[tb]\sidecaption
  \centering
  \subfloat[]{\label{fig:besGP}\includegraphics[width=\halfwidth]{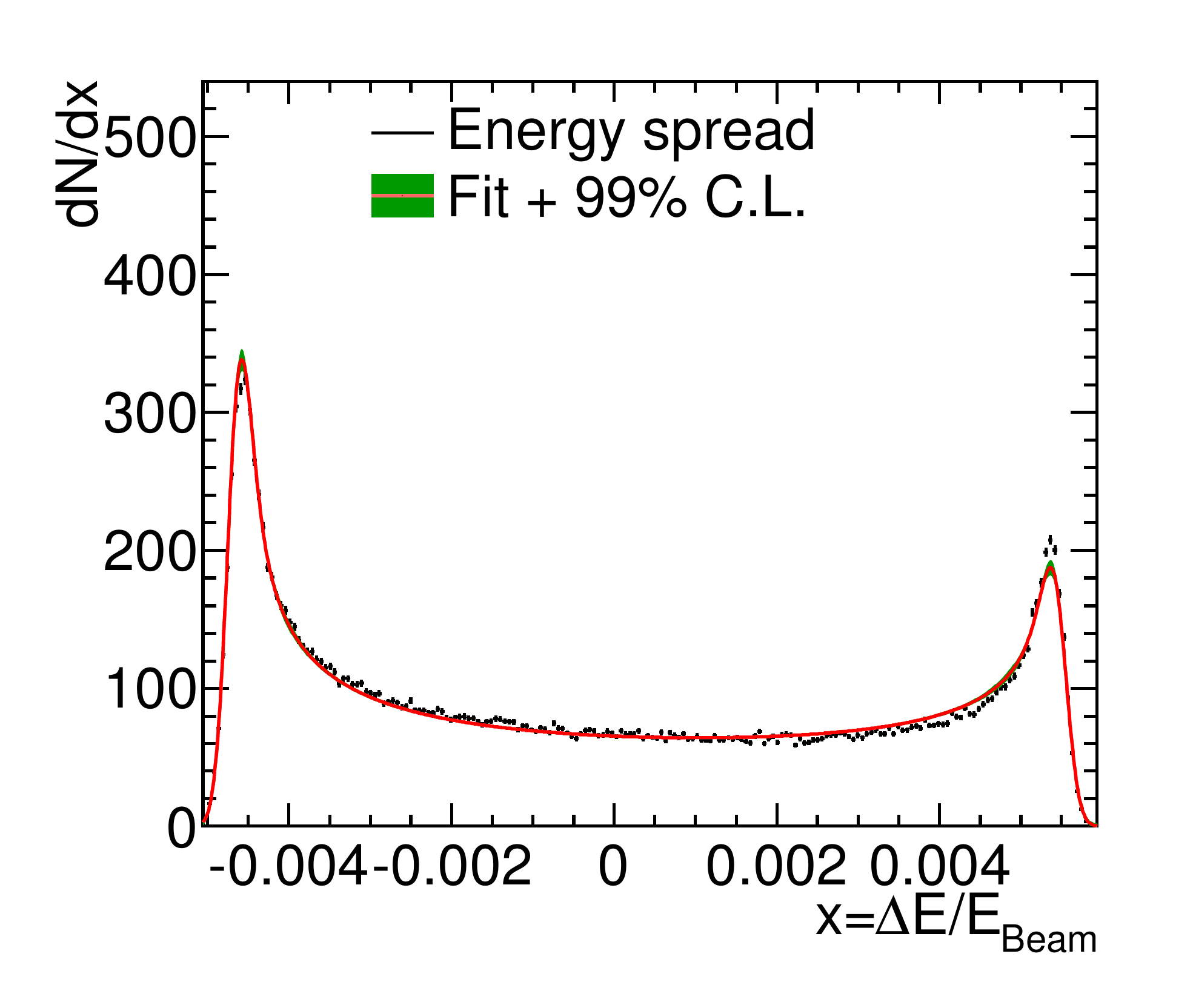}}
  \subfloat[]{\label{fig:besNL}\includegraphics[width=\halfwidth]{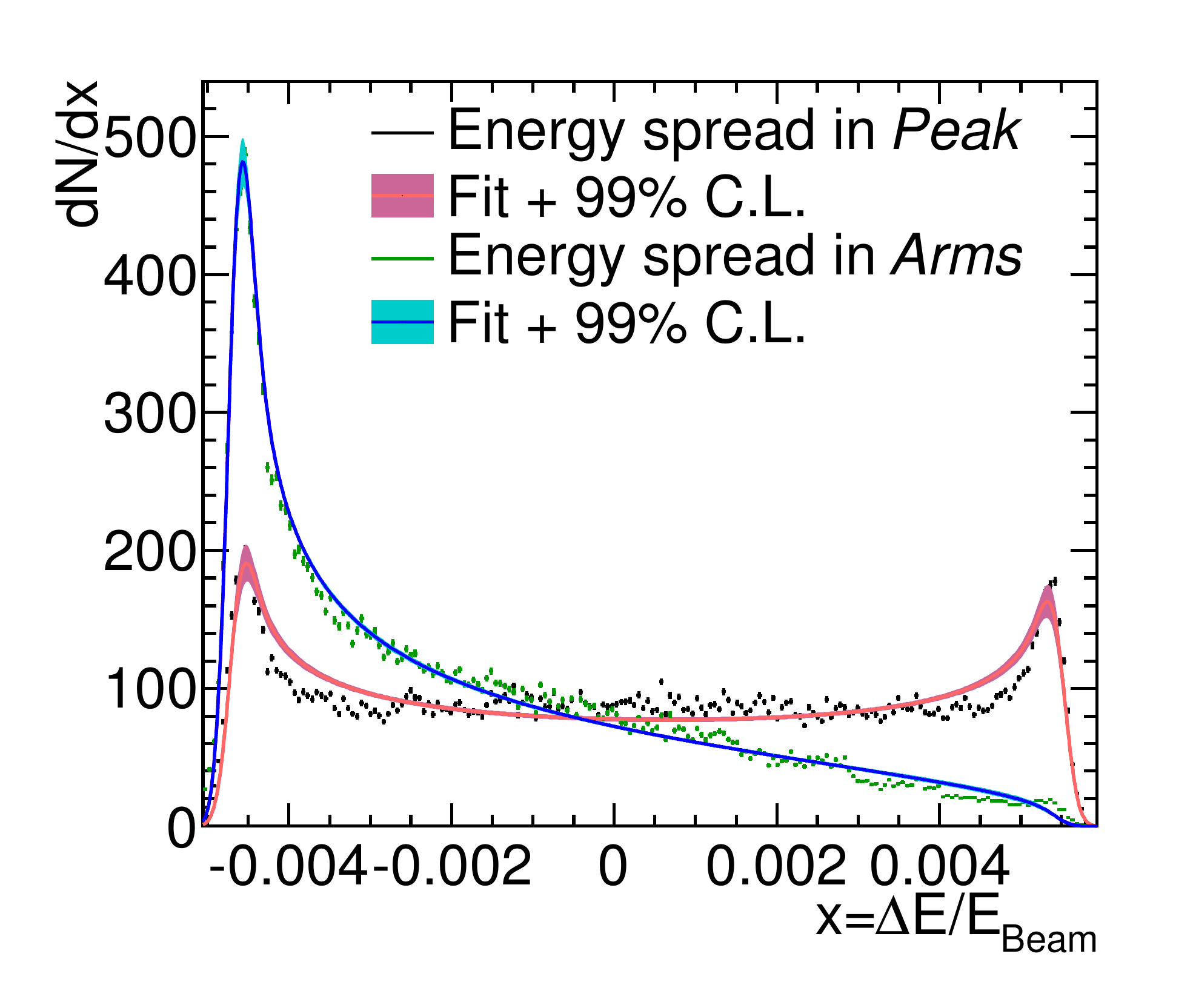}}
  \caption{\Subref{fig:besGP}~The beam-energy spread distribution from the
    accelerator simulation~\cite{Schulte:572820} and the best fit to the
    beam-energy spread with Equation~\eqref{eq:bes}. \Subref{fig:besNL}~Energy spread
    after the simulation in \guineapig. The distribution of the \peak requires
    that both particles have $E > 0.995 \EBeam$, and the distribution in the
    \arms requires that one particle has $E < 0.995 \EBeam$. The colour bands in
    both plots indicate the confidence interval at 99\%.}
  \label{fig:besFit}
\end{figure*}

The function $f_{\mathrm{Peak}}$ to describe the behaviour of the beam-energy spread
(Figure~\ref{fig:besGP}) has to rise very steeply at the two extremities. A
hyperbolic cosine, parabola, or higher order polynomials -- with a reasonable
number of parameters -- do not describe this energy distribution well. A
beta-distribution $\BetaD{t}=\frac{1}{N} t^{\beamspreadpara{\betaA}{}} (1-t)^{\beamspreadpara{\betaB}{}}$  is used to describe the
beam-energy spread. As the beta-distribu\-tion is limited between $0$ and $1$, a
variable transform
\begin{equation}
  t = \frac{x-\xmin}{\xmax-\xmin}
\end{equation}
is used to describe the beam-energy spread between the two endpoints $\xmin$
and $\xmax$ near the maximal values at the beginning and end of the
distribution. Here, the variable $x$ is the relative difference between a
particle's energy \EParticle and the nominal beam energy \EBeam,
\begin{equation}
  x = \frac{\EParticle-\EBeam}{\EBeam} = \frac{\Delta{E}}{\EBeam},
\end{equation}
which corresponds to $\Delta\xSpread{}$ from Equation~\eqref{eq:randomVariatesX}. To
also describe the particles with energies below the $\xmin$ and above \xmax, the
beta-distribution is convoluted with a Gaussian distribution \Gauss{x} with a
mean $\mu=0$ and a width $\sigma$. The Beam-Energy Spread (BES)
function is
\begin{equation}
  \label{eq:bes}
  \Bes{x; \beamspreadpara{\betaA}{}, \beamspreadpara{\betaB}{}, \sigma} = b(x; \beamspreadpara{\betaA}{}, \beamspreadpara{\betaB}{}) \conv  \Gauss{x; \sigma},
\end{equation}
where $\conv$ is the convolution operator $h(x) \equiv \left(f \conv g\right)(x)  \equiv  \int_{-\infty}^{\infty} f(\tau)  g(x-\tau) \dd{\tau}$. Due to Fubini's theorem the convolution of two probability
density functions always results in another probability density function~\cite{forster84:analysis3}.

The beam-energy spread histogram is fitted by the function \Bes{x} with a
binned log-likelihood fit with \textsc{Root} version
5.34.01~\cite{Brun1997}. Figure~\ref{fig:besGP} shows the best fit to the
beam-energy spread with this model, and the resulting parameters are given in
Table~\ref{tab:besFitPar}. The histogram contains 300\,000 entries.

The width of the Gaussian $\sigma$ and the boundaries of the beam-energy spread
beta-distributions ($\xmin, \xmax$) are fixed for all following fits. This
assumes an existing knowledge of the beam-energy spread coming from the
accelerator. Fixing these parameters can introduce a large systematic error, if
they are not measured correctly.
\begin{table*}[tb]
  \centering
  \caption{Parameters found by the fit of Equation~\eqref{eq:bes} to the beam-energy
    spread from the accelerator simulation and to the beam-energy spread for two
    different regions of the luminosity spectrum.}
  \label{tab:besFitPar}
  \begin{tabular}{l c *3{R{2}{3} R{2}{3}} }\toprule
                          &               & \tabtt{Energy Spread} & \tabtt{\peak} & \tabtt{\arms}                                 \\\cmidrule(lr){3-4}\cmidrule(lr){5-6}\cmidrule(lr){7-8}
    Parameter             & \tabt{Factor} & \tabt{Value}          & \tabt{Uncertainty} & \tabt{Value}  & \tabt{Uncertainty} & \tabt{Value} & \tabt{Uncertainty} \\\midrule
    $\beamspreadpara{\betaA}{}$ &               & -0.522338             & 0.00145089         & -0.332595      &   0.00172795      &  -0.469857  &   0.00129228 \\
    $\beamspreadpara{\betaB}{}$ &               & -0.409289             & 0.0018557          &  -0.29787      &   0.00182727      &   0.405222  &   0.00392909  \\
    $\xmin$               & $10^{-3}$     & -4.678887             & 0.00143641         & \tabtt{Fixed}                      & \tabtt{Fixed}               \\
    $\xmax$               & $10^{-3}$     & 5.495029              & 0.00183891         & \tabtt{Fixed}                      & \tabtt{Fixed}               \\
    $\sigma$              & $10^{-4}$     & 1.36742               & 0.00979374         & \tabtt{Fixed}                      & \tabtt{Fixed}               \\\midrule
    \chindf               &              & \tabtt{$764/195$}                            & \tabtt{$6032/198$}                   & \tabtt{$3803/198$} \\\bottomrule                   
  \end{tabular}
\end{table*}

\subsubsection{Luminosity Weighted Beam-Energy Spread}

The correlation between the particle energy and its position in the bunch causes
a change in the effective beam-energy spread. The probability to radiate
Beamstrahlung photons, and therefore the fraction of energy lost by particles,
increases with the distance travelled in the electromagnetic field of the
oncoming bunch.

Figure~\ref{fig:besNL} shows two energy distributions, one of the \peak-region,
where both particles posses an energy of more than 99.5\% of the nominal beam
energy, and one of the \arms-region, where only one of the particle
contains more than 99.5\% of the nominal beam energy. Both histograms contain 300\,000 entries.

The energy spread of the \peak-region is clearly flatter than the energy
spread coming from the accelerator (Figure~\ref{fig:besGP}). In the column
\peak, Table~\ref{tab:besFitPar} lists also the parameters
$\beamspreadpara{\betaA}{}$ and $\beamspreadpara{\betaB}{}$ found by fitting Equation~\eqref{eq:bes}
to the distribution. For this fit, the limits and the Gaussian width are
fixed. For the best fit both $\beamspreadpara{\betaA}{}$ and $\beamspreadpara{\betaB}{}$ are
closer to zero, but still negative, i.e., there are still two maxima at the
lower and upper end of the distribution. The peak at the lower end of the
spectrum is reduced, because the particles in the tail are less likely to
interact with particles that did not radiate Beamstrahlung.

The energy spread of the \arms-region, where one of the particles radiated
Beamstrahlung, shows a large peak at the lower energy, and almost no peak at the
upper end of the spectrum. This is also caused by the correlation between the
energy and the position in the bunch. Particles in the tail are more likely to
collide with a particle that already radiated Beamstrahlung, and therefore the
peak at the lower edge of the beam-energy spread is enhanced. Likewise, only very
few particles with the highest energy -- located near the front of the bunch --
interact with particles from the tail of the bunch, which leads to the
disappearance of the peak at the highest energies. The beam-energy spread for
the \arms-region is described by a beta-distribution for which
$\beamspreadpara{\betaB}{} > 0$ (see column \arms in Table~\ref{tab:besFitPar}).

The \chindf becomes larger for the fits to the luminosity weighted beam-energy
spreads than for the fit to the initial beam-energy spread. The chosen function
cannot perfectly model the distributions, however, the fits are only used to
check qualitatively if the model can represent the luminosity spectrum at this
stage. In addition, as there was only a single input file available to run
\guineapig, the macro-particles are re-used for luminosity events. This re-use
means that the fluctuations in the number of entries are larger than what can be
expected from the statistical uncertainties, which also increases the \chindf value.


%% file: beamstrahlung.tex
\subsection{Beamstrahlung}
\label{sec:beamstrahlung}

\begin{figure*}[tbp]\sidecaption
  \centering
  \subfloat[]{\label{fig:fitEPart00}\includegraphics[width=\halfwidth]{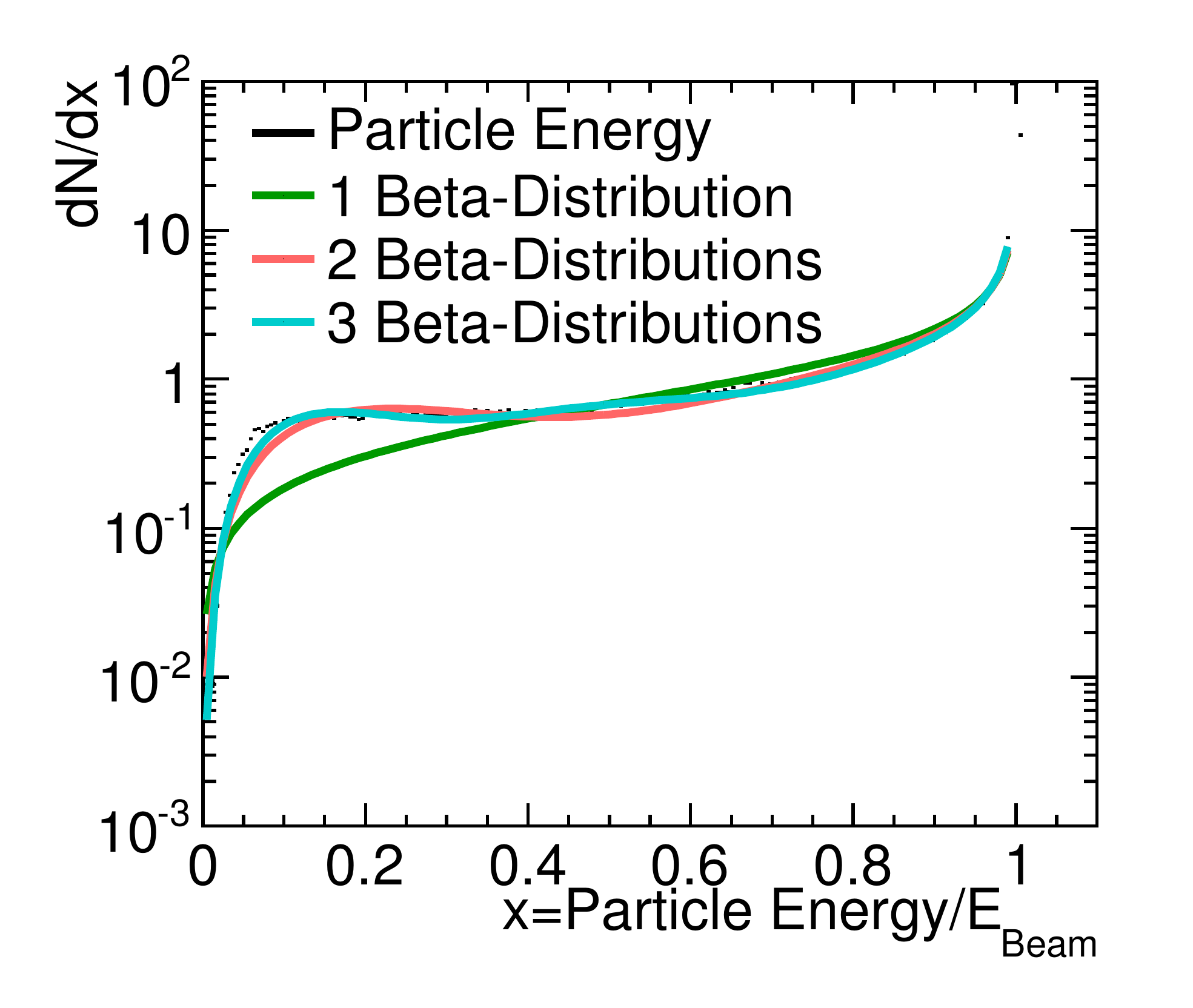}}
  \subfloat[]{\label{fig:fitEPart05}\includegraphics[width=\halfwidth]{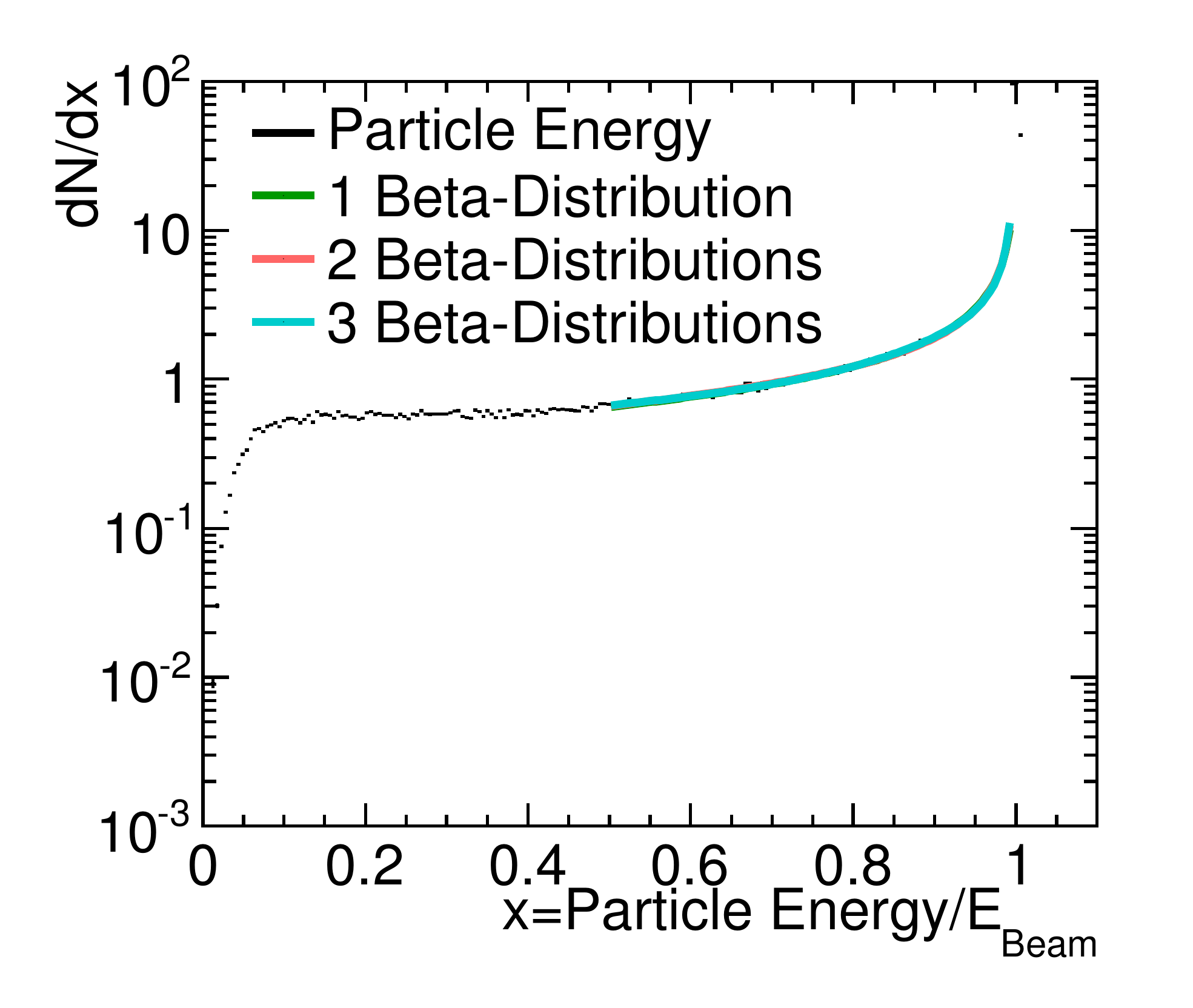}}
  \caption{Fit of the linear combination of one, two, and three
    beta-distributions (Equation \eqref{eq:linBeta}) to the particle energy spectrum after
    Beamstrahlung. \Subref{fig:fitEPart00}~Fit for $0.0 < x < 0.995$,
    \Subref{fig:fitEPart05}~fit for $0.5 < x < 0.995$.}
  \label{fig:beamFit}
\end{figure*}

Following Ohl's \textsc{Circe} model~\cite{Ohl:1996fi}, the energy
distribution of the particles after the emission of Beamstrahlung photons is
modelled with a beta-distribution. Beamstrahlung will always reduce the energy
of a particle, so that the random variate $\Delta\xStrahlung{}$ would be between $-1$ and $0$
(cf. Equation~\eqref{eq:randomVariatesX}). Beta-distributions are limited
between 0 and 1, so that the function describing the Beamstrahlung
effect is convoluted with the $\delta$-distribution from
Equation~\eqref{eq:deltaFunc}, which moves the range to \mbox{$ 0 <
  \xStrahlung{} = 1 + \Delta\xStrahlung{} < 1$} and no further variable transform is necessary for the
probability density function.

The parameters of the beta-distributions used to describe the energy distribution due to
Beamstrahlung are called $\beamstrahlungpara{\betaA}{}$ and $\beamstrahlungpara{\betaB}{}$. The
beta-distribution parameters must fulfil the conditions $ 0 <
\beamstrahlungpara{\betaA}{}$ and $ -1 < \beamstrahlungpara{\betaB}{} < 0$ for the
distribution to fall towards $x=0$ and rise towards $x=1$.

Previous studies by Daniel Schulte have shown that the tail of the CLIC centre-of-mass energy
distribution is better modelled by a sum of three
beta-distributions. Therefore, the energy distributions from
Beamstrahlung are initially fitted by linear combinations of $\nBeta$ incomplete beta-distribu\-tions
\begin{equation}
\label{eq:linBeta}
  b_{\mathrm{linear}}(x) = \sum_{i=1}^{\nBeta} p_{i}\BetaD{x;\parset{i},\blimit{}},
\end{equation}
and the constraint
\begin{equation}
  1 \overset{!}{=} \sum_{i=1}^{\nBeta} p_{i},
\end{equation}
where $p_{i}$ are the respective fractions of the individual beta-distribution contribution
and $\parset{i}=\{\beamstrahlungpara{\betaA}{i},\beamstrahlungpara{\betaB}{i}\}$ the parameter-set for each beta-distribution.
The beta-distributions are limited with an upper limit of \mbox{$\blimit{}=0.995$}. Above 0.955 the beam-energy spread
is dominant and would have to be included for the fit.

Figure~\ref{fig:beamFit} shows the fits with $\nBeta=1, 2, 3$ to the distribution of
the particle energy. In Figure~\ref{fig:fitEPart00} the fit to the histogram is
performed in the range of $0.0 < x < 0.995$; It is visible
that the function with three beta-distributions -- with eight free parameters --
shows a better agreement with the distribution than the other
functions. As all the beta-distributions cover the full range for the fit, there
are large correlations between the parameters of different beta-distri\-butions.

Figure~\ref{fig:fitEPart05} shows the same fit of linear combinations with a range
limited to $ 0.5 < x < 0.995$; all three fit-functions overlap. Therefore, a
single beta-distribution is enough to describe the particle energy between
half and 99.5\% of the beam-energy. For the \model, the Beamstrahlung is described by a
single beta-distribution to reduce the number of free parameters. However, this
will also limit the energy range in which our \model can be considered as valid.


%% file: model.tex
\subsection{The \model for the Full Luminosity Spectrum}
\label{sec:model-full-lumin}
\begin{figure*}[tbp]\sidecaption
  \centering
  \includegraphics[width=0.66\textwidth]{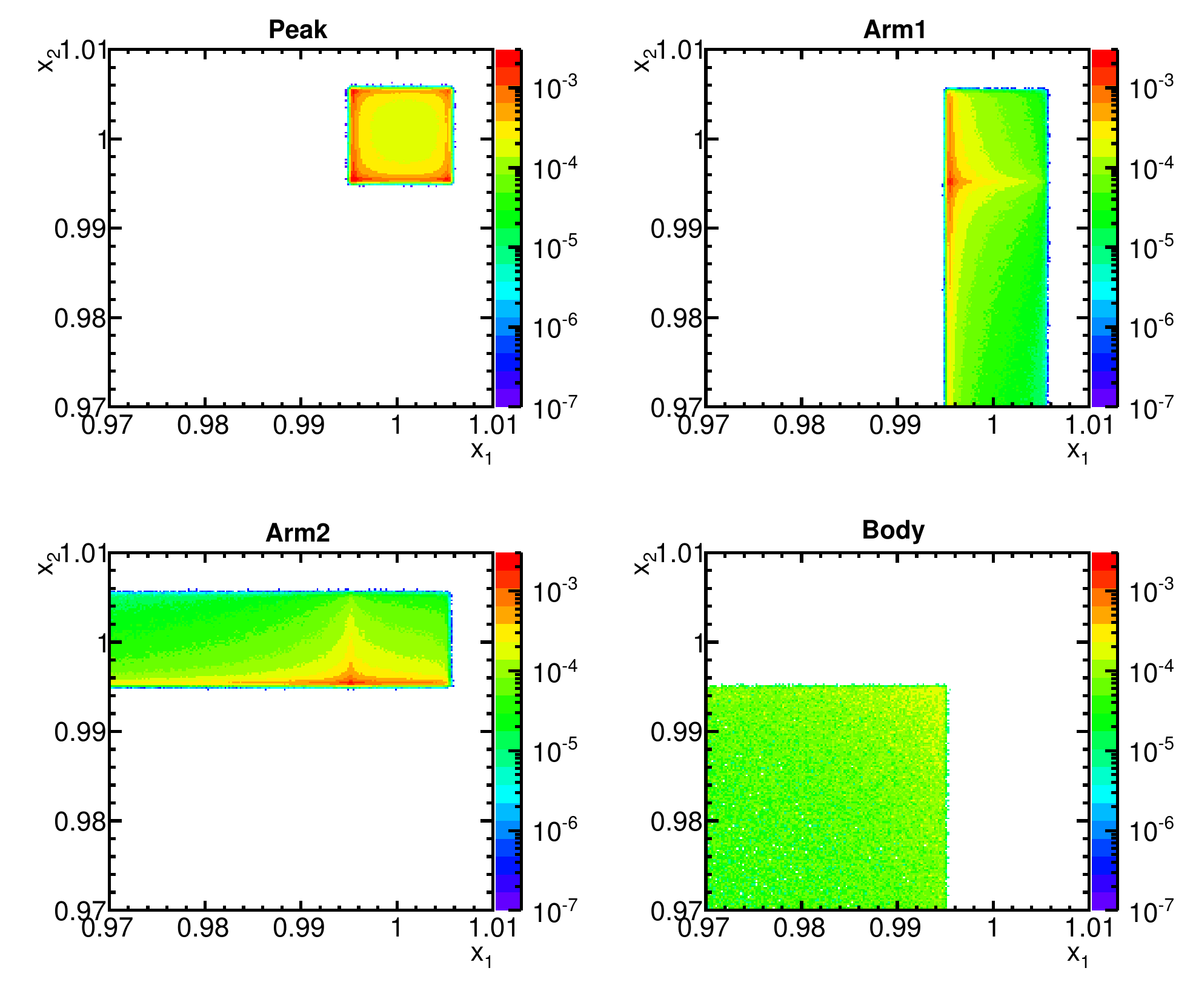}
  \caption{Different parts of the \model, described by Equation~\eqref{eq:mOverlap}.}\label{fig:models}
\end{figure*}

The individual contributions discussed in the previous sections are now used to
create the \model of the basic two-dimensional luminosity spectrum. As was discussed
in Section~\ref{sec:defining-the-bds-model} the beam-energy spread \Bes{x} is described
by a convolution of a Gaussian function $\Gauss{x}$ and a beta-distribu\-tion
$\BetaD{x}$
\begin{equation}
  \label{eq:bes2}
  \Bes{x}  =  \BetaGauss{x}.
\end{equation}
The beta-distribution for the beam-energy spread has a very narrow range.  The
particle energy distribution including the energy loss due to Beamstrahlung is
described by a convolution of the beam-energy spread with an incomplete
beta-distribution with the upper limit of
$\blimit{\mathrm{Arm}}=0.9999$,
\begin{equation}
  \label{eq:beamBes}
  \betabes{x}  =  \BetaBes{x}.
\end{equation}
The upper limit is chosen to be close to 1, so that the convolution with beam-energy spread
causes an overlap with the \peak-region (cf. Figure~\ref{fig:models}).

To describe particle energy distributions only negligibly affected by
the beam-energy spread, a beta-distribu\-tion with an upper limit of
$\blimit{\mathrm{Body}}=0.995$ convoluted with a Gaussian function is used
\begin{equation}
  \label{eq:betagauss}
  \betagauss{x} = \BetaGauss{x}.
\end{equation}
This upper limit separates the distribution from those more significantly affected by the
beam-energy spread. This function is different from \eqref{eq:bes2} due to the
different ranges of the beta-distributions.

As described in Equation~\eqref{eq:piecewise}, the distributions in the four different
regions are described by the product of two functions, one for each
particle. The explicit piecewise description shown in Equation~\eqref{eq:piecewise},
however, is replaced by the use of delta-distribution and implicit ranges of the
individual functions. The \peak region is described by two pure beam-energy
spread functions (Equations~\eqref{eq:bes} or \eqref{eq:bes2}) and delta-distributions to signify the
absence of Beamstrahlung; the \arms are modelled by one beam-energy spread
function and a delta distribution, and one function describing Beamstrahlung
convoluted with the beam-energy spread (Equation~\eqref{eq:beamBes}); the \body is
described by two functions describing only the Beamstrahlung
(Equation~\eqref{eq:betagauss}).
\begin{equation}
\label{eq:mOverlap}
  \begin{aligned}
    \lumispec{\xe,\xp} & =  \pPeak                           & \delta(1-\xe)\conv & \Bes{\xe;\pepeak}                  \\ & & \delta(1-\xp)\conv & \Bes{\xp;\pppeak}                \\
                       & \mspace{1mu}+  \mspace{1.7mu}\pArmA & \delta(1-\xe)\conv & \Bes{\xe;\pearmA}                  \\ & &                    & \betabes{\xp;\pparmA,\blimit{\mathrm{Arm}}} \\
                       & \mspace{1mu}+  \mspace{1.7mu}\pArmB &                    & \betabes{\xe;\pearmB,\blimit{\mathrm{Arm}}}   \\ & & \delta(1-\xp)\conv & \Bes{\xp;\pparmB}                \\
                       & \mspace{1mu}+  \mspace{1.7mu}\pBody &                    & \betagauss{\xe;\pebody,\blimit{\mathrm{Body}}} \\ & &                    & \betagauss{\xp;\ppbody,\blimit{\mathrm{Body}}} ,
  \end{aligned}
\end{equation}
with $\blimit{\mathrm{Arm}}=0.9999$, $\blimit{\mathrm{Body}}=0.995$. In addition, the
constraint
\begin{equation}
  \pBody = 1 - \pPeak - \pArmA - \pArmB
\end{equation}
has to be fulfilled, which results in
\begin{equation}
  \int \lumispec{\xe,\xp} \dd{\xe} \dd{\xp} = 1,
\end{equation}
as required for a probability density function. The function given in
Equation~\eqref{eq:mOverlap} will be used to describe the luminosity spectrum.  The random
variates according to the individual parts of Equation~\eqref{eq:mOverlap} are shown in
Figure~\ref{fig:models}. Each summand of Equation~\eqref{eq:mOverlap} corresponds to one of the
distributions. Due to the convolution of beam-energy spread and Beamstrahlung
functions, the region around the nominal beam energies ($\xe\approx1$, $\xp\approx1$) is
described by a superposition of individual contributions.


%% file: fit.tex
\section{Reweighting Fit}
\label{sec:reweighting-fit}

The separate one-dimensional parts of the luminosity spectrum were fitted to the
parts of the \model. Now the complete \model has to be fit to the
two-dimensional spectrum.

It is possible to fit Equation~\eqref{eq:mOverlap} to the basic luminosity
spectrum. The convolutions with the $\delta$-distribution can be performed
explicitly. The other convolutions have to be performed numerically, because the
convolution between the beta-distribution and the Gaussian function cannot be
expressed in a closed form\footnote{We have no formal proof of this
  statement. However, neither the integral of the Gaussian function (resulting in
  the error-function) nor the integral of the beta-distribution (yielding
  Gamma-functions) can be expressed in a closed form with a
  finite number of elementary functions.}.

For the implementation of the function the numerical convolutions are evaluated
with the QAG\footnote{Quadrature Adaptive General integrand} integration
algorithm~\cite{gsl} interfaced via the \texttt{GSLIntegrator} from
\Root \texttt{MathMore}. The evaluation of the function takes about 160 seconds
for the full range. A direct fit with the function, requiring multiple
iterations, would be slow. The fitting procedure can be sped up by using a
reweighting fit and by exploiting the fact that the random variates according to
Equation~\eqref{eq:mOverlap} can also be described by the sum of the random variates of
the individual functions~\cite{grinstead12:_introd_probab}.

\begin{figure*}[tbp]\sidecaption
  \centering
  \includegraphics[width=0.8\textwidth]{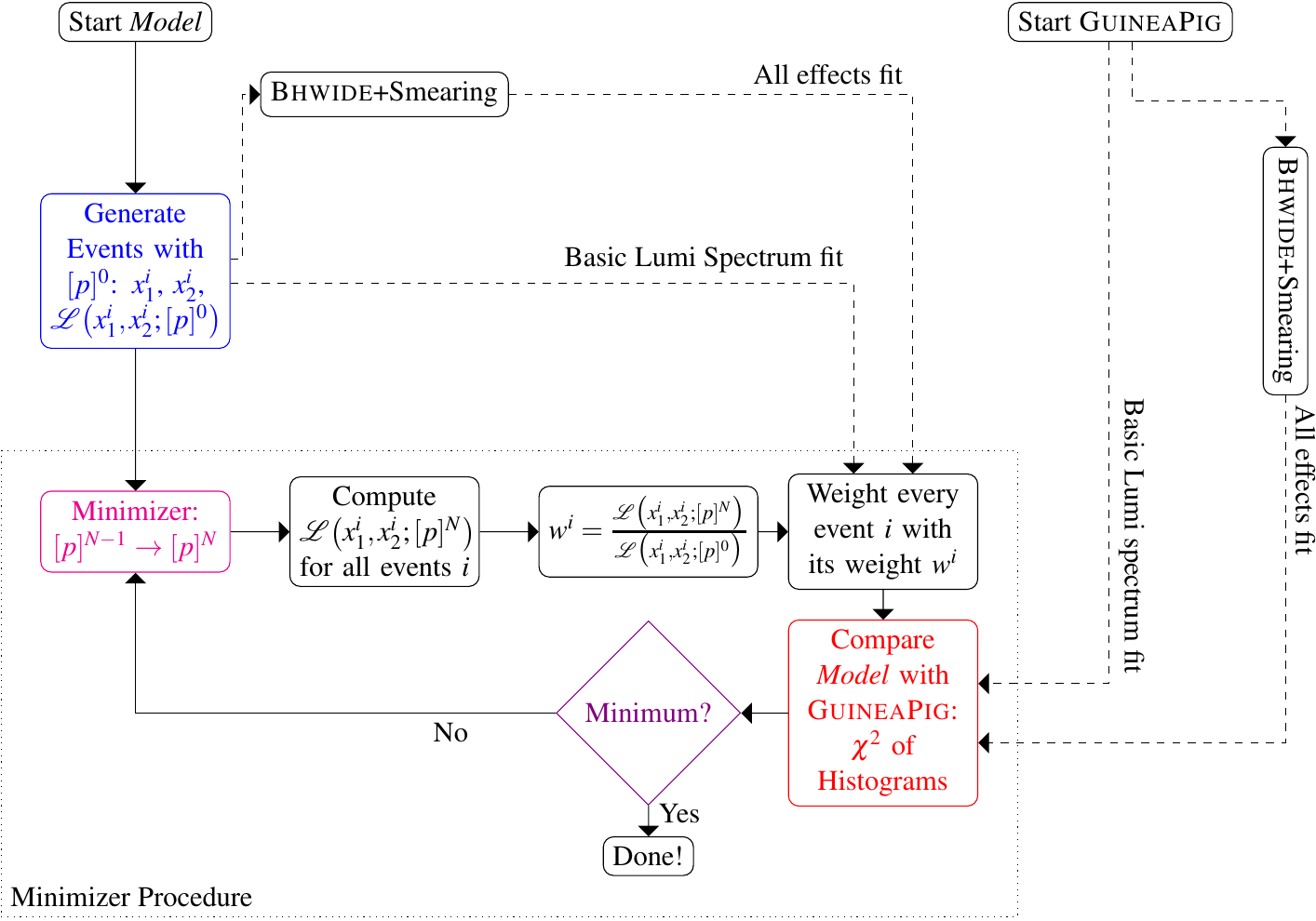}
  \caption{Flowchart diagram of the reweighting fit procedure. The
    dashed arrows indicate the different event samples used for the 
    different fits.}\label{fig:flowchart_reweighting}
\end{figure*}

The principle of the reweighting technique is shown in
Figure~\ref{fig:flowchart_reweighting}.
A \chisquare minimization, utilizing \minuit~\cite{James1975} implemented in
\Root, is used to fit a sample of \model events to the \guineapig
sample. The procedure starts with the generation of a large number of events
according to the \model. This produces events consisting of pairs of beam
energies $(\xe^{\varEv}, \xp^{\varEv})$ and the corresponding probability
$\lumispec{\xe^{\varEv}, \xp^{\varEv}; \parset{0}}$ to obtain a given event. The
probability depends on the initial set of parameter values \parset{0}.

The minimizer is used to obtain a new set of parameter values
\parset{N} that results in new probabilities $\lumispec{\xe^{\varEv},\xp^{\varEv}; \parset{N}}$
for each event. The event weight
\begin{equation}\label{eq:eventweight}
    w^{\varEv} = \frac{\lumispec{\xe^{\varEv},\xp^{\varEv}; \parset{N}}}{\lumispec{\xe^{\varEv},\xp^{\varEv}; \parset{0}}}
\end{equation}
is used to weight each event of the Monte Carlo distribution. For every set of
parameter values \parset{N} a reweighted Monte Carlo distribution is
obtained. The minimum \chisquare between the distribution of the \model and the
distribution from \guineapig corresponds to the optimal parameter values
matching the \guineapig sample. 

The $\chi^{2}$ between the two histograms is calculated from the number of
entries in bin $\varBin$ of the \guineapig sample $\nGuinea^{\varBin}$ and its
uncertainty $\dGuinea^{\varBin}$, the sum of the weights in bin $\varBin$ of the \model
sample $\nModel^{\varBin}$, and the uncertainty $\dModel^{\varBin}$, which are
calculated from the event samples according to
\begin{equation}
  \label{eq:chi2variables}
  \begin{aligned}
  \nGuinea^{\varBin} & = \sum_{\text{GP Events \varEv in Bin \varBin}} 1\, ,           \\
  \dGuinea^{\varBin} & = \sqrt{\nGuinea^{\varBin}} ,                                   \\
  \nModel^{\varBin}  & = \sum_{\text{Model Events \varEv in Bin \varBin}} w^{\varEv} , \\
  \dModel^{\varBin}  & = \sqrt{\sum_{\text{Model Events \varEv in Bin } \varBin} (w^{\varEv})^{2}}.
\end{aligned}
\end{equation}
The $\chi^{2}$ to be minimized is then calculated with
\begin{equation}
  \chi^{2} = \sum_{\text{Bins \varBin}} \frac{\bigl( \nGuinea^{\varBin} - \scaleFac\cdot\nModel^{\varBin} \bigr)^{2}}{(\dGuinea^{\varBin})^{2} + (\scaleFac\cdot\dModel^{\varBin})^{2}},
\end{equation}
where
\begin{equation}
\label{eq:scaleFac}
\scaleFac=\frac{\displaystyle\sum_{\text{GP Events}}{1}}{\displaystyle\sum_{\text{Model Events}~i}{w^{\varEv}}}
\end{equation}
is a scaling factor that takes into account the difference in the sample sizes
and the normalisation of the event weights due to the limited number of
\model-events. The entire procedure has the advantage that only one sample
of \model-events is needed, contrary to traditional template-fit procedures that
require generating new Monte Carlo samples for every parameter-set.

However, by itself this requires even more evaluations of
Equation~\eqref{eq:mOverlap} -- one for every Monte Carlo event used in the
generated sample -- but the random variates according to this function can also
be described by the sum of the random variates of the individual
functions. The particle
energy can be built up from the individual contributions
\begin{equation}
  \label{eq:randVars}
  x_{\mathrm{Particle}} = \xStrahlung{} + \xSpread{} + \xGauss{},
\end{equation}
where \xStrahlung{} is the random variate from the beta-distribution for the
Beamstrahlung, \xSpread{} the random variate from the beta-distribution for the
beam-energy spread, and \xGauss{} the random variate from the Gaussian-function of
the beam-energy spread. Each random variate can be generated according to its
probability density function.  Equation~\eqref{eq:randVars} is connected to
Equation~\eqref{eq:randomVariatesX}: $\xStrahlung{}=x_{\EBeam}+\Delta\xStrahlung{}$,
and $\xSpread{} + \xGauss{}= \Delta\xSpread{}$. During the generation of events,
the combination of functions is chosen according to the probability given by the
parameters for each region $p_{\mathrm{Peak/Arm1/Arm2/Body}}$.  The probability
for a particle's energy in an event is given by the product of all individual
probabilities
\begin{equation}
  P(\xStrahlung{},\xSpread{},\xGauss{}) = \BetaD{\xStrahlung{}} \cdot \BetaD{\xSpread{}} \cdot \Gauss{\xGauss{}},
\end{equation}
and the product of the probabilities for the individual particles multiplied by
the probability for the region gives the probability for the event
\begin{multline}
  \lumispec{\xStrahlung{\ele},\xSpread{\ele},\xGauss{\ele},\xStrahlung{\pos},\xSpread{\pos},\xGauss{\pos}}
  \equiv\\
  p_{\mathrm{Region}} \cdot  P(\xStrahlung{\ele},\xSpread{\ele},\xGauss{\ele}) \cdot P(\xStrahlung{\pos},\xSpread{\pos},\xGauss{\pos}).
\end{multline}
Thus Equation~\eqref{eq:eventweight} becomes
\begin{multline}\label{eq:eventweight2}
    w^{\varEv}=\\
    \frac{\pregion{N}\BetaD{\xStrahlung{\varEv,\ele},\parset{N}} \BetaD{\xSpread{\varEv,\ele},\parset{N}} \Gauss{x^{\varEv,\ele}_{\mathrm{G}},\parset{N}}}%
         {\pregion{0}\BetaD{\xStrahlung{\varEv,\ele},\parset{0}} \BetaD{\xSpread{\varEv,\ele},\parset{0}} \Gauss{x^{\varEv,\ele}_{\mathrm{G}},\parset{0}}}\,\cdot\\
         \frac{\BetaD{\xStrahlung{\varEv,\pos},\parset{N}} \BetaD{\xSpread{\varEv,\pos},\parset{N}} \Gauss{x^{\varEv,\pos}_{\mathrm{G}},\parset{N}}}%
         {\BetaD{\xStrahlung{\varEv,\pos},\parset{0}} \BetaD{\xSpread{\varEv,\pos},\parset{0}} \Gauss{x^{\varEv,\pos}_{\mathrm{G}},\parset{0}}}
\end{multline}
and no numerical convolutions have to be calculated.

The probability for obtaining energies \xe and \xp is not the same as the probability to
obtain a specific group of variates, even if
$\xStrahlung{\ele}+\xSpread{\ele}+\xGauss{\ele}=\xe$ and
$\xStrahlung{\pos}+\xSpread{\pos}+\xGauss{\pos}=\xp$.
There are many combinations of the variates \xSpread{}, \xStrahlung{}, and \xGauss{},
which can lead to the same \xe or \xp. To estimate the probability for any given
pair of energies \lumispec{\xe,\xp} the convolutions have to be performed
either numerically or via Monte Carlo generation.

\subsection{Application of the Reweighting Fit to Other Distributions}
\label{sec:rewfit2}

The reweighting fit is also used to fit the distributions after the inclusion of
the Bhabha scattering, Initial and Final State Radiation, and detector
resolutions.  The individual events are passed through the Bhabha Monte Carlo
generator and detector simulation, which can be understood as additional
convolutions of the existing distribution. As can be seen in
Equation~\eqref{eq:eventweight2}, if the parameter governing one of the contributions does
not change, the contribution does not affect the new weight. This enables the
use of the reweighting fit also for the reconstruction of the spectrum from the
Bhabha events, because the Bhabha scattering and detector resolutions
$\Det{O_{k}}$ are not varied during the fit.
A measured distribution $f$ of Observables $O_{k}$ can be approximately written as
\begin{multline}
  \label{eq:rewfit2}
  f(O_{1},O_{2},\ldots) \approx\\
  \sigma(E_{\ele},E_{\pos};O_{1},O_{2},\ldots) \times \lumispec{E_{\ele},E_{\pos}} \conv
  \ISR{E_{\ele},E_{\pos}} \conv\\
  \FSR{O_{1},O_{2},\ldots} \conv \Det{O_{1}}\Det{O_{2}}\ldots,
\end{multline}
where $\sigma$ represents the centre-of-mass energy dependence of the Bhabha
scattering and the observables, \ISR{E_{\ele},E_{\pos}} represents the probability
for the energy distribution after Initial State Radiation, and
\FSR{O_{1},O_{2},\ldots} represents the probability for the energy distribution after Final
State Radiation.  If the cross-section and detector resolutions are well enough
known, the only difference between the measured and generated distributions is
the luminosity spectrum. For this study the same Bhabha generator and detector
simulations are used for both samples, so the additional effects are
statistically the same. Any difference for the contributions can lead to a
systematic error in the reconstruction of the luminosity spectrum

\subsection{Equiprobability Binning}
\label{sec:binning}

The \chisquare-fit requires binned histograms. To obtain an unbiased estimator
of the compatibility in a \chisquare-fit, all bins should contain at least seven
entries, and the number of events in all bins should be
similar~\cite[p.304]{FJames2006}. These requirements can be fulfilled when an
\emph{equiprobability} binning is generated based on the respective \guineapig
sample used in the fits. With equal-size bins either a large number of bins
could be used -- where most would contain very few or no entries and would have
to be rejected for the $\chi^{2}$ calculation -- and the peak substructure could
be resolved, or fewer bins with larger dimensions could be used, but then the
peak could not be resolved. Therefore, the equiprobability binning can make
better use of the available events.

Following the recipe of James~\cite[p.305]{FJames2006}, the events are first
evenly separated along one axis, and then all events falling in the range on
this first axis are again evenly separated in the second axis. If additional
dimensions are used, the separation is repeated. In this way each bin has
different dimensions along each axis, but the number of events per bin is
constant. 

\begin{figure*}[tbp]\sidecaption
  \centering
  \captionsetup[subfloat]{captionskip=4pt}
  \subfloat[]{\label{fig:binning2d}\includegraphics[width=\halfwidth,height=\halfwidth,keepaspectratio=false]{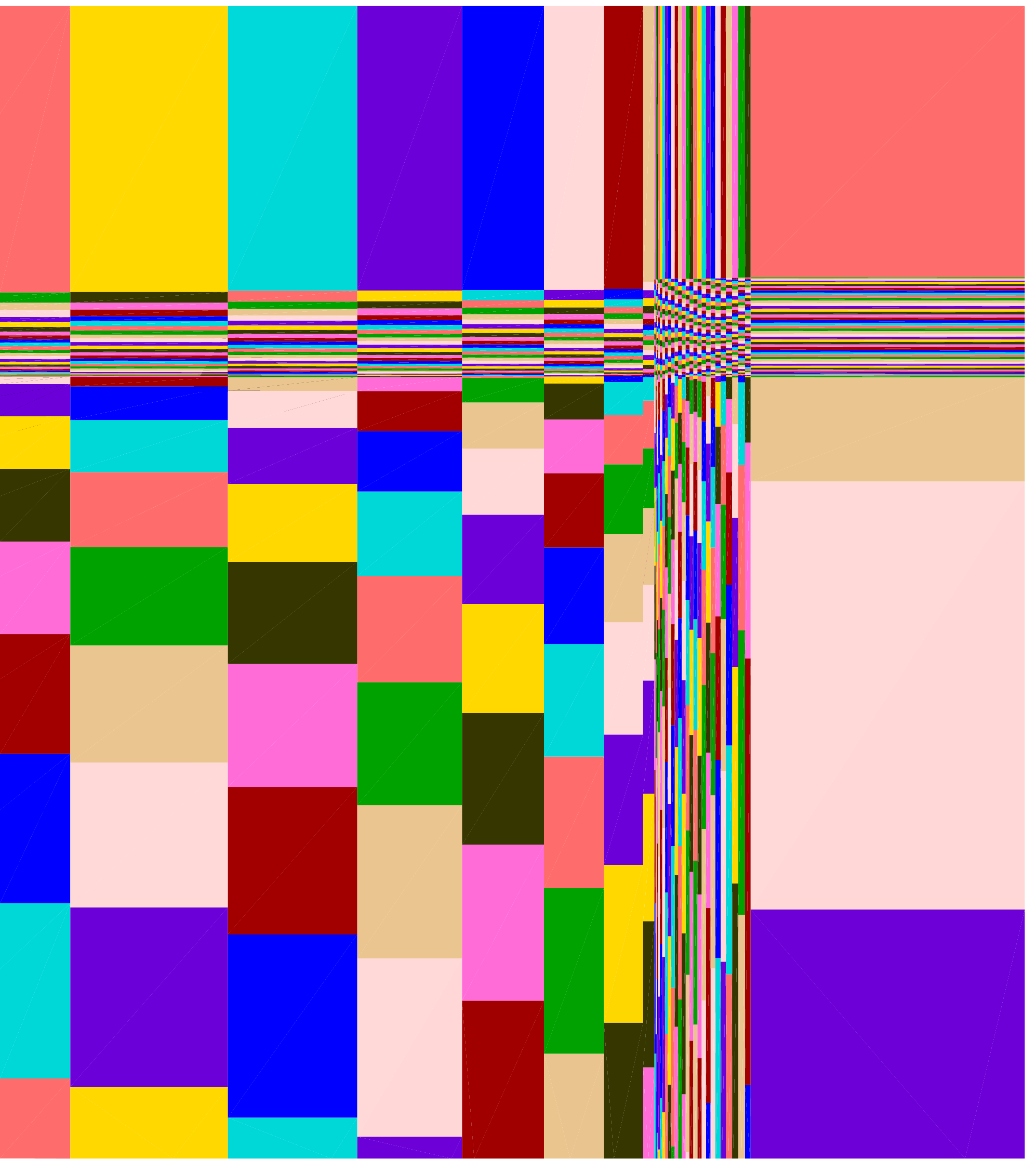}}
  \subfloat[]{\label{fig:binning3d}\includegraphics[width=\halfwidth]{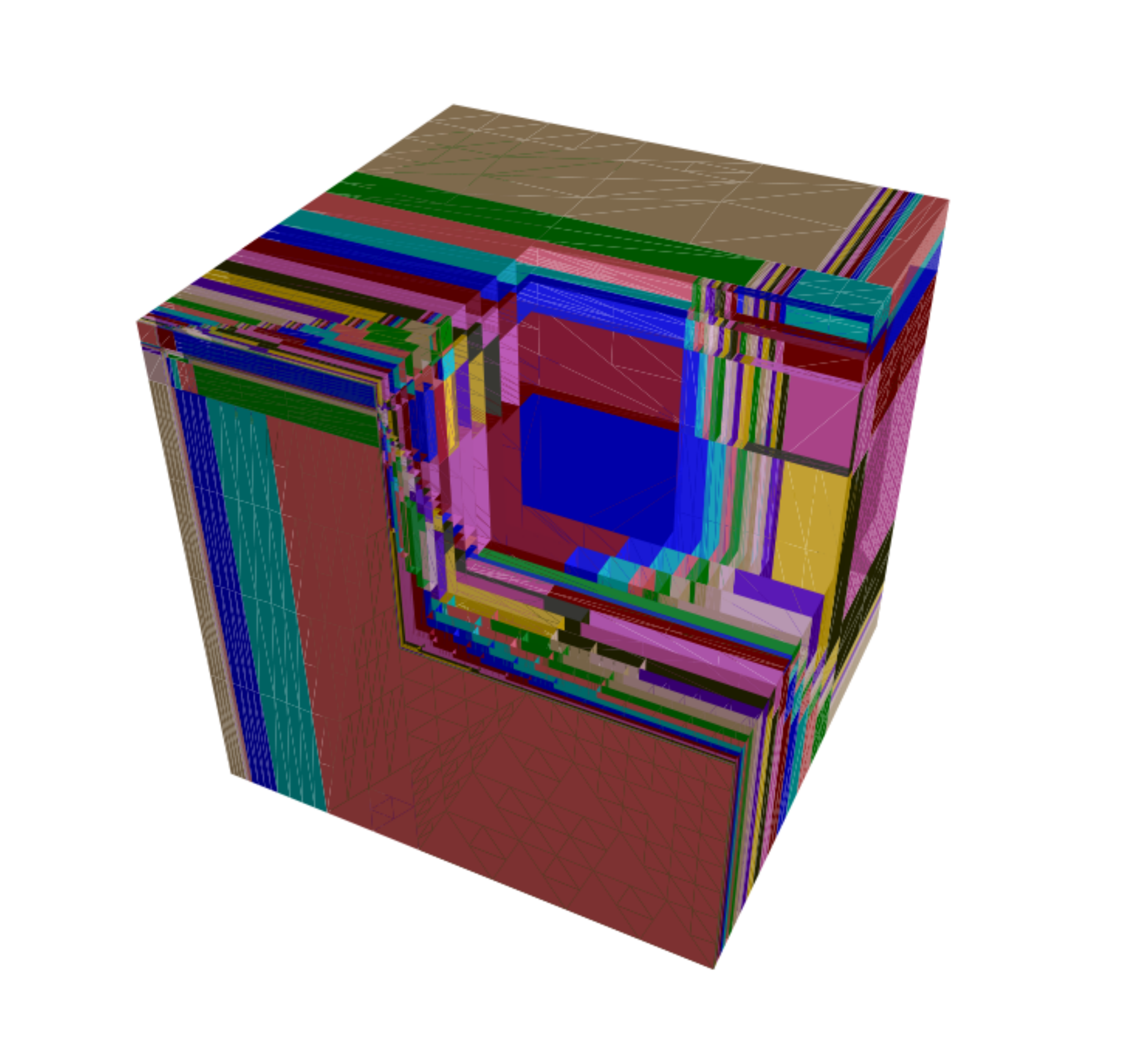}}
  \caption{Example of the equiprobability binning in
    \Subref{fig:binning2d}~two dimensions (zoomed to the peak region)
    \Subref{fig:binning3d}~and three dimensions. The colours are arbitrary. By
    construction every cell contains a similar number of
    events.}\label{fig:binning}
\end{figure*}

For the fit to the \emph{basic} luminosity spectrum, as discussed in Section~\ref{sec:genfit}, the distribution
of the two particle energies is stored in a two-dimensional histogram. For the
reconstruction of the spectrum from the Bhabha events the energy of the scattered
electron and positron, and the relative centre-of-mass energy
reconstructed from the acollinearity $\sfrac{\rootsaco}{\rootsnom}$ are filled into a three-dimensional equiprobability
histograms. Figure~\ref{fig:binning} shows examples for a two- and three-dimensional bin
structure. It can be seen that around the nominal beam energies the size of the
bins becomes smaller. Because the separation of events is done individually
along each axis, the bin structures are not symmetric.

\subsection{Uncertainty Estimation}\label{sec:mcmcfit}

In order to ensure that the \model is unbiased and consistent, a large
number of \model vs.\ \model fits were performed with a varying number
of bins. In each case, the procedure is as
follows: two sets of events are created according to the \model of the \emph{basic}
luminosity spectrum. The samples are then used in the fit procedure described
before. In each fit a cut on the centre-of-mass energy of $\rootsprime >
1.5~\mathrm{TeV}$ is applied.

The pull distribution of every free parameter is obtained and fitted with a
Gaussian function. The \model is unbiased, if the mean of every pull
distribution is close to zero. The uncertainty is correctly estimated, if the pull
width is compatible with unity. This is called the Normality
condition~\cite[p.310]{FJames2006}.

For all parameters the pull distributions are independent of the binning. Most
parameters are unbiased (i.e., the mean is zero). Exceptions are the \abodyAA
and \abodyBA parameters, whose pulls are not normally distributed. These
parameters describe the behaviour of the beta-distribution at the lower edge of
the respective particle energy distribution. Therefore, the bias is caused by
the cut on the centre-of-mass energy, which reduces the sensitivity to the lower
energy part of the \body. The lower limit of these parameters is zero, which is
often found by the minimizer instead of the nominal value. When the cut on the
centre-of-mass energy is removed, the pulls are symmetric, and the parameters
are correctly estimated. It is also found that the width of the Gaussian
function is consistent with unity, so the uncertainties are correctly estimated
by the minimization procedure.


%% file: initialElectronFit.tex
\section{Luminosity Spectrum Reconstruction}
\label{sec:lumin-spectr-reconst}

All ingredients for the reconstruction of the luminosity spectrum -- the \model
and the reweighting procedure -- are now available. 

The fits based on the \emph{basic} luminosity spectrum
(Section~\ref{sec:genfit}) are used to assess the similarity between the \model
and the \guineapig spectrum. The energies of the electron and positron pairs are
filled into the two-dimensional equiprobability structure used in the
reweighting fit. In the next step, the cross-section scaling, the Bhabha
scattering, and the smearing for the detector-resolutions are applied and the
reweighting fit is done with the observables defined in
Section~\ref{sec:observables-resolutions}. The step-by-step inclusion of the
intermediate steps and their impact on the reconstruction is detailed elsewhere~\cite{LCD-2013-008}.

The initial values of the parameters, used to generate the \model events, are
given in Table~\ref{tab:iniPara}. All the regions are chosen to start with a
similar number of events (25\%). The starting \beamspreadpara{}{} parameters are
taken from the fit to the beam-energy spread before the collisions
(Table~\ref{tab:besFitPar}). The other parameters are chosen arbitrarily in a
way to cause a behaviour similar to the \guineapig luminosity spectrum. The
position of the two boundaries for the beam-energy spread (\xmin and \xmax) are
also taken from Table~\ref{tab:besFitPar}. Table~\ref{tab:iniPara} also lists
the lower and upper bounds limiting the values for the minimizer.

\begin{table}[tbp]
  \centering
  \caption{Initial parameter values used for the generation of the events. Also
    listed are the lower and upper bounds used in the reweighting fits.}
  \label{tab:iniPara}
  \begin{tabular}{l R{2}{1} R{2}{6} R{2}{1} }\toprule
    Parameter        & \tabt{Lower} & \tabt{Nominal Value} & \tabt{Upper} \\\midrule
    \tabt{\pPeak}   & 0.0                & 0.25                 &  0.4                \\
    \tabt{\pArmA}   & 0.0                & 0.25                 &  0.3                \\
    \tabt{\pArmB}   & 0.0                & 0.25                 &  0.3                \\
    \tabt{\bPeakAA} & -1.0               & -0.522336            &  0.0                \\
    \tabt{\bPeakAB} & -1.0               & -0.409289            &  0.0                \\
    \tabt{\bPeakBA} & -1.0               & -0.522336            &  0.0                \\
    \tabt{\bPeakBB} & -1.0               & -0.409289            &  0.0                \\
    \tabt{\barmAA}  & -1.0               & -0.522336            &  0.0                \\
    \tabt{\barmAB}  & -1.0               & 0.35                 & 10.0                \\
    \tabt{\barmBA}  & -1.0               & -0.522336            &  0.0                \\
    \tabt{\barmBB}  & -1.0               & 0.35                 & 10.0                \\
    \tabt{\aarmAA}  & 0.0                & 2.5                  & 10.0               \\
    \tabt{\aarmAB}  & -1.0               & -0.75                &  0.0                \\
    \tabt{\aarmBA}  & 0.0                & 2.5                  & 10.0               \\
    \tabt{\aarmBB}  & -1.0               & -0.75                &  0.0                \\
    \tabt{\abodyAA} & 0.0                & 0.15                 & 10.0               \\
    \tabt{\abodyAB} & -1.0               & -0.55                &  0.0                \\
    \tabt{\abodyBA} & 0.0                & 0.15                 & 10.0               \\
    \tabt{\abodyBB} & -1.0               & -0.55                &  0.0                \\\bottomrule
  \end{tabular}
\end{table}

\subsection{Fit to the Basic Luminosity Spectrum}\label{sec:genfit}

To verify that the \model can represent the basic luminosity spectrum from \guineapig,
the distribution of the initial particle energies are used in the
\chisquare-fit. The data histogram is shown in Figure~\ref{fig:2dSpectra}. The Monte
Carlo sample is shown in Figure~\ref{fig:models}. The \guineapig sample consists of 3
million events and the \model provides 10 million events. Fits are done with a binning varied
from $50\times50$ bins to $300\times300$ bins in steps of 10 bins. Only events with
$\rootsprime > 1.5~\mathrm{TeV}$ are used in the fit. The cut is applied because
the \model has limited validity range, and the events below half the nominal
centre-of-mass energy would have a negative impact on the fit result.

\begin{figure*}[tbp]\sidecaption
  \centering
  \subfloat[]{\label{fig:binsGen}\includegraphics[width=\halfwidth]{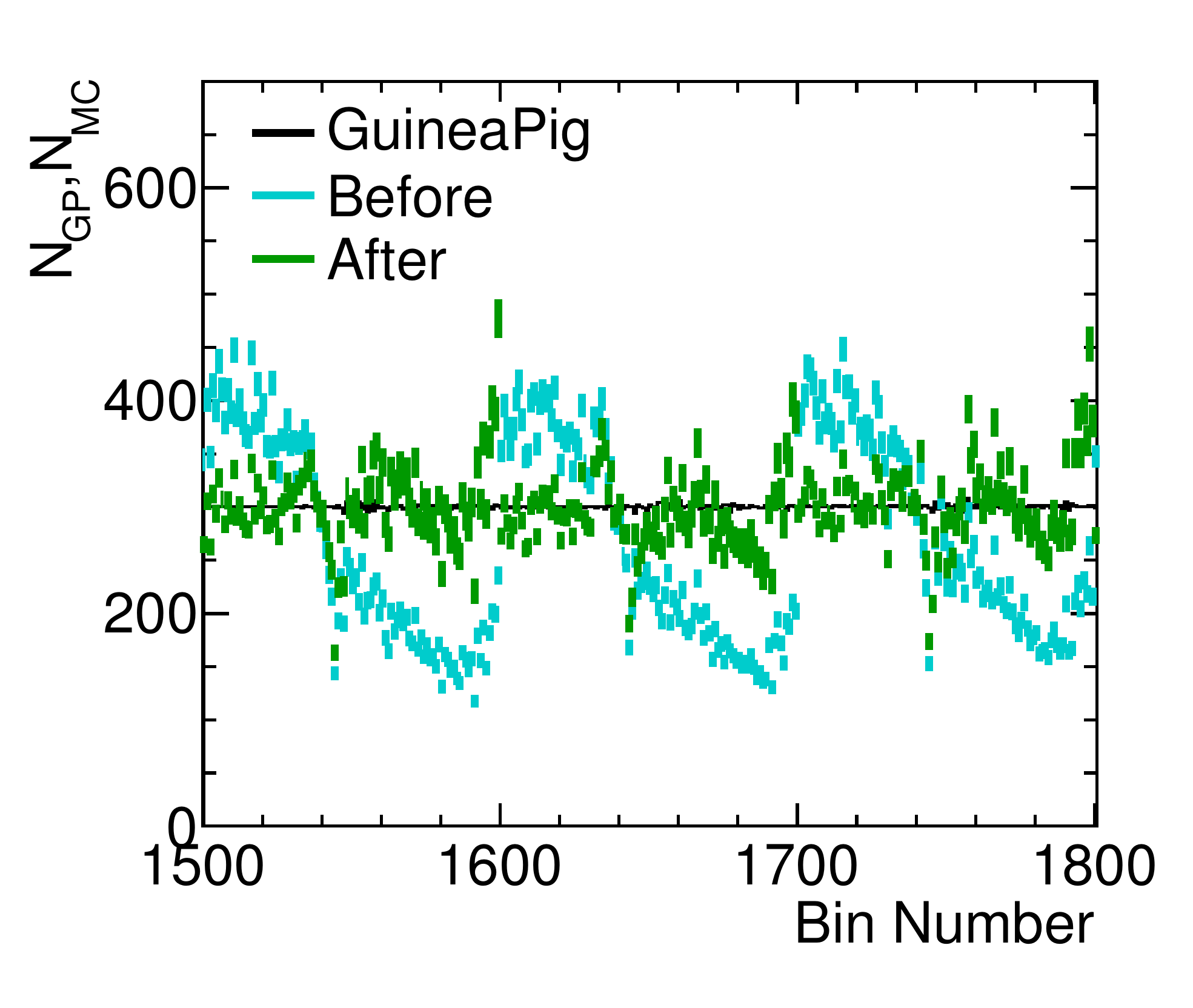}}
  \subfloat[]{\label{fig:pullGen}\includegraphics[width=\halfwidth]{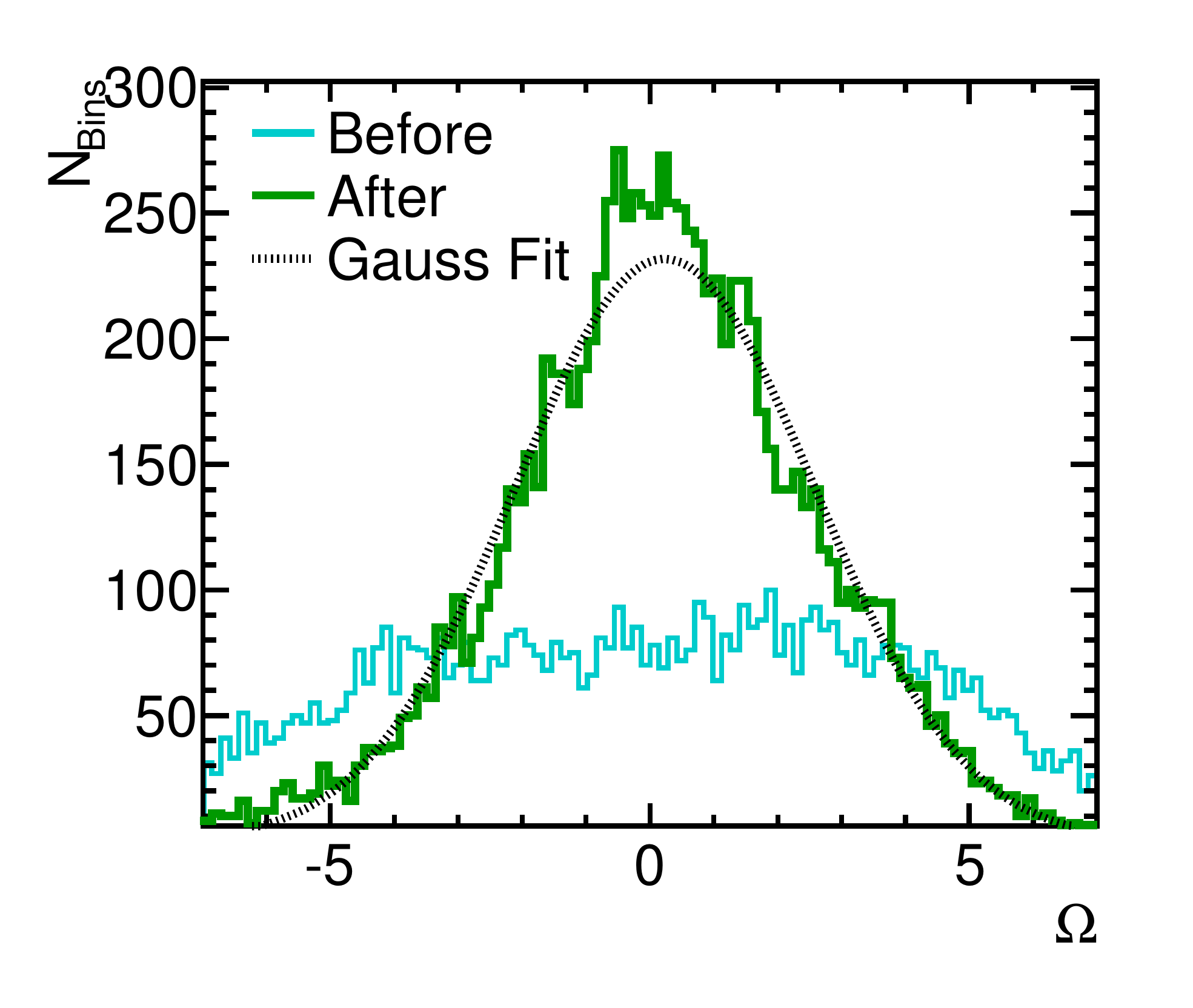}}
  \caption{\Subref{fig:binsGen}~Blow-up of a small section of the bins used in
    the re-weighting fit of the initial electron energies. The histogram for
    \guineapig (black) has, by construction, a constant number of events per
    bin. Also shown are the histograms for the \model with the initial parameter
    values before the fit and after the fit. \Subref{fig:pullGen} Distribution
    of the pulls $\Omega=\sfrac{\bigl(\nGuinea^{\varBin} -
      \scaleFac\nModel^{\varBin} \bigr)}{\bigl((\dGuinea^{\varBin})^{2} +
      (\scaleFac\dModel^{\varBin})^{2}\bigr)^{\sfrac{1}{2}}}$ (symbols are
    defined in Equations~\eqref{eq:chi2variables} and \eqref{eq:scaleFac}) between the \guineapig and
    \model samples before and after the re-weighting fit. A Gaussian function is
    fitted to the distribution of pulls after the fit.}
  \label{fig:binpulls}
\end{figure*}

As an example for the result of the reweighting fit, Figure~\ref{fig:binpulls} shows
a small section of the histogram mapped onto one dimension and the pull
distribution for all the bins before and after the fit. The data histogram has a
constant number of events per bin, as designed by the equiprobability
binning. The pull distribution after the fit converged is well centred around 0
with a width of 2.3. The width of the pull distribution is not equal to
1, because the \chindf is larger than 1. This means that the \model is not
completely identical to the \guineapig distribution.  Two of the differences
are the limited number of beta-distributions used to model the tail of the
spectrum (see Section~\ref{sec:beamstrahlung}), where deviations appear, and the
differences in the peak of the spectrum (see Section~\ref{sec:defining-the-bds-model}),
where a much larger number of parameters would be needed. As the \chindf is not
equal to unity, additional parameters would enable a better description of the spectrum.

For one fixed binning, the fit to the luminosity spectrum was done 198 times
with the same \guineapig sample and independent \model samples. All the
parameter values vary within their uncertainties. Therefore, the \model and the
fit procedure are consistent.


%% file: bhabhaFit.tex
\subsection{Fits to the Observables}
\label{sec:fits-observables}

The observables are defined in
Section~\ref{sec:observables-resolutions}. Binnings\footnote{For the number of bins
  given, the first number represents the number of bins for the relative
  centre-of-mass energy, and the second and third number represents the number
  of bins for the two particle energies.} from $10\times10\times10$ bins to
$80\times50\times50$ bins were used in the fits. The binning step is 5 bins for
the relative centre-of-mass energy and 10 bins for the particle energies.

For the last step, the Bhabha events generated with the scaled luminosity spectrum
are smeared with the detector resolutions as described in
Section~\ref{sec:observables-resolutions}. The selection cuts had to be modified, and
the cut is applied on the centre-of-mass energy calculated from the smeared
four-vectors $\rootsvec > 1.5~\mathrm{TeV}$ and in addition on the individual
particle energies \mbox{$E_{\ele} > 150~\mathrm{GeV}$} and \mbox{$E_{\pos} >
150~\mathrm{GeV}$}. To recover Final State Radiation, the energy of all photons
in a 3\degrees cone around an electron is summed up.

\subsection{Discussion of the Results}
\label{sec:discuss}
\input{results}

For the two stages of the reconstruction multiple fits with different binnings were
done. However, as the reconstructed spectra are fairly similar, only one
reconstructed luminosity spectrum per stage is shown in detail. In addition, the
parameter dependence on the number of bins is shown. For the fits to the basic
and scaled luminosity spectrum the results with $100\times100$ bins are
shown. For the reconstruction from the observables the fits with
$40\times50\times50$ bins are shown.

In Table~\ref{tab:overlapResults} the \chindf and parameters extracted by the selected
fit stages are listed. The reconstructed parameters are far away from the
initial values of the parameters (cf.~Table~\ref{tab:iniPara}), therefore the fit results are
not artificially improved by using a good starting point. The final values of
the beam-energy spread parameters \beamspreadpara{}{} are close to the values
found by the one-dimensional fit to the beam-energy spread distributions
detailed in Section~\ref{sec:defining-the-bds-model}. Because of the cut on the minimum
centre-of-mass energy, the sensitivity on the lower Beamstrahlung parameter
$\beamstrahlungpara{\betaA}{}$ is lost, and both fits give a result of
$\approx0$ with large uncertainty for these parameters. The reconstruction of the upper Beamstrahlung parameter
$\beamstrahlungpara{\betaB}{}$ is consistent.

The largest variation in the parameters is observed for the beam-spread
parameters \beamspreadpara{}{}. This increase is mostly due to the detector
effects. In total the uncertainty increases by a factor ten, and the values are
significantly different.

There are significant correlations between the parameters. The largest
correlations are between parameters from the same beta-distribution. The
correlations also increase when the additional effects are taken into
account. Some of the changes of the parameter values could, therefore, be due to increased
correlations.


\begin{figure*}[ptb]
  \centering%
    \subfloat[40 GeV Bins]{\label{fig:genS00}\includegraphics[width=0.33\textwidth,clip,trim=0 5 0 12]{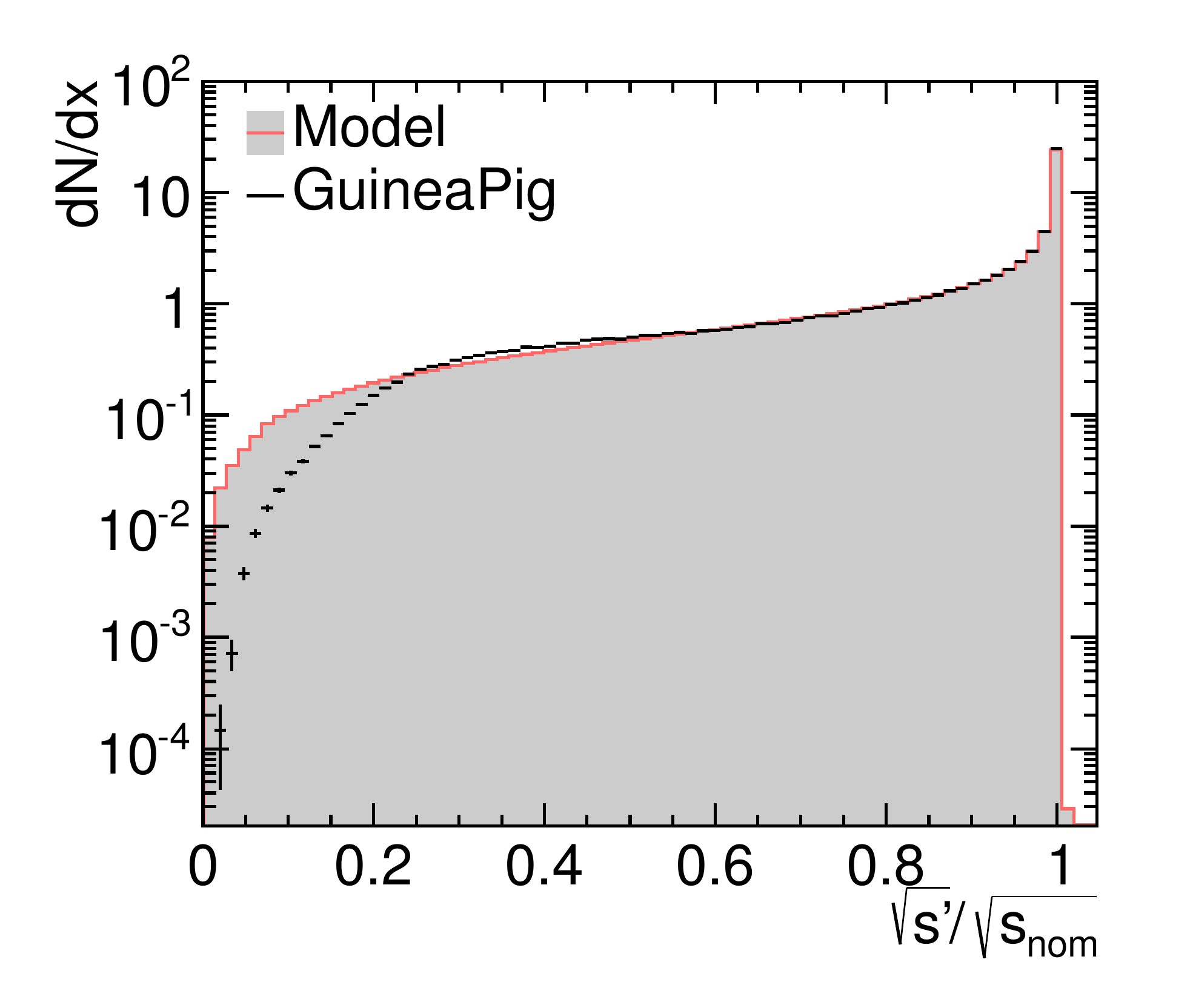}}
    \subfloat[20 GeV Bins]{\label{fig:genS50}\includegraphics[width=0.33\textwidth,clip,trim=0 5 0 12]{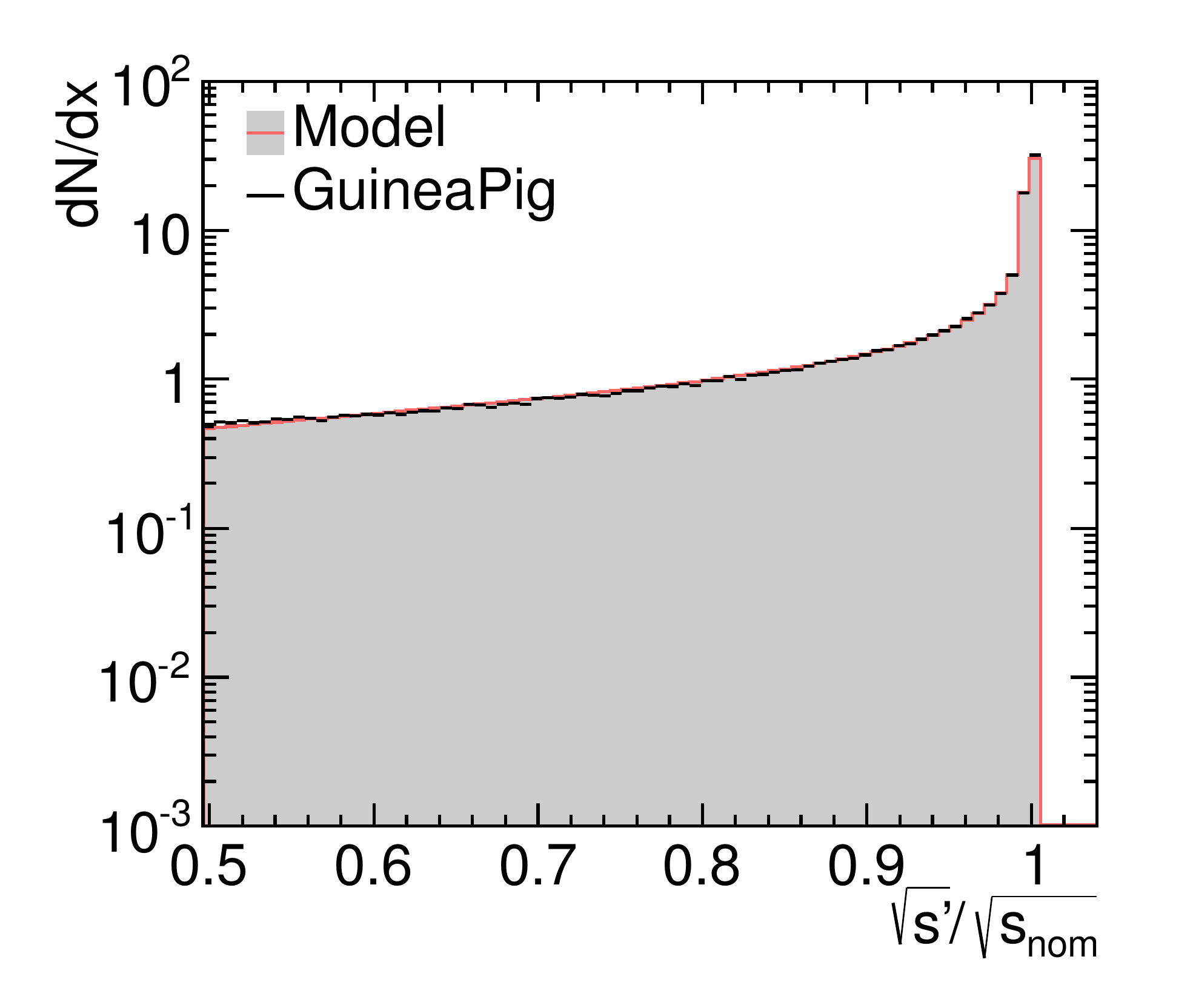}}
    \subfloat[2 GeV Bins]{\label{fig:genS90}\includegraphics [width=0.33\textwidth,clip,trim=0 5 0 12]{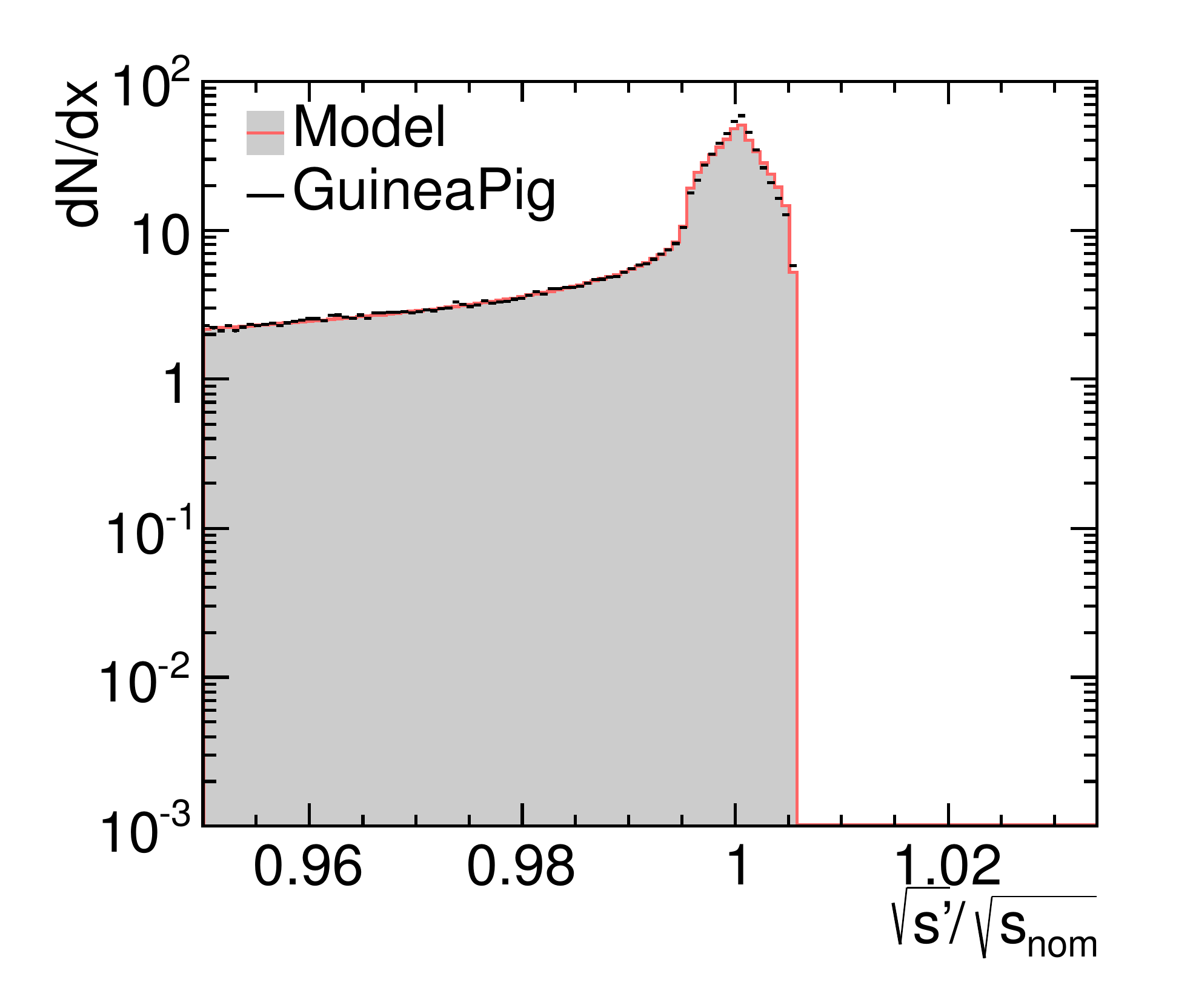}}\\\vspace{-12pt}
    \subfloat[40 GeV Bins]{\label{fig:genR00}\includegraphics[width=0.33\textwidth,clip,trim=0 5 0 12]{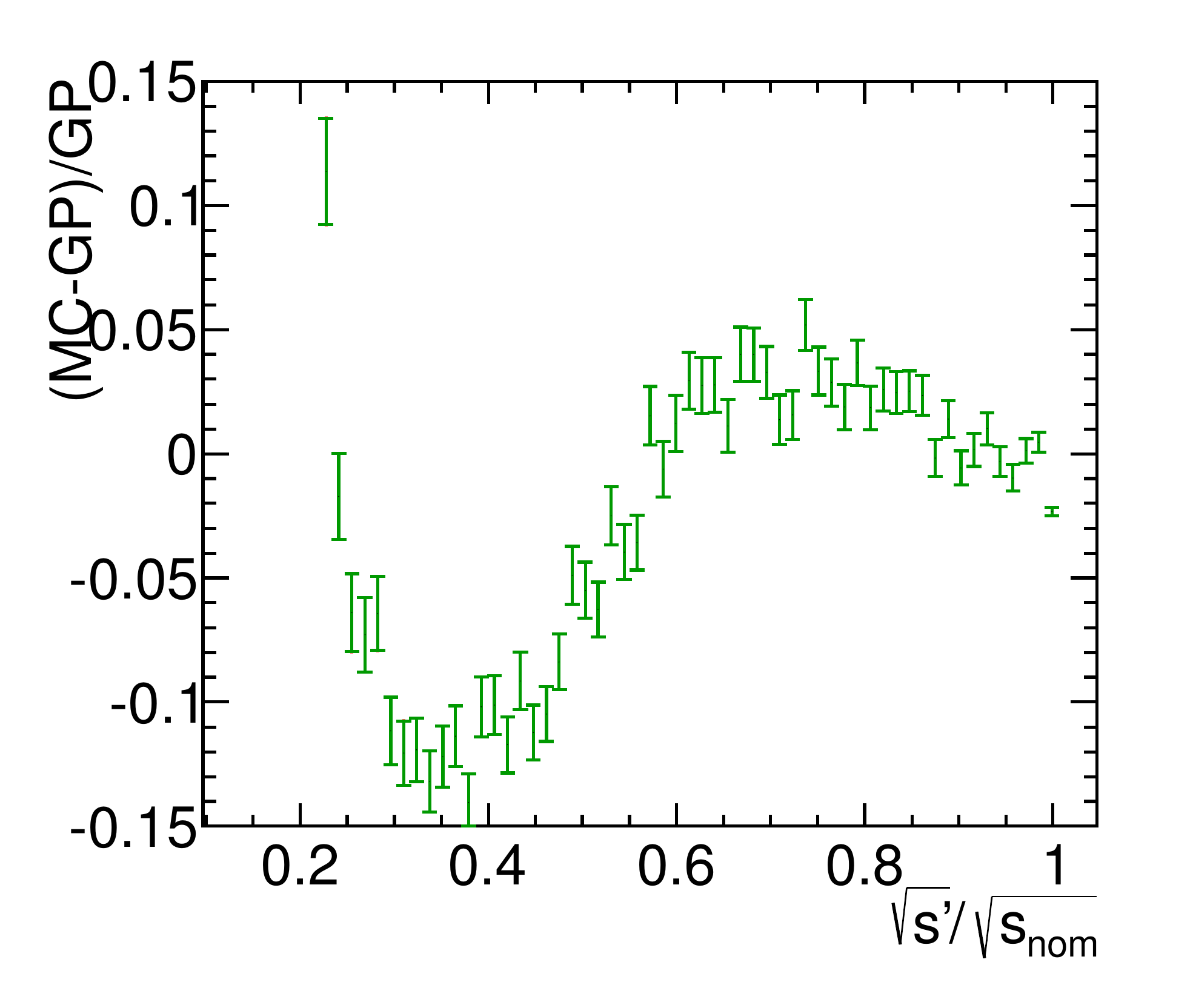}}
    \subfloat[20 GeV Bins]{\label{fig:genR50}\includegraphics[width=0.33\textwidth,clip,trim=0 5 0 12]{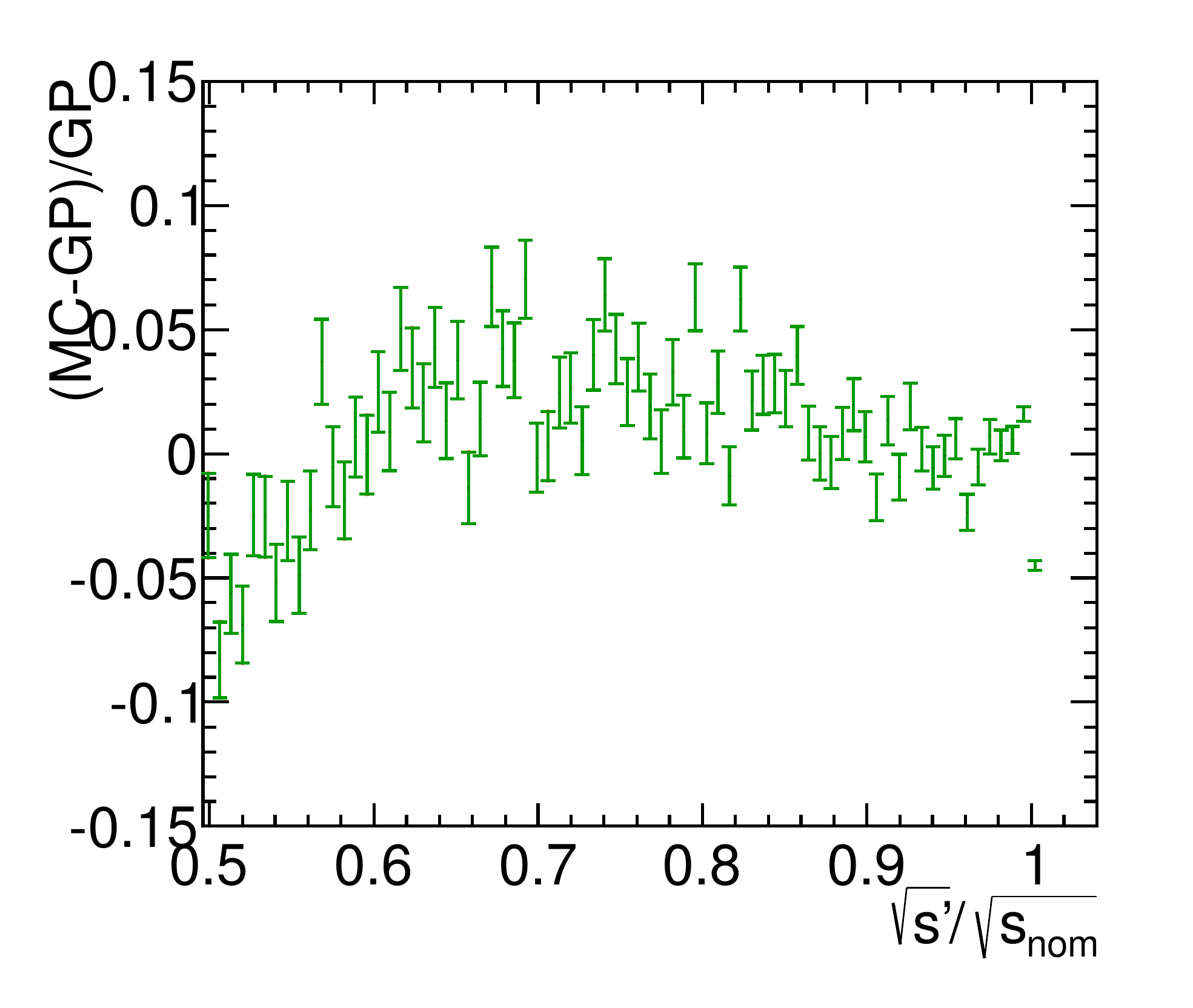}}
    \subfloat[2 GeV Bins]{\label{fig:genR90}\includegraphics [width=0.33\textwidth,clip,trim=0 5 0 12]{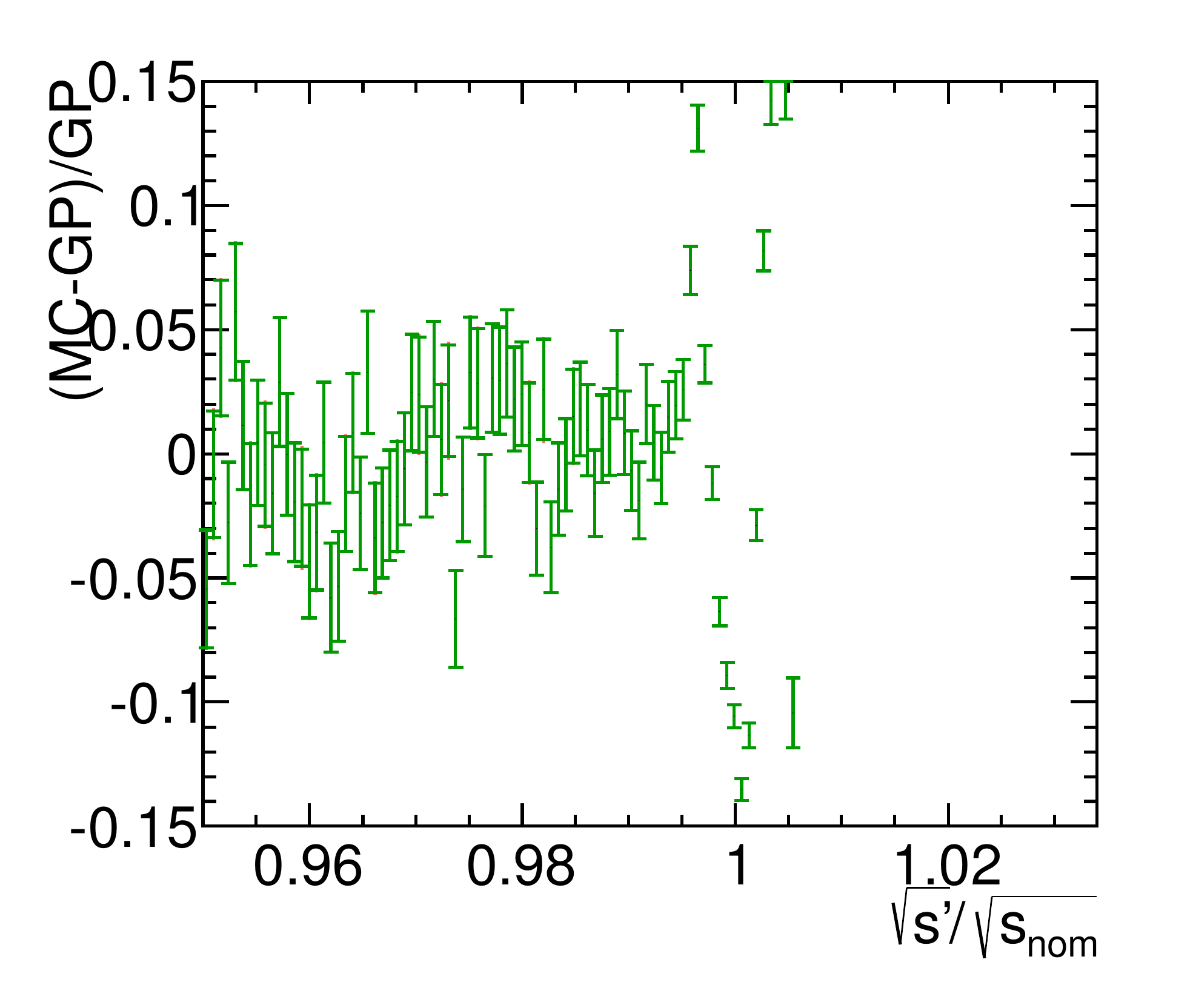}}
  \caption{Resulting spectra for the fit to the basic luminosity spectrum and a
    binning of \mbox{$100\times100$}. \Subref{fig:genS00}--\Subref{fig:genS90}~\spectraPhrase
    \Subref{fig:genR00}--\Subref{fig:genR90}~\ratioPhrase}
  \label{fig:fitgen}
\end{figure*}

\begin{figure*}[pbt]\sidecaption
  \includegraphics[width=0.8\textwidth]{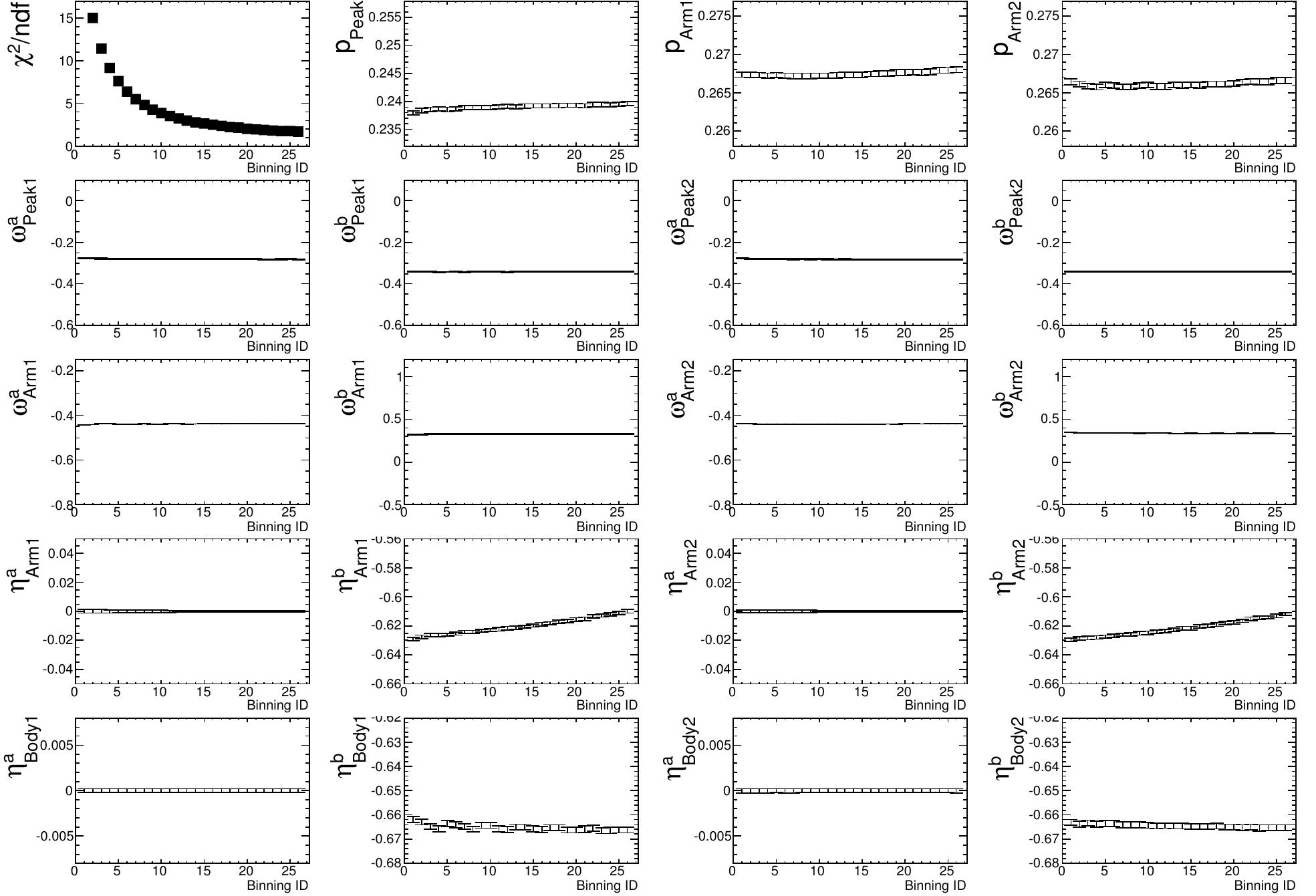}
  \caption{Parameter variation with respect to
  the Binning ID for the fits to the basic luminosity spectrum. The entries are
  sorted by falling \chisquare.}\label{fig:parSpreadGen}
\end{figure*}%

\begin{figure*}[ptb]
  \centering%
    \subfloat[40 GeV Bins]{\label{fig:BHWS_S00}\includegraphics[width=0.33\textwidth,clip,trim=0 5 0 12]{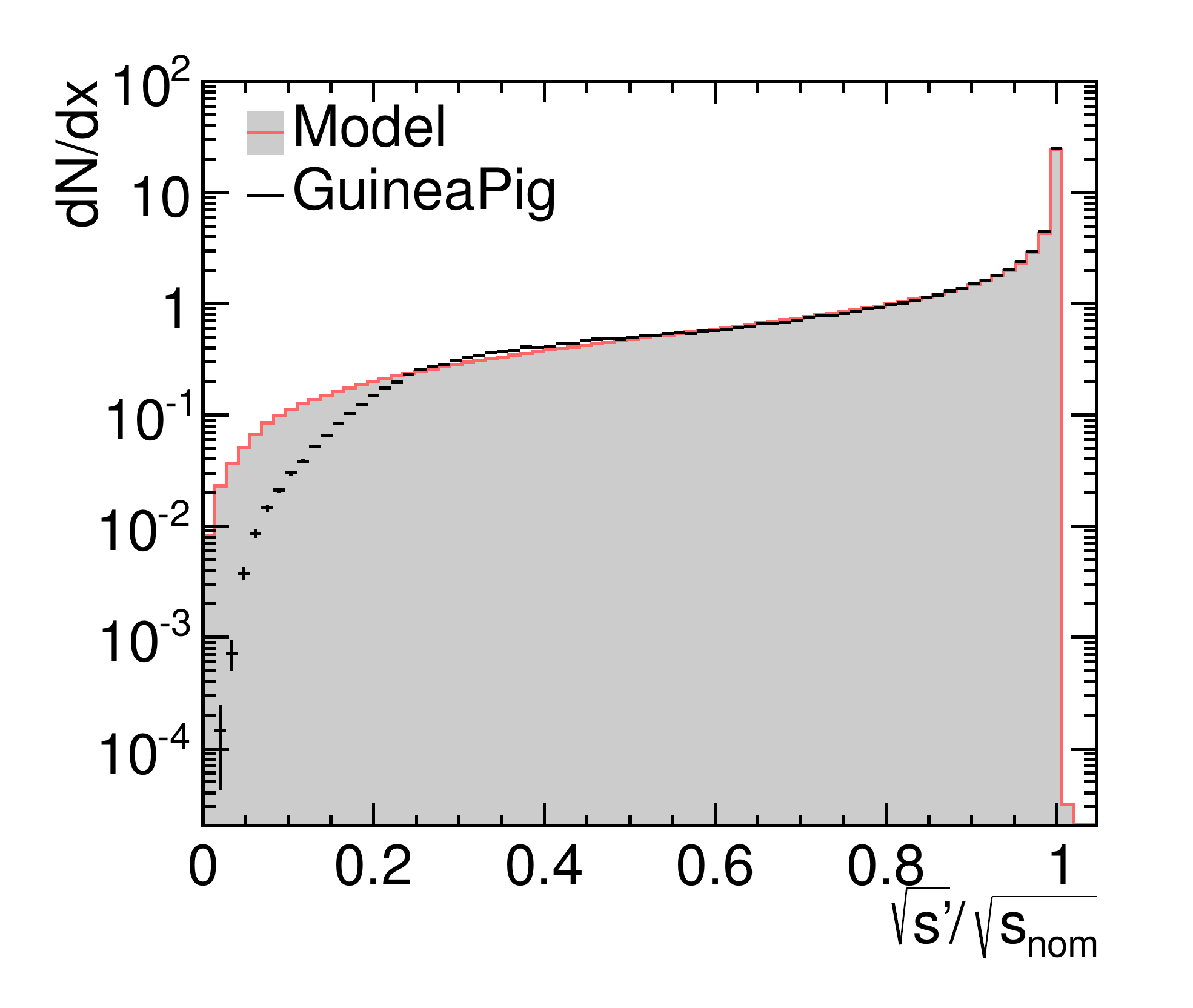}}
    \subfloat[20 GeV Bins]{\label{fig:BHWS_S50}\includegraphics[width=0.33\textwidth,clip,trim=0 5 0 12]{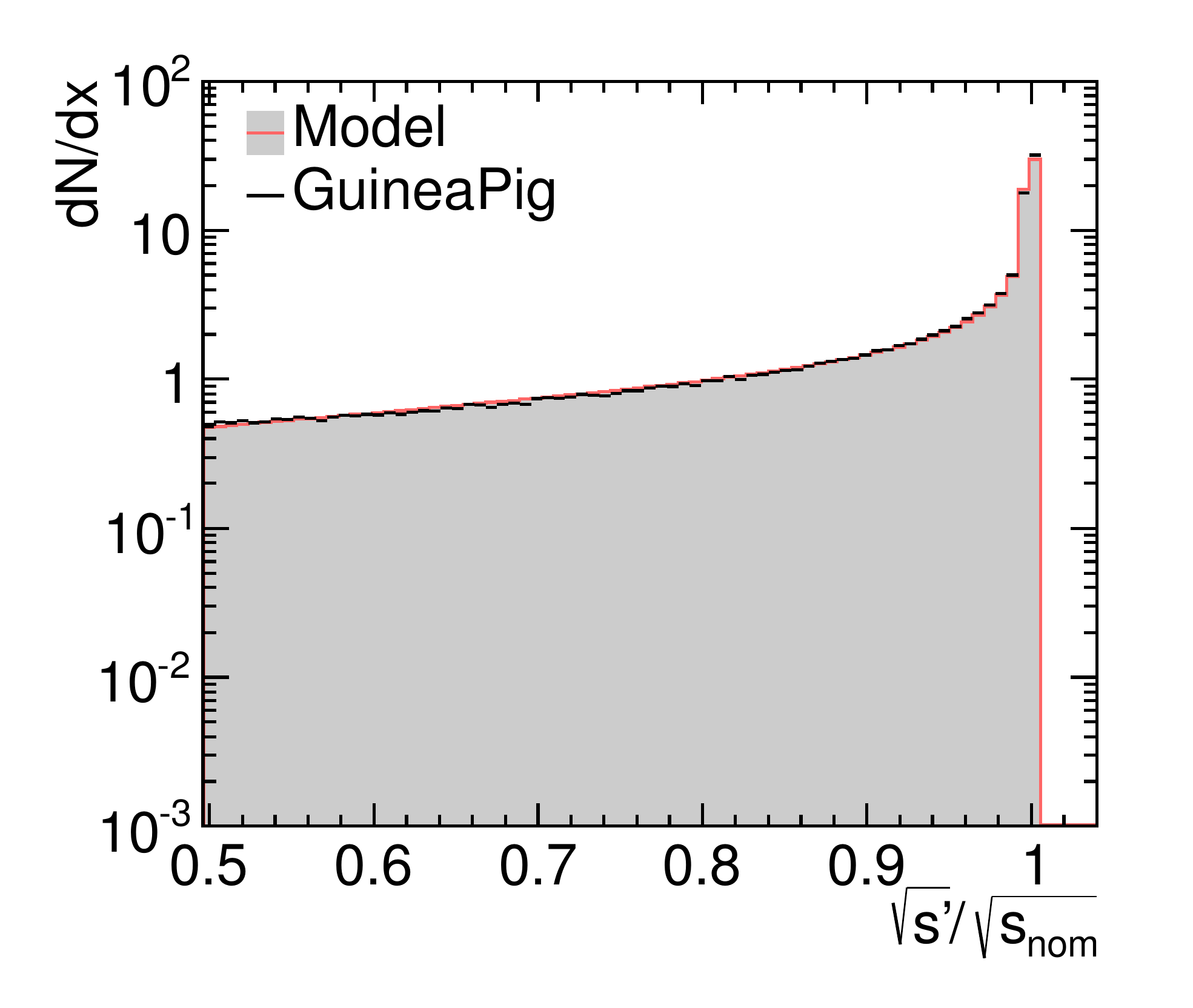}}
    \subfloat[2 GeV Bins]{\label{fig:BHWS_S90}\includegraphics [width=0.33\textwidth,clip,trim=0 5 0 12]{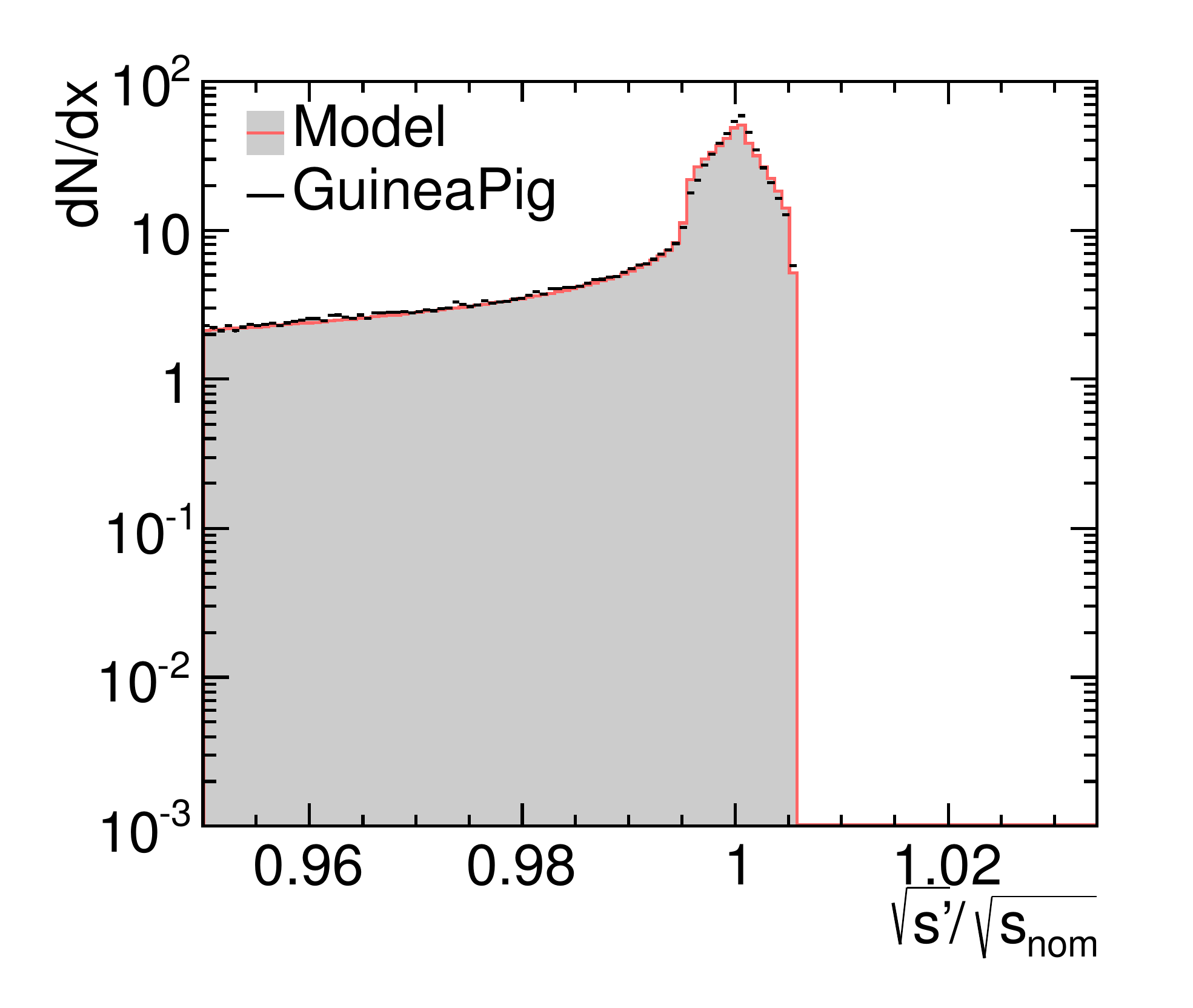}}\\\vspace{-12pt}
    \subfloat[40 GeV Bins]{\label{fig:BHWS_R00}\includegraphics[width=0.33\textwidth,clip,trim=0 5 0 12]{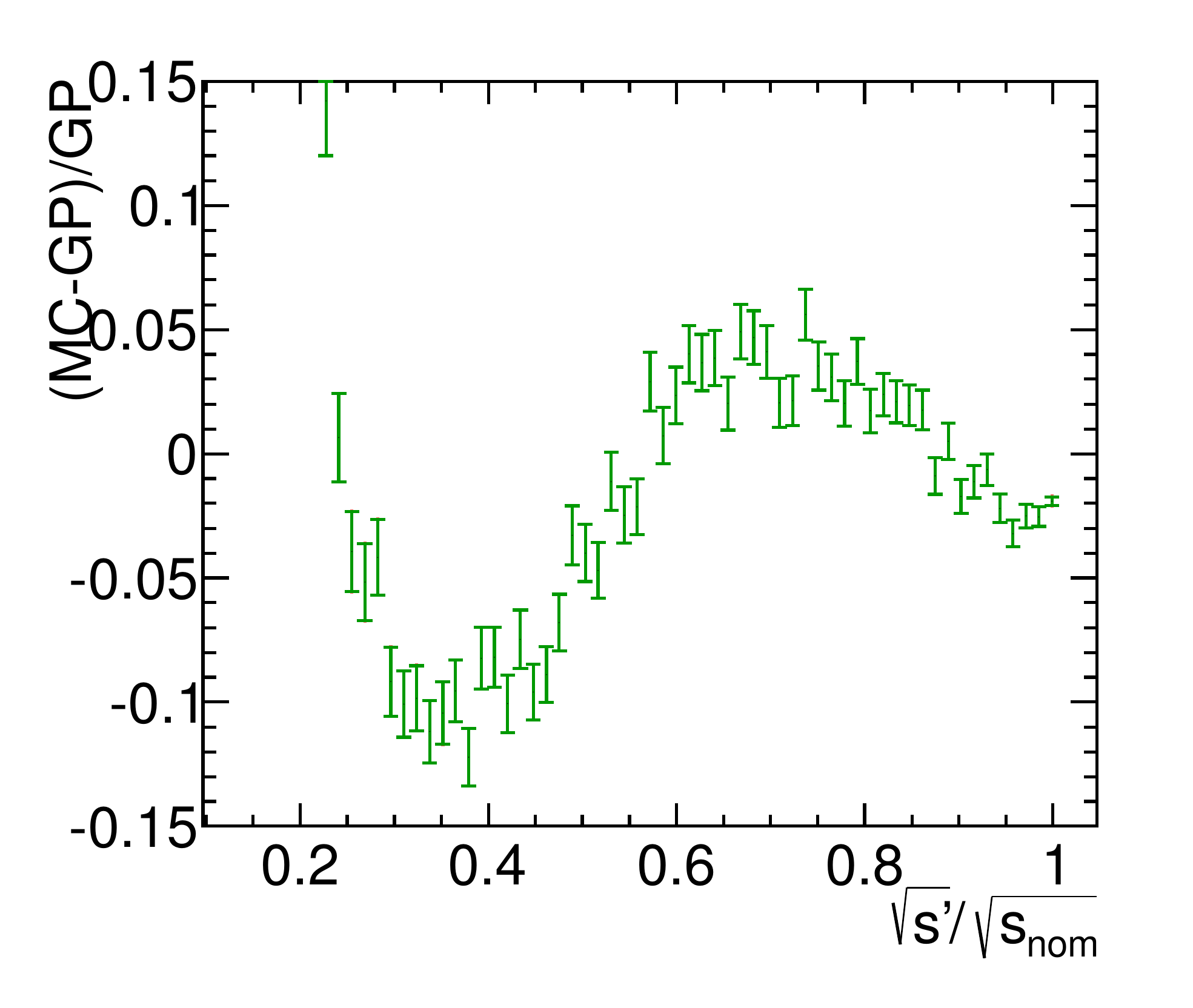}}
    \subfloat[20 GeV Bins]{\label{fig:BHWS_R50}\includegraphics[width=0.33\textwidth,clip,trim=0 5 0 12]{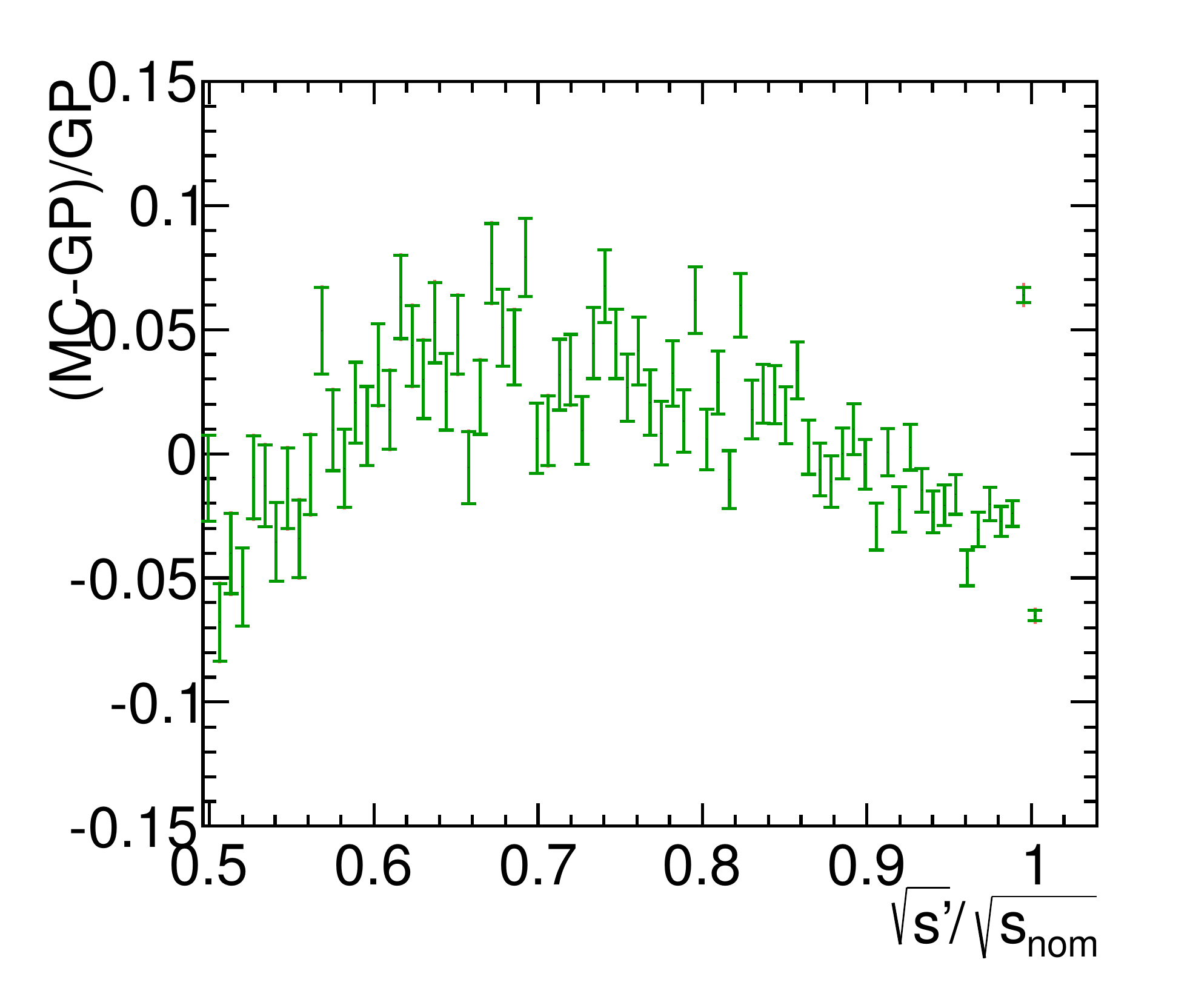}}
    \subfloat[2 GeV Bins]{\label{fig:BHWS_R90}\includegraphics [width=0.33\textwidth,clip,trim=0 5 0 12]{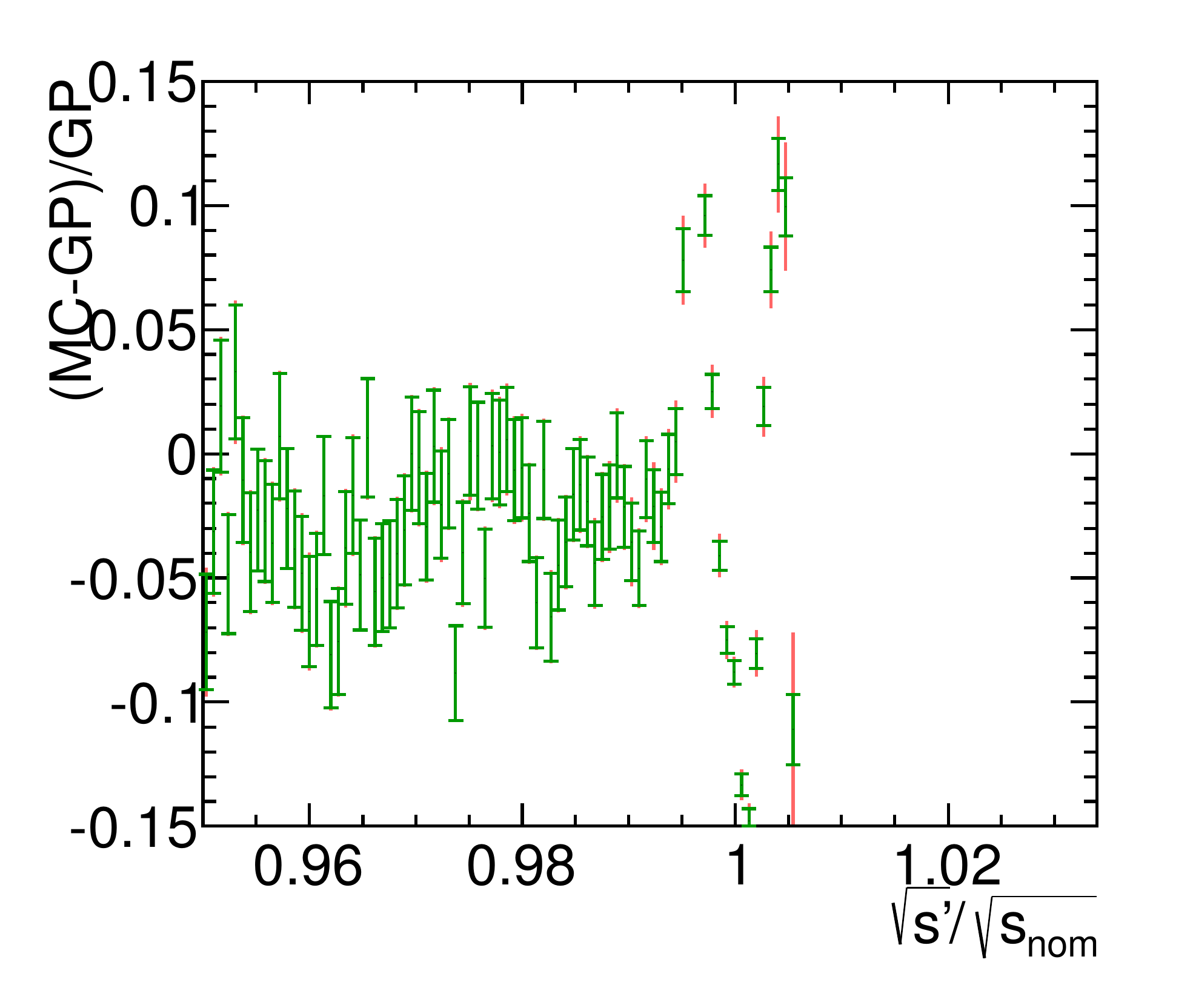}}
  \caption{Resulting spectra for the fit to the observables with the scaled
    luminosity spectrum, detector resolutions, and a binning of
    $40\times50\times50$. \Subref{fig:BHWS_S00}--\Subref{fig:BHWS_S90}~\spectraPhrase
    \Subref{fig:BHWS_R00}--\Subref{fig:BHWS_R90}~\ratioPhrase}
  \label{fig:fitBHWS}
\end{figure*}

\begin{figure*}[pbt]\sidecaption
  \includegraphics[width=0.8\textwidth]{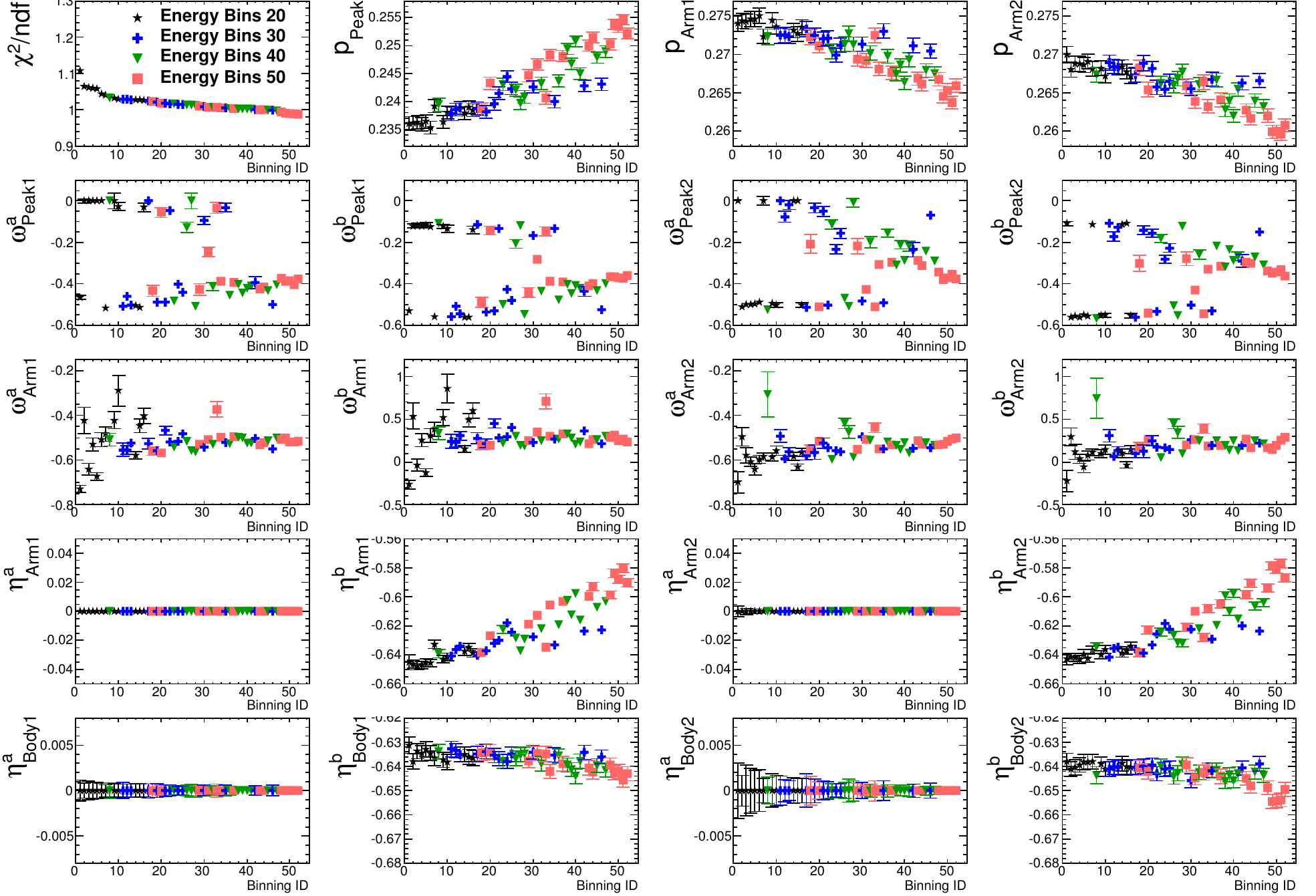}
  \caption{Parameter variation with respect to the Binning ID for the fits to
  the observables with the scaled luminosity spectrum and including detector
  resolutions. The entries are sorted by decreasing \chisquare and the colours and markers give
  the number of bins used for the energy observables.}\label{fig:parBHWS}
\end{figure*}%

Table~\ref{tab:summtop} lists the fraction of events with a centre-of-mass energy
larger than $0.99\rootsnom$ from \guineapig and from selected fits of the
different fit stages. The uncertainty of the \guineapig value is the statistical uncertainty
from one million events. The uncertainty for the fits is calculated from the
uncertainty of the individual parameters and accounts for the correlation between them. 

\begin{table}[tbp]
  \centering
  \caption{Summary of the fraction of events with $\rootsprime > 0.99\rootsnom$ from \guineapig
    and the reconstructed luminosity spectra from the different fit stages.}
  \label{tab:summtop}
  \begin{tabular}{l R{2}{2}@{$\pm$}R{1}{2}}\toprule
                                                    & \tabtt{Fraction [\%]} \\\midrule
    \guineapig sample                               & 35.4106 & 0.0595      \\
    Basic Luminosity Spectrum                       & 34.61   & 0.0108461   \\
    \hspace{1em}with Scaled Spectrum and Det.\ Res. & 34.7243 & 0.0714585   \\
    \bottomrule
  \end{tabular}
\end{table}

The difference of the fractions between \guineapig and the \model is less than
one percentage point. Given the size of the uncertainties the difference is
significant. However, processes with lower cross-section will effectively use
smaller samples from the luminosity increasing the uncertainty to around one
percentage point. The difference in the fraction of events in the top 1\% might
therefore be insignificant for other measurements at 3~TeV.

The basic luminosity spectrum from \guineapig compared with the reconstructed
basic luminosity spectra from the two fit stages for the selected fits are shown
in Figures~\ref{fig:fitgen} and \ref{fig:fitBHWS}. For the ratios the green
error bars show the statistical uncertainty for one million \guineapig events
and the barely visible red error bars show the uncertainty coming from the
parameterisation.

In both cases the luminosity spectrum is reconstructed within
5\% between $0.55\rootsnom$ and $0.995\rootsnom$.  Close to the peak, above $0.995\rootsnom$, the
beam-energy spread is the dominant effect and the difficulty of modelling this
peak becomes visible. Still, this difference is seen only, when looking at small
bin sizes (e.g., compare the bins around 1 in Figures~\ref{fig:genR00} or
\ref{fig:genR50} with Figure~\ref{fig:genR90}). As Table~\ref{tab:summtop}
shows, the average fraction around the peak is reconstructed within 1 percentage
point. Improved parameterisations should be able to better describe and
reconstruct the shape of the peak, at the cost of longer run-time for the fit.

Below $0.5\rootsnom$, the \model is much more inconsistent with \guineapig, but
this is given by the design of this \model and the cut on the centre-of-mass
energy applied for the fits.

Some of the reconstructed parameter values depend on the number of bins used in
the fit. Figure~\ref{fig:parSpreadGen} shows the dependence of the reconstructed
parameters on the number of bins used in the fit. Fits with a binning of
$50\times50$ bins to $300\times300$ bins with the same number of events were
done. In Figure~\ref{fig:parSpreadGen} the results are sorted by \chindf, or
increasing number of bins. The Binning ID corresponds to the number of bins.

The parameters \aarmAB and \aarmBB, which represent the upper edge of the
beam-energy spread of the \arms, show a significant dependence on the
binning. For the other parameters the change is below one sigma. It is also
visible that with more bins the parameter \pPeak rises, while the two
parameters \pArmA and \pArmB fall, which is also visible in the correlation
matrix and their correlation coefficient of about $-0.4$.

Figure~\ref{fig:parBHWS} shows the parameters obtained in the fit to the
observables. The results are again sorted by decreasing \chindf, which defines
the Binning ID\@. In the figure the different markers give the number of bins used
for the energy observables. As the \chindf falls with increasing number of bins
the larger the Binning ID the large is also the number of bins used for the relative
centre-of-mass energy observable.

The parameter values depend much stronger on the number of bins. This is mostly due to
the inclusion of the detector resolutions. Without a minimum number of bins
the peak structure cannot be resolved, and the \beamspreadpara{}{}-parameters are
completely different from the previous results and show large fluctuations in
their values. If a large enough number of bins is used, the results are only a
few sigma different from the previous fit results. The detector resolutions have
a strong impact on resolving the structure of the luminosity peak.


%% file: results.tex
\begin{table*}[tbp]\centering
  \caption{The parameter values found in selected fits to the initial electron
    and positron energies (first rows) and to the observables (second rows). The
    details of the fits are given in the text.}
  \label{tab:overlapResults}
  \begin{tabular}{*{4}{R{2}{4} @{~$\pm$~} R{1}{4}} } 
\toprule%
                  \tabtt{\chindf} &                    \tabtt{\pPeak} &                    \tabtt{\pArmA} &                    \tabtt{\pArmB} \\\midrule%
       \tabtt{\hspace{4.5pt}63832 /\hspace{5.75pt}10000} &        0.238731 &        0.000371 &        0.267209 &        0.000361 &        0.265935 &        0.000362  \\
       \tabtt{100593 /  100000} &        0.248286 &        0.001025 &        0.268107 &        0.000883 &        0.263190 &        0.000884  \\
\midrule[1pt]
                 \tabtt{\bPeakAA} &                  \tabtt{\bPeakAB} &                  \tabtt{\bPeakBA} &                  \tabtt{\bPeakBB} \\\midrule%
      -0.278775 &        0.001613 &       -0.342512 &        0.001319 &       -0.280546 &        0.001614 &       -0.341713 &        0.001316  \\
      -0.387927 &        0.014918 &       -0.388227 &        0.013480 &       -0.305774 &        0.017512 &       -0.328315 &        0.015342  \\
\midrule[1pt]%
                  \tabtt{\barmAA} &                   \tabtt{\barmAB} &                   \tabtt{\barmBA} &                   \tabtt{\barmBB} \\\midrule%
      -0.439940 &        0.001219 &        0.324319 &        0.003689 &       -0.439851 &        0.001200 &        0.336439 &        0.003594  \\
      -0.499447 &        0.010728 &        0.305437 &        0.030524 &       -0.550060 &        0.009844 &        0.184225 &        0.029151  \\
\midrule[1pt]%
                  \tabtt{\aarmAA} &                   \tabtt{\aarmAB} &                   \tabtt{\aarmBA} &                   \tabtt{\aarmBB} \\\midrule%
       0.000000 &        0.000818 &       -0.625326 &        0.001132 &        0.000000 &        0.000744 &       -0.626799 &        0.001134  \\
       0.000000 &        0.000349 &       -0.605403 &        0.002730 &        0.000000 &        0.000429 &       -0.607994 &        0.002772  \\
\midrule[1pt]%
                 \tabtt{\abodyAA} &                  \tabtt{\abodyAB} &                  \tabtt{\abodyBA} &                  \tabtt{\abodyBB} \\\midrule%
       0.000000 &        0.000194 &       -0.664036 &        0.001203 &        0.000000 &        0.000216 &       -0.663609 &        0.001194  \\
       0.000000 &        0.000365 &       -0.642061 &        0.002936 &        0.000000 &        0.000522 &       -0.641502 &        0.002923  \\
\bottomrule
\end{tabular}
\end{table*}


%% file: physperfs.tex
\section{Systematic Impact on Smuon Mass Measurement}\label{sec:physperfs}

There are significant differences between the reconstructed luminosity spectrum and the one
from \guineapig when looking at large event samples. Typical cross-sections for
New Physics phenomena will be much smaller than that of Bhabha scattering, and the
luminosity spectrum sampled for a specific process will therefore have larger
statistical fluctuations, so that the difference between the reconstructed and
actual spectrum might not be significant. To estimate the impact of the
difference between \guineapig and the reconstructed spectrum, the measurement of
the smuon mass \msmu and neutralino mass \mneutr from smuon pair production is used. 
In this model the masses are $\msmu=1011~\mathrm{GeV}$ and
$\mneutr=340~\mathrm{GeV}$.

The smuon decays into a muon and a neutralino, so that the energy spectrum of
the muons $f(\Emu)$ can be used to extract the smuon and neutralino masses. The
details of the analysis are described elsewhere~\cite{blaising_slepton_13}, here
only the parts directly concerning the systematic uncertainty from the luminosity
spectrum are repeated. There are some differences in the treatment of the
statistical uncertainty between the version of the fitting program used here,
and the one used in the original paper.

In an ideal situation -- with a single centre-of-mass energy \rootsnom~-- the muon
energy spectrum is a uniform distribution $\Uni{\Emu}$ with the boundaries~\cite{PhysRevD.49.2369}
\begin{equation}
  \label{eq:ehl}
  E_{\mathrm{H,L}} = \frac{\rootsnom}{4} \left( 1 - \frac{\mneutr^{2}}{\msmu^{2}} \right)
        \left( 1 \pm \sqrt{1 - 4 \frac{\msmu^{2}}{s_{\mathrm{nom}}}}\;\right).
\end{equation}
The uniform distribution therefore depends on the smuon and neutralino masses.

\begin{figure*}[tbp]\sidecaption
  \captionsetup[subfloat]{captionskip=0pt}
  \centering
  \subfloat[]{\label{fig:smufitEx}\includegraphics[width=\halfwidth]{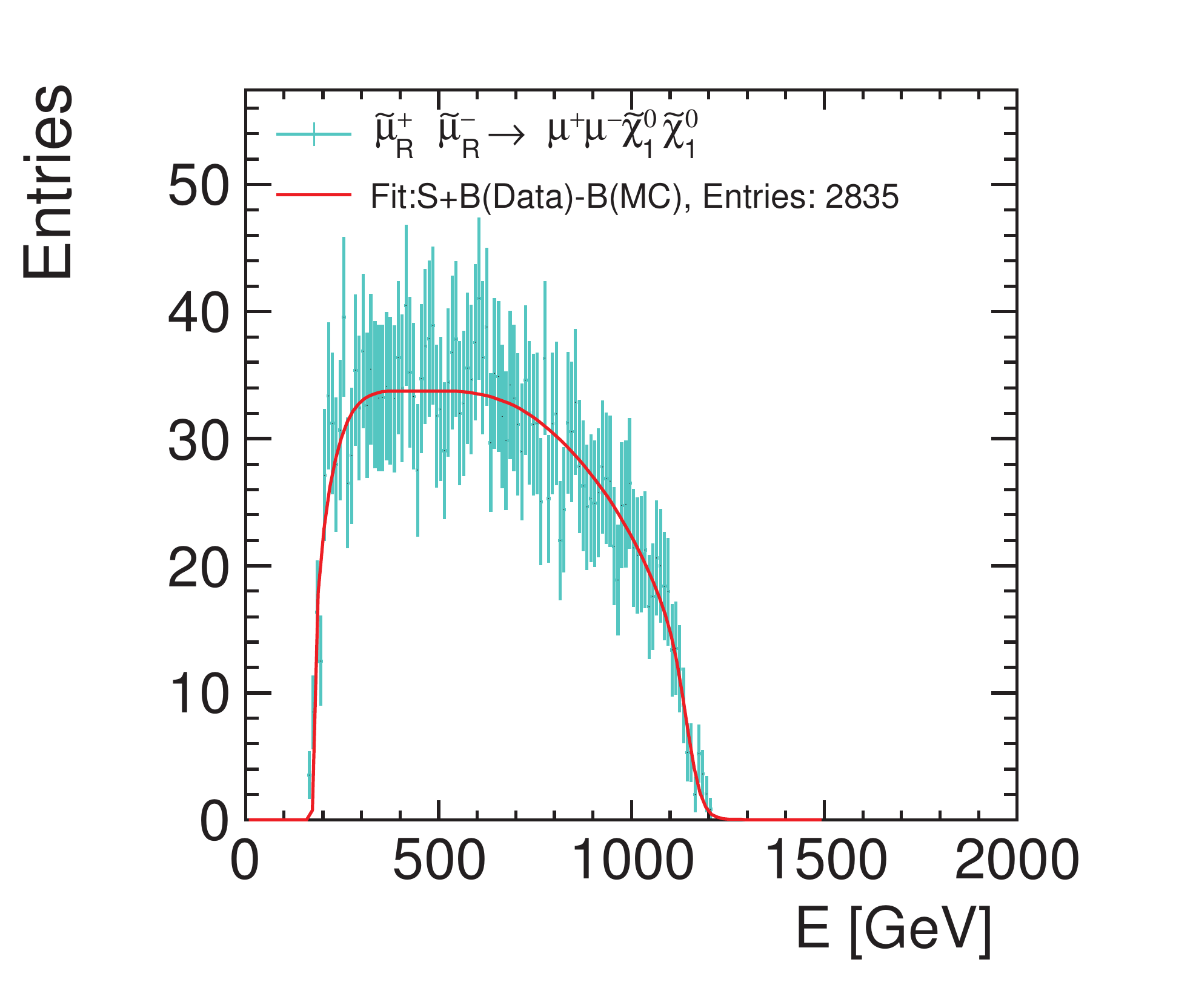}}
  \subfloat[]{\label{fig:smures}\includegraphics[width=\halfwidth]{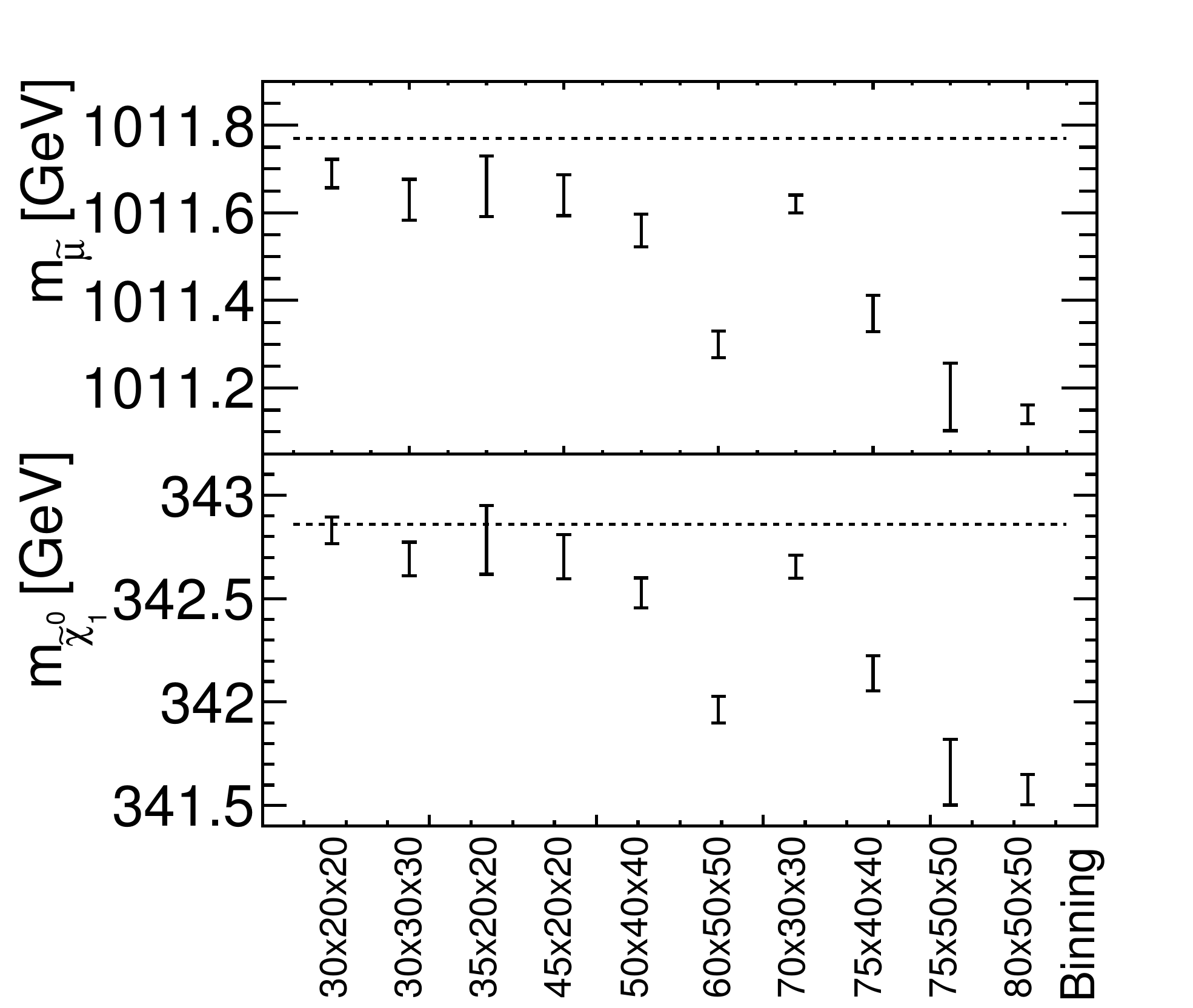}}
  \caption{\Subref{fig:smufitEx}~Background subtracted signal sample and the
    best fit to extract the smuon and neutralino mass.
    \Subref{fig:smures}~Reconstructed masses with the luminosity spectra taken
    from the fits to the Bhabha observables with different binnings. The dashed
    lines mark the result obtained with the \guineapig spectrum.}
\label{fig:smufit}
\end{figure*}

In reality, there is not a single centre-of-mass energy, and for every
centre-of-mass energy the uniform distribution has different limits. Therefore,
the measured muon-energy spectrum is affected by the basic luminosity spectrum,
the Initial State Radiation, the cross-section, and the detector
resolution \Det{\Emu}. The luminosity spectrum \lumispec{\xs}, Initial
State Radiation \ISR{\xs}, and cross-section \Sigmasmu{\roots} can be
combined into the number of events per centre-of-mass energy
\Neff{\xs}. The Initial State Radiation and
luminosity spectrum are convoluted and the resulting function is multiplied with
the smuon-pair production cross-section and with the total integrated luminosity
\Lint
\begin{equation}
  \label{eq:lEff}
  \Neff{\xs} = \Lint \cdot \Bigl(\lumispecNoArg\conv\mathrm{ISR}\Bigr)\bigl(\xs\bigr)\cdot \Sigmasmu{\xs\rootsnom}.
\end{equation}
The Initial State Radiation function $\mathrm{ISR}$ describes the distribution
of the energy after the radiation of initial state radiation. In this case, the
distribution is obtained from the Monte Carlo generator used to generate the
smuon events. It is here assumed to be independent of the nominal centre-of-mass
energy. The function to fit the muon energy spectrum is then the convolution of
the uniform energy spectrum with the detector resolution weighted by the respective number
of events
\begin{multline}
  \label{eq:smuFit}
  f(\Emu) = \int\limits_{0}^{\infty}\;\Neff{\xs} \cdot\\
  \int\limits_{\EL{\rootsprime}}^{\EH{\rootsprime}}
          \Uni{\msmu,\mneutr,\xs\rootsnom,\tau} \cdot \Det{\Emu-\tau}\;\dd{\tau}\;\dd{\xs}.
\end{multline}
Figure~\ref{fig:smufitEx} shows the background-subtracted signal sample and an
example fit with Equation~\eqref{eq:smuFit}. To estimate the impact of the
reconstruction, the fit results when the luminosity spectrum is taken directly from \guineapig
are compared with those, when the spectrum is coming from the reconstruction.

The masses extracted from the fit with Equation~\eqref{eq:smuFit} become a
function of the parameters $\vec{p}$ from the spectrum reconstruction $m =
m(\vec{p})$ with the luminosity spectrum reconstructed from the Bhabha
events. To estimate the systematic uncertainty due to the reconstruction of the
spectrum, the fit is performed with the nominal set of parameters $\vec{p}$ and
with each parameter $p_{i}$ increased or decreased by half of a standard
deviation $\sigma_{p_{i}}$
\begin{equation}
  \mplus{i} = m\left( \vec{p} + \vec{e_{i}}\frac{\sigma_{p_i}}{2} \right)\;,\qquad
  \mminus{i} = m\left( \vec{p} - \vec{e_{i}}\frac{\sigma_{p_i}}{2} \right).
\end{equation}
The systematic uncertainty on the fitted value is then given by
\begin{equation}
  \sigma_{m} = \left(\sum_{i,j} \delta_{i} C_{ij} \delta_{j}\right)^{1/2}
\end{equation}
with $\delta_{i} = \mplus{i} - \mminus{i}$, and the correlation matrix $C$.

Table~\ref{tab:smuFit} lists the smuon and neutralino masses from the fit when
the luminosity spectrum in equation (\ref{eq:lEff}) is directly taken from
\guineapig and when the luminosity spectrum is obtained from the reconstruction
with the observables with the scaled luminosity spectrum and detector
resolutions with a binning of $50\times40\times40$ bins. The difference in the
reconstructed masses for these two luminosity spectra is smaller than the
statistical uncertainty. However, as the reconstructed luminosity spectrum shows
a dependence on the binning, so do the reconstructed
masses. Figure~\ref{fig:smures} shows the reconstructed masses for the spectra
reconstructed with different binnings. There is a dependence of the
reconstructed masses on the number of bins, but the spread of the reconstructed
masses is smaller than the statistical uncertainty (cf.\ Table~\ref{tab:smuFit}).

As the difference between the obtained masses and the spread of masses is
smaller than the statistical uncertainty, the reconstruction of the luminosity
spectrum does not introduce a significant bias compared with the statistical
uncertainty. The systematic uncertainty due to the luminosity spectrum
reconstruction is also much smaller than the statistical uncertainty, so that
the total uncertainty on the reconstructed mass is not increased significantly.

\begin{table*}[tbp]
  \centering
  \caption{Extracted smuon and neutralino masses from the fits to the signal
    sample using different (effective) luminosity spectra.}
  \label{tab:smuFit}
  \begin{tabular}{l *2{R{4}{2} @{~$\pm$~} R{1}{2} R{1}{2}}}\toprule
    \tabt{Spectrum}                & \tabttt{Smuon Results [GeV]} & \tabttt{Neutralino Results [GeV]}                                                                                                                 \\\cmidrule(lr){2-4}\cmidrule(lr){5-7}
                                   & \tabt{Mass}                  & \tabt{$\sigma_\mathrm{Stat}$} & \tabt{$\sigma_{\mathrm{Syst}}$} & \tabt{Mass} & \tabt{$\sigma_{\mathrm{Stat}}$} & \tabt{$\sigma_{\mathrm{Syst}}$} \\\midrule
    \guineapig-spectrum            & 1011.77                      & 3.05                          & \tabempty                       & 342.86      & 6.98                            & \tabempty                       \\
    Fit $50\times40\times40$       & 1011.56                      & 3.05                          & $\pm$0.04                       & 342.528     & 6.82                            & $\pm$0.07                       \\
    \bottomrule
    \end{tabular}
\end{table*}


%% file: summary.tex
\section{Summary, Conclusions, and Outlook}
\label{sec:sco}

A framework has been developed for the reconstruction of the basic luminosity
spectrum at future linear colliders. The spectrum can be reconstructed from
Bhabha events measured with the tracking detectors and calorimeters. All
important effects were included: the 
luminosity spectrum from beam-beam simulations -- including the non-Gaussian CLIC beam-energy
spread -- the \rootsprime-dependence of the Bhabha cross-section, Initial and Final State
Radiation, and the detector resolutions.

The \model of the 3~TeV CLIC luminosity spectrum, required for the reweighting
fit, has some limitations. For technical reasons the energy range to describe
the tail of the Beamstrahlung is limited to $\rootsprime > 1500~\mathrm{GeV}$, and the
peculiar beam-energy spread cannot be modelled precisely with few
parameters. The reweighting fit itself does not impair the reconstructed
spectrum. The differences between \guineapig and the reconstructed spectrum do
not significantly change between the fit to the basic luminosity spectrum and
the fit to the observables with the scaled luminosity spectrum and including
detector resolutions. With an improved model, and increased processing power, an
improved reconstruction of the CLIC 3~TeV spectrum should be possible.

The fraction of events above 99\% of the nominal centre-of-mass energy is
reconstructed within 1 percentage point. The centre-of-mass energy distribution
is reconstructed to better than 5\% between the nominal and about half the nominal
centre-of-mass energy, the validity limit of our \model. These results are
obtained regardless of the included level of details, so that one can conclude
that the limitations of the \model cause most of the discrepancies to the
simulated spectrum, and if a better model is used, the discrepancies should be
reduced.

To estimate the systematic impact on other physics measurements, the
reconstructed spectrum was used in the study of smuon decays, one of the CLIC
3~TeV benchmark processes. The reconstructed spectrum does not
induce a significant bias on the measured mass, nor does it cause a significant
systematic uncertainty. The systematic uncertainty from the spectrum
reconstruction is two orders of magnitude smaller than the statistical uncertainty.

The spectrum is well enough reconstructed for the chosen physics channel. In
this case a good reconstruction of the tail of the spectrum is tested. The
reconstruction of the peak is less important, because the process is far above
threshold and the cross-section does not change significantly over the \emph{peak} region. More work
is needed to evaluate and possibly improve the reconstruction of the \emph{peak}.

\subsection{Outlook}
\label{sec:outl}

The framework can also be applied for the reconstruction of the luminosity
spectrum at other centre-of-mass energies and linear electron--positron
colliders than CLIC\@. Depending on the beam-energy spread and the demanded range of the
reconstruction, the \model has to be adapted, but this will not increase the
computational complexity of the reconstruction. 

The energy range of the current \model can be increased by replacing the single
Beamstrahlung beta-distributions by linear combination of
beta-distributions. Improving the description of the beam-energy spread is less
obvious without a large increase in the number of parameters. 

The boundaries of the beam-energy spread -- the parameters \xmin and \xmax\ --
were fixed during the fit. It should be evaluated how much the measurement is
affected, when these parameter values differ from those of the beam-energy
spread. It should also be tried to vary the boundaries of the beam-energy spread
during the re-weighting fit. For varying these parameters during the reweighting fit
the initial samples have to be produced with overlapping regions. For example, the \emph{peak}
region would be produced with an \xmin smaller than the upper limit of the \emph{arm}
or \emph{body} regions. During the re-weighting the value for \xmin or \xmax
would be given by the minimizer, and events in the \emph{peak} below \xmin or
above \xmax would be dropped, as would events in the \emph{arm} or \emph{body}
above their respective upper limit.

The observables from the Bhabha events can also be exchanged for other suitable
choices, always keeping the detector resolutions in mind. The impact of the
detector resolutions on the reconstructed spectrum can be easily studied by
changing the resolutions used in the four-vector smearing. The same detector
resolutions and Bhabha generator were used for the \guineapig and \model
events. Differences in the predicted detector resolution and Bhabha scattering
to the actual events can introduce systematic errors into the
reconstruction. These effects could be studied by varying the detector
resolutions or the Bhabha cross-section independently for the two samples used
in the fit.

Only Bhabha events -- and no other physics processes -- were considered. It
should be checked if multi-peripheral two-photon events, in which the spectator
electrons scatter at large angles, are a background.

As the luminosity spectrum depends on the accelerator, the impact of possible
variations of the beam parameter on the reconstruction of the luminosity
spectrum should be studied with realistic variations of the beam parameters.

\begin{acknowledgements}
We are grateful to Klaus M{\"o}nig for proposing the reweighting fit to
reconstruct the luminosity spectrum at future linear colliders; Barbara Dalena
for providing the input files for \guineapig and luminosity spectra; Daniel
Schulte for useful discussions about the peculiarities of the CLIC luminosity
spectrum; Jean-Jacques Blaising for the estimate of the detector resolutions
and for providing the fitting code for the smuon studies; and Konrad
Elsener for careful reading of this document. Our thanks also go to the members
of the FCal-collaboration and the LCD analysis working group for useful
discussions, inputs, and comments.
\end{acknowledgements}
